\documentclass[aps,prb,reprint,twocolumn,longbibliography]{revtex4-1}

\usepackage{amssymb,amsfonts,amsmath}
\usepackage{graphicx}
\usepackage{dcolumn}
\usepackage{rotating} 
\usepackage{changepage} 
\usepackage{color}      
\usepackage{MnSymbol}
\usepackage{mathtools}
\usepackage{enumitem}
\usepackage[normalem]{ulem}

\usepackage{xr} 
\AtBeginDocument{\nocite{achemso-control}}
\begin{document}

\title[A microscopic perspective on heterogeneous catalysis]
  {A microscopic perspective on heterogeneous catalysis}

\author{Miguel~A.~Gosalvez$^{1,2,3}$}
\email[]{miguelangel.gosalvez@ehu.es \\ \\ {\bf References:} }

\author{Joseba~Alberdi-Rodriguez$^{2}$}

\affiliation{$^1$
Dept. of Materials Physics, University of the Basque Country UPV/EHU, 20018 Donostia-San Sebasti\'an, Spain
}
\affiliation{$^2$
Donostia International Physics Center (DIPC), 20018 Donostia-San Sebasti\'an, Spain
}
\affiliation{$^3$
Centro de F\'isica de Materiales CFM-Materials Physics Center MPC, centro mixto CSIC -- UPV/EHU, 20018 Donostia-San Sebasti\'an, Spain
}

\date{\today}

\begin{abstract}
A general formalism is presented to describe the turnover frequency (TOF) during heterogeneous catalysis beyond a mean field treatment. For every elementary reaction we define its multiplicity as the number of times the reaction can be performed in  the current configuration of the catalyst surface, divided by the number of active sites. 
It is shown that any change in the multiplicity with temperature can be directly understood as a modification in configurational entropy. 
Based on this, we determine the probability of observing any particular elementary reaction,
leading to  a procedure for identifying any Rate Controlling Step (RCS) as well as the Rate Determining Step (RDS), if it exists. Furthermore, it is shown that such probabilities provide a thorough description of the overall catalytic activity, enabling a deep understanding of the relative importance of every elementary  reaction. Most importantly, we formulate a simple expression to describe accurately the apparent activation energy of the TOF, valid even when adsorbate-adsorbate interactions are included, and compare it to previous, approximate expressions, including the traditional Temkin formula for typical reaction mechanisms (Langmuir-Hinshelwood, Eley-Rideal, etc...). To illustrate the validity of our formalism beyond the mean field domain we present Kinetic Monte Carlo simulations for two widely-studied and industrially-relevant catalytic reactions, namely, the oxidation of CO on RuO$_2$(110) and the selective oxidation of NH$_{3}$ on the same catalyst. 
\end{abstract}

\maketitle

\noindent
{\bf Keywords}: apparent activation energy, rate determining step, degree of rate control, rate sensitivity, multiplicity, kinetic Monte Carlo, CO oxidation, NH$_3$ oxidation.

\section{Introduction}
\label{Introduction}

Enabling life through enzymatic acceleration of biochemical processes, catalytic reactions are also a key element of modern society, speeding up the production of a wide variety of chemical, pharmaceutical, petrochemical and fertilizing compounds. In a typical heterogeneous reaction, many elementary reactions continuously compete with each other at the catalyst surface. This includes elementary adsorption, desorption, diffusion and recombination reactions, with temperature dependent rate  constants, $k_{\alpha} \propto e^{-E_{\alpha}^{k}/k_{B}T}$, where $k_{B}$ is Boltzmann's constant, $T$ is the temperature and $E_{\alpha}^{k}$ is the activation energy for reaction $\alpha$. On the other hand, the turnover frequency ($TOF$) measures the  overall number of molecules of the product of interest generated per active site per unit time.  Interestingly, the $TOF$ typically increases with temperature according to an Arrhenius behavior, $TOF \propto e^{-E_{app}^{TOF}/k_{B}T}$, where $E_{app}^{TOF}$ is referred to as the {\em apparent activation energy}  ---usually constant within some temperature range. Thus, the overall catalytic reaction occurs as if a single reaction would be in control.

Traditionally, this is accounted for by considering every elementary reaction as an elementary step and the overall reaction as a {\em sequence} of such elementary  steps, assuming that the rate of one particular elementary reaction (say $\lambda$) is sufficiently low so that it acts as a bottleneck or rate-determining step (RDS) \cite{Stegelmann2009, Campbell2017, Chorkendorff2003, Lynggaard2004}. 
Based on this, traditional descriptions of surface reactions using standard models, such as the Langmuir-Hinshelwood mechanism---for reactions between two adsorbed molecules---lead to expressions for the $TOF$ in terms of the adsorbate coverages (See Section \ref{Apparent_activation_energy_in_the_Langmuir_Hinshelwood_model} of the Supporting Information for some examples).
For instance, if the recombination of the adsorbates, A and B, is the RDS, with rate $r_{\lambda}$ and rate constant $k_{\lambda}$,
one writes: $TOF = r_{\lambda} \approx k_{\lambda} \theta_A \theta_B$, where the coverage product $\theta_A \theta_B$ assumes A and B are highly mobile/freely intermix (random homogeneous mixing or mean field approximation).

By assuming Langmuir adsorption-desorption equilibria for all adspecies ($A$, $B$ and $AB$) and their gaseous counterparts ($A_{(g)}$, B$_{2(g)}$ and $AB_{(g)}$), the coverages are traditionally expressed in terms of the partial pressures ($p_A$, $p_B$ and $p_{AB}$):
$\theta_{A} = K_{A}p_{A} / D $, $\theta_{B} = \sqrt{K_{B}p_{B}} / D$ and $\theta_{AB} = K_{AB}p_{AB} / D$, with $D = 1 + K_{A}p_{A} + \sqrt{K_{B}p_{B}} + K_{AB}p_{AB}$.
Here, $K_{X} = k_{a}^{X} / k_{d}^{X} \propto e^{\Delta H_{X} /k_BT}$ is the equilibrium constant for the adsorption of $X$, with $\Delta H_{X} = E_{d}^{X} - E_{a}^{X}$ the formation enthalpy (or heat of adsorption) of $X$ \cite{Chorkendorff2003, Lynggaard2004, Bond2006}, and $E_{a}^{X}$ ($E_{d}^{X}$) is the activation barrier for adsorption (desorption) of $X$, with $k_{a}^{X} \propto e^{-E_a^{X} /k_BT}$ ($k_{d}^{X} \propto e^{-E_d^{X} /k_BT}$) the rate constant for adsorption (desorption). This leads to: $TOF \approx k_{\lambda} \theta_A \theta_B =  k_{\lambda} (K_{A}p_{A})(K_{B}p_{B})^{1/2} / D^2$, which is re-written as: $TOF = k_{\lambda} (K_{A}p_{A})^x (K_{B}p_{B})^y (K_{AB}p_{AB})^z $, where $x = \frac{ \partial \log TOF }{ \partial \log p_A }$, $y = \frac{ \partial \log TOF }{ \partial \log p_B }$ and $z = \frac{ \partial \log TOF }{ \partial \log p_{AB} }$ are the partial reaction orders, effectively transferring the original coverage dependence into them. Since the temperature dependence is $TOF \propto e^{-E_{app}^{TOF}/k_{B}T}$ while $k_{\lambda} \propto e^{-E_{\lambda}^{k} / k_B T}$ and $K_{X} \propto e^{\Delta H_{X} /k_BT}$, for $X = A, B, AB$, this gives rise to the familiar Temkin formula \cite{Chorkendorff2003, Lynggaard2004, Bond2005, Bond2006}:
\begin{eqnarray}
\hspace{-5mm}
E_{app}^{TOF} & = & E_{\lambda}^{k} - x \Delta H_{A} - y \Delta H_{B} - z \Delta H_{AB}. \label{E_app_temkin}
\end{eqnarray} 

Eq. \ref{E_app_temkin} provides a traditional explanation to the convoluted nature of $E_{app}^{TOF}$, departing from the activation barrier of the RDS, $E_{\lambda}^{k}$, due to a weighted sum of formation enthalpies with coverage-dependent reaction orders as weights.
Beyond the mean field treatment ($TOF = r_{\lambda} \approx k_{\lambda} \theta_A \theta_B$), Eq. \ref{E_app_temkin} remains valid in the presence of correlated configurations on the catalyst surface, since in this case one may still write $r_{\lambda} = k_{\lambda} \theta_A^{x'} \theta_B^{y'}$, which preserves the general form of Eq. \ref{E_app_temkin}. Once more, this transfers the details about the dependence on the spatial configuration (including any possible correlations) to the pressure-dependent reaction orders, thus diverting the focus from the actual surface configuration. Nevertheless, this has proved very useful in practice, since the reaction orders can be determined experimentally with relative ease. 

In this study, however, we stress the importance of considering the spatial structure of the surface, explicitly describing the presence of correlated configurations via an alternative formulation:
$TOF = k_{\lambda} M_{\lambda}$.
Here, the general phenomenological term $\theta_A^{x'} \theta_B^{y'}$ is replaced by the {\em multiplicity}, $M_{\lambda}$, 
which directly accounts for the actual number of locations where reaction $\lambda$ can be performed per active site. 
To our best knowledge, the presence of a quantity like $M_{\lambda}$ has been traditionally obviated, directly replacing it by simple/sophisticated functions of the coverages and, correspondingly, of the pressures through Langmuir-type adsorption equilibria. However, here we assign $M_{\lambda}$ a central role, directly relating it to configurational entropy in Section \ref{Theory}.  
Amongst other benefits, the use of the multiplicity enables an alternative description of the complex behavior of $E_{app}^{TOF}$.

Turning away from Eq. \ref{E_app_temkin}, $E_{app}^{TOF}$ is sometimes attributed to (i) the elementary  reaction with the largest activation energy (slowest rate constant), $E_{app}^{TOF}=\{E_{\alpha}^{k}\}_{max}$, or (ii) the activation energy of the bottleneck itself (slow enough rate constant), $E_{app}^{TOF}=E_{\lambda}^{k}$, without any modifying contribution in either case. The idea that $E_{app}^{TOF}$ corresponds to the largest $E_{\alpha}^{k}$ contradicts careful computational studies outside the mean field formulation, where $E_{app}^{TOF}$ deviates (usually by large) from any of the $E_{\alpha}^{k}$'s present in the system \cite{Meskine2009, Hess2017}. To describe the surface anisotropy and lateral interactions outside the mean field treatment, those studies use the Kinetic Monte Carlo (KMC) method \cite{Pogodin2016, Reuter2006, Voter2007, Chatterjee2007, Reuter2011, Jansen2012, Temel2007, Gosalvez2017, Shah2014}. By accounting for fluctuations, correlations and the spatial distribution of the reaction intermediates--even including adsorbate clustering/islanding intrinsically--KMC provides a thorough picture of the ongoing competition between the various elementary reactions, whose modeling within a rate-equation approach would be rather complex. 
Within this framework, detailed consideration of the {\em degree of rate sensitivity\cite{Meskine2009}   } ($\xi_{\alpha}^{} $), originally referred to as the {\em rate sensitivity} \cite{Dumesic2008}  , concludes that $E_{app}^{TOF}$ can be formally described as an {\em average} over all forward and backward elementary activation energies \cite{Meskine2009}:
\vspace{-1mm}
\begin{eqnarray}
E_{app}^{TOF} & = & \Sigma_{\alpha} \xi_{\alpha}^{} E_{\alpha}^{k}  , \label{E_app_using_sensitivities}
\end{eqnarray}
where $\xi_{\alpha}^{} = \frac{k_{\alpha}}{TOF} \frac{\partial TOF}{\partial k_{\alpha}} |_{ k_{\alpha' \ne \alpha } } $ and the partial derivative with respect to rate constant $k_{\alpha}$ is taken by keeping fixed all other rate constants $k_{\alpha' \ne \alpha}$.
In fact, a closely related quantity, the {\em degree of rate control} ($\chi_{\alpha_*}^{}=\xi_{\alpha_+}^{} + \xi_{\alpha_-}^{}$, where $\alpha_*$ designates the combined forward-and-backward reaction) has been successfully and repeatedly used in many systems to identify (i) the RDS, which is defined as the elementary reaction for which $\chi_{\alpha_*}^{}=1$, if it exists, and (ii) the Rate Controlling Steps (RCSs), which are defined as those elementary reactions for which $\chi_{\alpha_*}^{}$ significantly departs from 0 \cite{Lynggaard2004, Meskine2009, Hess2017, Campbell1994}.
Furthermore, a combined analysis of both $\chi_{\alpha_*}^{}$ and $\xi_{\alpha}^{}$ provides crucial knowledge on the relative importance of the various elementary reactions \cite{Meskine2009, Hess2017}, giving valuable guidance as to which reactions need to be determined with higher accuracy \cite{Choksi-Greeley2016, Zhao-Greeley2017}. 

In practice, however, the determination of $\chi_{\alpha_*}^{}$ and $\xi_{\alpha}^{}$ outside a mean field formulation requires a formidable effort \cite{Meskine2009, Hess2017}. Not only these quantities form a high-dimensional space, but every value needs to be determined by carefully analyzing the numerical derivative of the $TOF$ for various values of $k_{\alpha}$, while every $TOF$ value must be obtained by averaging over several stochastic KMC simulations after reaching the steady state, which in turn is achieved at the long time limit on computationally inefficient stiff systems (where some reactions are executed many orders of magnitude less frequently than others). Thus, in practice the description of $E_{app}^{TOF}$ by Eq. \ref{E_app_using_sensitivities} is time-consuming and relatively inaccurate (see the Discussion for details). Indeed, the computational effort required to determine $\chi_{\alpha_*}^{}$ and $\xi_{\alpha}^{}$  is so large that alternative 'practical approaches' are being sought \cite{Hoffmann2017}. 
\begin{figure*}[htb!]
  \begin{center}
    \includegraphics[width=1.6\columnwidth]{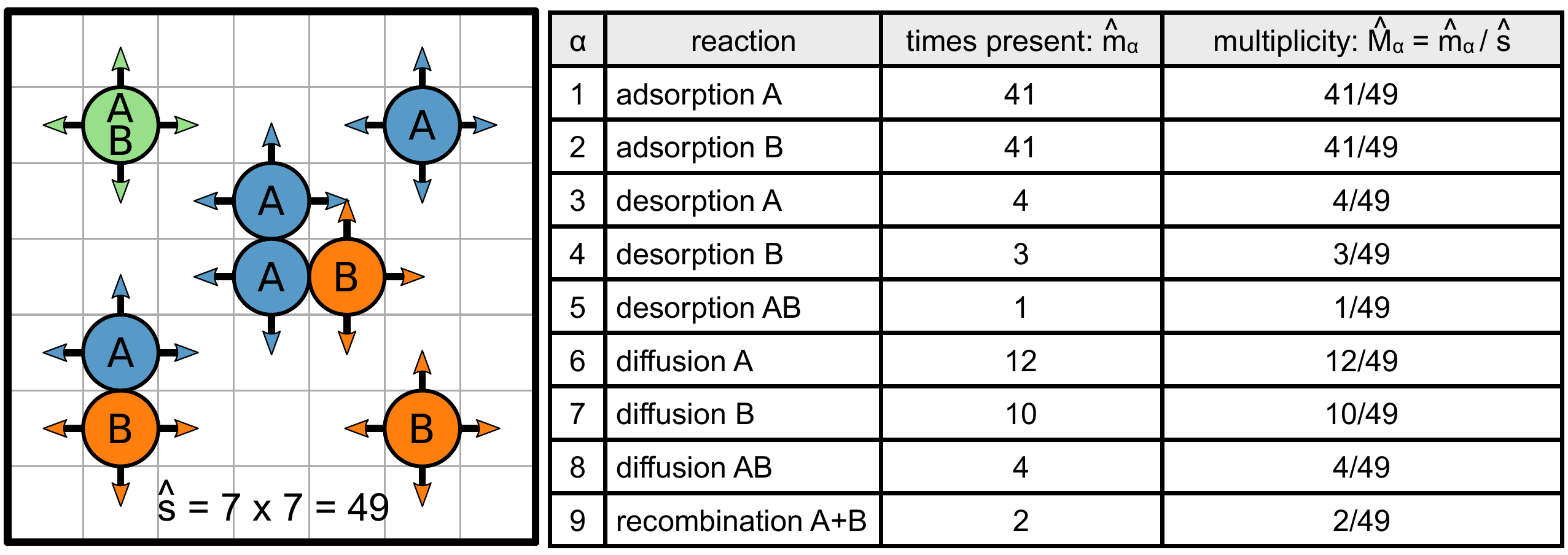}
  \end{center}
  \caption{
  Simplistic example of an instantaneous configuration on a catalist, showing the instantaneous multiplicity values for a reaction mechanism with nine elementary reactions. Diffusion and recombination are limited to nearest neighbor active sites. The system has $\hat{s} = 7 \times 7 = 49$ sites.}
  \label{Figure01}
\end{figure*}

In addition, Eq. \ref{E_app_using_sensitivities} does not formally fit the requirements of a {\it weighted average}. Although the sensitivities sum one ($\Sigma_{\alpha} \xi_{\alpha}^{} = \Sigma_{\alpha_*} \chi_{\alpha_*}^{} = 1$,  see Ref. \cite{Campbell2017}), they are unbounded (taking any possible value: positive, negative or zero) \cite{Meskine2009}. 
While this is a valuable feature for sensitivity analysis, with positive (negative) values denoting promotion (hindering) of the TOF, a problem appears when $\xi_{\alpha}^{}$ and $\chi_{\alpha_*}^{}$ are used effectively as weights to describe the most dominant contributions to the apparent activation energy, as in Eq. \ref{E_app_using_sensitivities} for the case of $\xi_{\alpha}^{}$ (or Eq. \ref{E_app_jorgensen_3} below, for $\chi_{\alpha_*}^{}$). Mathematically, the weights in a weighted average are probabilities and, thus, they should be non-negative, between 0 and 1. This enables a simple interpretation of the dominant/vanishing contributions. From the perspective that an average is a middle value, negative weights may lead to a result outside the range of the data, in which case one will be confronted with a linear combination, not a weighted average. Unfortunately, linear combinations in general, and Eqs. \ref{E_app_using_sensitivities} and \ref{E_app_jorgensen_3} in particular, are not the most suitable approach to describe dominance. If one truly wishes to find out which elementary reactions have a dominant role, then the weights need to be positive and, thus, $\xi_{\alpha}^{}$ and $\chi_{\alpha_*}^{}$ need to be reconsidered.

Given such limitations in the use of Eqs. \ref{E_app_temkin} and \ref{E_app_using_sensitivities}, we propose a different approach to analyze heterogeneous catalytic reactions in general. 
Simply stated, we present the idea that, at any given instant, every elementary reaction occurring on a catalyst can be performed at different locations and, thus, every elementary reaction has an associated multiplicity.
In this manner, while tradition considers the adsorbate coverages as the natural (irreducible) variables required to describe the evolution of the system, we put forward the idea that it is the collection of these multiplicities--for each and every elementary reaction--that provides the natural description of the configurational structure of the surface and, thus, the evolution of the system. 

Compared to Eq. \ref{E_app_using_sensitivities}, explicit use of the multiplicities provides access to an alternative, more accurate weighted average for $E_{app}^{TOF}$ (Eq. \ref{EappTOF_5} below). The new expression is both simpler to use in practice and theoretically robust, incorporating always-positive-and-properly-normalized probabilities as weights. Compared to Eq. \ref{E_app_temkin}, when a RDS exists, the corresponding new expression (Eq. \ref{EappTOF_6} below) describes how the elementary activation energy of the RDS, $E_{\lambda}^{k}$, contributes to $E_{app}^{TOF}$ with a modified value due to changes in configurational entropy, remaining valid even when adsorbate-adsorbate interactions are taken into account. Furthermore, we show below that the proposed multiplicities also provide an alternative route in order to determine the RDS as well as the sensitivity  of the $TOF$ to the different elementary reactions. In this manner, the proposed multiplicities enable an alternative perspective for the analysis of heterogeneous catalysis in general. 

We finally stress that, for other surface processes, such as two-dimensional epitaxial growth and three-dimensional anisotropic etching, the origin of the apparent activation energy has been previously explained via similar multiplicity-based formulations \cite{Gosalvez2017, Gosalvez_Ea_NJP}.

\section{Theory}
\label{Theory}

\subsection{Multiplicity of an elementary reaction}
\label{Multiplicity of an elementary reaction}

Let us consider a general heterogeneous catalytic system evolving in time. The system consists of a surface with a number of active sites as well as various adsorbates and their respective gases, all of them acting as reactants/products in a complex network of elementary reactions. Starting from a given initial configuration, the system evolves in time and currently, at time $t$, it displays some specific configuration. Note that $t$ denotes any instant along the initial transient or during the final steady state.

In this context, elementary reaction $\alpha$ (with rate constant $k_{\alpha}$) is associated an instantaneous multiplicity, $\hat{M}_{\alpha} = \hat{m}_{\alpha} / \hat{s}$, which denotes the number of times the reaction can be performed in the current configuration, $\hat{m}_{\alpha}$, divided by the number of active sites, $\hat{s}$ (see Fig. \ref{Figure01}). In other words, the instantaneous multiplicity describes the number of locations where the elementary reaction can occur (at the current instant and per active site), {\it i.e.} the actual abundance of the reaction per active site. Beyond the simplistic, periodic array of active sites depicted in Fig. \ref{Figure01}, the proposed multiplicity remains valid for more general scenarios, {\it e.g.} for randomly distributed active sites on a complex, three dimensional support. 

Although the number of active sites $\hat{s}$ typically remains constant, the value of $\hat{m}_{\alpha}$ (and, thus, that of $\hat{M}_{\alpha}$) changes dynamically as new configurations of the surface are visited during the transient, eventually settling down to some value and fluctuating around it at the steady state. In this context, the average value of any instantaneous variable $\hat{A}$ 
is defined as $A = \langle \overline{ \hat{A} } \rangle$, where $\overline{X} = \frac{\int \hat{X} dt}{\int dt} = \frac{\Sigma_{n} \hat{X} \Delta t_{n}}{\Sigma_{n} \Delta t_{n}}$ is the time average of $X$, and $\langle X \rangle$ is the mean value of $X$ for a total of $K$ evolutions from the initial state, in the limit of large $K$. 
Below, we focus on performing the time average $\overline{X}$ within the steady state, since most catalytic systems are of interest in that condition. In addition, it is implied below that any variable not preceded by the word 'instantaneous' and/or not displayed with the 'hat' symbol (\^{}) is either a constant or it designates the steady-state average value, even if the word 'average' is not mentioned.
While the instantaneous values (such as $\hat{M}_{\alpha}$, $\hat{m}_{\alpha}$ or $\hat{s}$) apply to a particular configuration of the system, the steady-state averages (such as $M_{\alpha}$, $m_{\alpha}$ or $s$) describe features of the macroscopic state (or 'average' configuration).

\newcount\colveccount
\newcommand*\colvec[1]{
        \global\colveccount#1
        \begin{smallmatrix} 
        \colvecnext
}
\def\colvecnext#1{
        #1
        \global\advance\colveccount-1
        \ifnum\colveccount>0
                \\
                \expandafter\colvecnext
        \else
                \end{smallmatrix} 
        \fi
}

\subsection{Rate equations and master equation}
\label{Rate equations and master equation}

Typical rate equations in heterogeneous catalysis describe the time evolution of the coverage for every adsorbate ($\theta_{X}$) in terms of (i) the coverage of the other adsorbates ($\theta_{Y}$) and (ii) the rate constants of the elementary reactions where $\theta_{X}$ is modified. For instance, for a standard Langmuir-Hinshelwood mechanism,
\begin{equation}
\hspace{-8mm}
\begin{tabular}{ccccc}
$A_{(g)}$ & & $\frac{1}{2} {B_2}_{(g)}$ & & $C_{(g)}$ \\
${\scriptstyle \colvec{2}{k_{d}^{A}}{[R1]} } \updownarrows {\scriptstyle  \colvec{2}{k_{a}^{A}p_{A}}{[R2]} }$ & & ${\scriptstyle \colvec{2}{k_{d}^{B}}{[R3]} } \updownarrows {\scriptstyle \colvec{2}{k_{a}^{B}p_{B}}{[R4]} }$ & & ${\scriptstyle \colvec{2}{k_{d}^{C}}{[R5]} } \updownarrows {\scriptstyle \colvec{2}{k_{a}^{C}p_{C}}{[R6]} }$ \\
$A$ & $+$ & $B$ & $\xrightarrow[RDS]{k_{r} [R7]}$ & $C$ , 
\label{LH_eq}
\end{tabular}
\end{equation}
where the irreversible reaction between adsorbates $A$ and $B$ is considered as the Rate Determining Step (RDS), the rate equations are:
\begin{eqnarray}
\hspace{-5mm}
\theta_{*} & = & 1 - \theta_{A} - \theta_{B} - \theta_{C} \label{Eq_LH_1_e1} \\ 
\scriptstyle \frac{ d \theta_{A} }{ dt } & = & p_{A} k_{a}^{A} \theta_{*} - k_{d}^{A} \theta_{A} - k_{r} \theta_{A} \theta_{B}  \label{Eq_LH_1_e2} \\ 
\scriptstyle \frac{ d \theta_{B} }{ dt } & = & p_{B} k_{a}^{B} \theta_{*}^2 - k_{d}^{B} \theta_{B}^2 - k_{r} \theta_{A} \theta_{B}  \label{Eq_LH_1_e3} \\ 
\scriptstyle \frac{ d \theta_{C} }{ dt } & = & p_{C} k_{a}^{C} \theta_{*} - k_{d}^{C} \theta_{C} + k_{r} \theta_{A} \theta_{B}  . \label{Eq_LH_1_e4}
\end{eqnarray}
Here, $\theta_{*}$ is the coverage by all the empty sites while $\theta_{*}^2$ is the coverage by all empty site pairs (in the mean field approximation) and $\theta_{A} \theta_{B}$ is the coverage by all site pairs occupied by $A$ and $B$ (also under random mixing). 

In this study, however, we stress the view that the rate equations can be written in terms of the multiplicities. 
For systems with a spatial representation (an important feature for the study of correlations beyond mean field), this seems more natural. Not only one has direct access to the multiplicities themselves, as shown below, but also the resulting equations remain valid beyond the mean field picture.

For this purpose, let us consider a spatial representation of a catalytic system evolving according to the reactions in Eq. \ref{LH_eq}:
\vspace{-2mm}
\begin{equation}
\begin{tabular}{ccccccccc}
  \scriptsize $\{s_i\}$ & & &  &     & \scriptsize $\{s_j\}$ &  &  &   \\ 
* & * & * & * &     & * & * & * & *  \\ 
* & A & B & * &     & * & C & * & *  \\ 
A & A & A & B &   $\rightarrow$    & A & A & A & B  \\ 
* & A & B & * &     & * & A & B & *  \\ 
* & * & B & * &     & * & * & B & *   
\end{tabular}
\label{Eq_drawing}
\end{equation}
Here, configuration $\{s_i\}$ has changed into configuration $\{s_j\}$ due to the elementary reaction $A +B \rightarrow C +*$ (under the assumption that $C$ always replaces $A$ and $*$ replaces $B$; the reverse leads to simple modifications). Traditionally, the time evolution of the system is described by the master equation:
\begin{equation}
\hspace{-6mm}
\frac{ d p_{ \{s_i\} } }{ dt } =  \sum_{ \{s_j\} } k_{ \{s_j\} \rightarrow \{s_i\} } p_{ \{s_j\} } - \sum_{ \{s_j\} } k_{ \{s_i\} \rightarrow \{s_j\} } p_{ \{s_i\} } , 
 \label{Eq_LH_4}
\end{equation}
where $p_{ \{s_i\} }$ is the probability to observe configuration $\{s_i\}$ at time $t$ and $k_{ \{s_j\} \rightarrow \{s_i\} }$ is the transition rate (= rate constant) for the elementary reaction that transforms $\{s_i\}$ into $\{s_j\}$. 

Because an elementary reaction can be performed only if the correct local configuration of the adsorbates and/or empty sites is present on one or more locations in the current configuration, the multiplicity of an elementary reaction corresponds to the multiplicity of that particular local configuration of the adsorbates and empty sites. Thus, for any given configuration $\{s_i\}$, we consider the instantaneous multiplicity of {\em local configuration} $\{l\}$, $\hat{M}_{ \{l\} } =  \hat{m}_{ \{l\} } / \hat{s} $, where $\hat{s}$ is the number of active sites (as before) and $\hat{m}_{\{l\}}$ is the number of times the local configuration $\{l\}$ appears on $\{s_i\}$. Here, $\{l\} = \{A,B,C,...,Z\}$ refers to any collection of sites, such that one site is occupied by adsorbate $A$, which has a neighbor site occupied by adsorbate $B$, which in turn has a neighbor site occupied by adsorbate $C$ and so on. Thus, local configurations $\{A,B,C,...,Z\}$ and $\{Z,...,C,B,A\}$ are the same, and empty sites are included by using the symbol $*$. This way, $\hat{M}_{ \{X\} }$ refers to the instantaneous coverage by adsorbate $X$ while $\hat{M}_{ \{X,Y\} }$ and $\hat{M}_{ \{X,Y,Z\} }$ indicate the instantaneous concentration of adsorbate pairs and adsorbate trios (per active site), respectively. Note that $\sum_{ \{X\} } \hat{M}_{ \{X\} } = 1$ and, similarly, $\sum_{ \{X,Y\} } \hat{M}_{ \{X,Y\} } = b$, $\sum_{ \{X,Y,Z\} } \hat{M}_{ \{X,Y,Z\} } = c$, $\sum_{ \{S,T,U,V\} } \hat{M}_{ \{S,T,U,V\} } = d$,..., where $b$, $c$, $d$,... depend on the actual spatial representation (see some specific values below). Eventually, the focus is on monitoring the multiplicities of the elementary reactions, $\hat{M}_{\alpha}$, each corresponding to a particular $\hat{M}_{ \{l\} }$.

Any change in the spatial configuration $\{s_i\}$ of the system due to an elementary reaction leads to modifications in the multiplicities. For instance, considering the system of Eq. \ref{Eq_drawing} and restricting the formation of neighbor pairs to the (periodic) horizontal and vertical directions, the multiplicities of the seven elementary reactions in Eq. \ref{LH_eq} ($\alpha = R1, R2,..., R7$) have changed as follows: $\hat{M}_{R1} \equiv \hat{M}_{\{A\}} = 5 \rightarrow 4 $, $\hat{M}_{R2} \equiv \hat{M}_{\{*\}} = 11 \rightarrow 12 $, $\hat{M}_{R3} \equiv \hat{M}_{\{B,B\}} = 1 \rightarrow 1 $, $\hat{M}_{R4} \equiv \hat{M}_{\{*,*\}} = 15 \rightarrow 17 $, $\hat{M}_{R5} \equiv \hat{M}_{\{C\}} = 0 \rightarrow 1 $, $\hat{M}_{R6} \equiv \hat{M}_{\{*\}} = 11 \rightarrow 12 $, $\hat{M}_{R7} \equiv \hat{M}_{\{A,B\}} = 6 \rightarrow 4 $. 
Although we may monitor many other local configurations ({\it e.g.} $\hat{M}_{\{*,A\}} = 6 \rightarrow 5 $, $\hat{M}_{\{*,B\}} = 8 \rightarrow 6 $, $\hat{M}_{\{A,B,*\}} = 3 \rightarrow 1 $, $\hat{M}_{\{B,B,A\}} = 1 \rightarrow 1 $,...), it is important to realize that none of these is strictly required to determine the $\hat{M}_{\alpha}$'s, since these can be directly obtained from the spatial configuration itself.

The previous definitions allow rewriting Eqs. \ref{Eq_LH_1_e2}-\ref{Eq_LH_1_e4} as:
\begin{eqnarray}
\hspace{-12mm}
\scriptstyle \frac{ d \hat{M}_{ \{A\} } }{ dt } &=& p_{A} k_{a}^{A} \hat{M}_{ \{*\} } - k_{d}^{A} \hat{M}_{ \{A\} } - k_{r} \hat{M}_{ \{A,B\} }  
\label{Eq_LH_2_e2} \\
\hspace{-12mm}
\scriptstyle \frac{ d \hat{M}_{ \{B\} } }{ dt } &=& p_{B} k_{a}^{B} \hat{M}_{ \{*,*\} } - k_{d}^{B} \hat{M}_{ \{B,B\} } - k_{r} \hat{M}_{ \{A,B\} }  
\label{Eq_LH_2_e3} \\
\hspace{-12mm}
\scriptstyle \frac{ d \hat{M}_{ \{C\} } }{ dt } &=& p_{C} k_{a}^{C} \hat{M}_{ \{*\} } - k_{d}^{C} \hat{M}_{ \{C\} } + k_{r} \hat{M}_{ \{A,B\} } . 
 \label{Eq_LH_2_e4}
\end{eqnarray}
The corresponding equation for $\frac{d\hat{M}_{\{*\}}}{dt}$ is redundant, since $\sum_{ \{X\} } \hat{M}_{ \{X\} } = 1$. Note that, in general, $\frac{d\hat{M}_{\{X\}}}{dt}$ depends on $\hat{M}_{\{U,V\}}$. Thus, these equations need to be completed by rate equations for $\frac{d\hat{M}_{\{U,V\}}}{dt}$:
\begin{eqnarray}
\hspace{-12mm}
\scriptstyle \frac{ d \hat{M}_{ \{A,B\} } }{ dt } &=& p_{A} k_{a}^{A} \hat{M}_{ \{*,B\} } + p_{B} k_{a}^{B} \hat{M}_{ \{*,A\} } \\ 
&& - k_{d}^{A} \hat{M}_{ \{A,B\} } - k_{d}^{B} \hat{M}_{ \{A,B,B\} } - k_{r} \hat{M}_{ \{A,B\} }  \nonumber 
\label{Eq_LH_3_e1} \\
\hspace{-12mm}
\scriptstyle \frac{ d \hat{M}_{ \{*,*\} } }{ dt } &=& - p_{A} k_{a}^{A} \hat{M}_{ \{*,*\} } - p_{B} k_{a}^{B} \hat{M}_{ \{*,*\} } \\ 
&& - p_{C} k_{a}^{C} \hat{M}_{ \{*,*\} } + k_{d}^{A} \hat{M}_{ \{*,A\} } \nonumber \\
&& + k_{d}^{B} \hat{M}_{ \{*,B,B\} } + k_{d}^{C} \hat{M}_{ \{*,C\} } \nonumber \\ 
&& + k_{r} \hat{M}_{ \{A,B,*\} } \nonumber
\label{Eq_LH_3_e2} \\
\hspace{-12mm}
\scriptstyle \frac{ d \hat{M}_{ \{B,B\} } }{ dt } &=& p_{B} k_{a}^{B} \hat{M}_{ \{*,*\} } - k_{d}^{B} \hat{M}_{ \{B,B\} } \\ 
&& - k_{r} \hat{M}_{ \{B,B,A\} } , \nonumber
 \label{Eq_LH_3_e3}
\end{eqnarray}
and similar equations for the derivatives of $\hat{M}_{ \{A,C\} }$, $\hat{M}_{ \{B,C\} }$, $\hat{M}_{ \{C,C\} }$, $\hat{M}_{ \{A,A\} }$, $\hat{M}_{ \{*,A\} }$, $\hat{M}_{ \{*,B\} }$ and $\hat{M}_{ \{*,C\} }$, with one of them redundant, since $\sum_{ \{X,Y\} } \hat{M}_{ \{X,Y\} } = 2$ in this example. 
As before, $\frac{d\hat{M}_{\{X,Y\}}}{dt}$ depends on $\hat{M}_{\{U,V,W\}}$. Thus, additional equations are written for $\frac{d\hat{M}_{\{U,V,W\}}}{dt}$ (with $\sum_{ \{X,Y,Z\} } \hat{M}_{ \{X,Y,Z\} } = 6$), and for $\frac{d\hat{M}_{\{P,Q,R,S\}}}{dt}$ (with $\sum_{ \{S,T,U,V\} } \hat{M}_{ \{S,T,U,V\} } = 36$) and so on.

Accordingly, for a general reaction mechanism, containing elementary reactions of different types, including adsorption ($a$), desorption ($d$), diffusion ($h$) and recombination ($r$), the generic rate equation for $\hat{M}_{ \{l_i\} }$ is:
\begin{eqnarray}
\hspace{-12mm}
\scriptstyle \frac{ d \hat{M}_{ \{l_i\} } }{ dt } &=&  \sum_{g=a,d,h,r} \sum_{ \{l_j\} } k^{g}_{ \{l_j\} \rightarrow \{l_i\} } \hat{M}_{ \{l_j\} } \nonumber \\
& & - \sum_{g=a,d,h,r} \sum_{ \{l_i\}_{j}^{ \sim } } k^{g}_{ \{l_i\}_{j}^{ \sim } \rightarrow \{l_j\} } \hat{M}_{ \{l_i\}_{j}^{ \sim } } ,
 \label{Eq_LH_4}
\end{eqnarray}
where $k^{g}_{ \{l_j\} \rightarrow \{l_i\} }$ is the rate constant for an elementary reaction of type $g$ that transforms local configuration $\{l_j\}$ into local configuration $\{l_i\}$, and local configuration $\{l_i\}_{j}^{ \sim }$ contains $\{l_i\}$ in such a way that the reaction $\{l_i\}_{j}^{ \sim } \rightarrow \{l_j\}$ destroys $\{l_i\}$ inside $\{l_j\}$. For instance,  $\{l_i\}_{j}^{ \sim } = \{B,B,A\}$ contains $\{l_i\} = \{B,B\}$ and the recombination of $A$ and $B$ will lead to $\{l_j\} = \{B,*,C\}$, thus destroying $\{l_i\} = \{B,B\}$ and decreasing $\hat{M}_{ \{B,B\} }$ (see Eq. \ref{Eq_LH_3_e3}).

Eq. \ref{Eq_LH_4} is the master equation considered in this study, written in terms of the time evolution of occupation variables, {\it i.e.} the instantaneous multiplicities of local configurations. Together with the expressions linking the multiplicities ($\sum_{X} \hat{M}_{ \{X\} } = 1$, $\sum_{X,Y} \hat{M}_{ \{X,Y\} } = b$, $\sum_{X,Y,Z} \hat{M}_{ \{X,Y,Z\} } = c$, etc...), Eq. \ref{Eq_LH_4} represents a large system of equations. However, it is important to realize that we only need to solve it if the spatial configuration of the surface is not accessible. In this case, knowledge of the initial values of the $\hat{M}_{ \{l_i\} }$'s will enable obtaining their future values and, thus, the values for the multiplicities of the elementary reactions. For extended catalytic systems, however, it is easier to monitor the multiplicities of the elementary reactions directly from the visited spatial configurations. 
Thus, in practice, the use of a spatial representation enables solving the master equation for the instantaneous multiplicities (Eq. \ref{Eq_LH_4}). 
After this, the average values are easily determined (Section \ref{Multiplicity of an elementary reaction}).

The KMC simulations presented in this study demonstrate that monitoring a small number of relevant multiplicities works well in practice. Note that such monitoring is applicable to other methods ({\it e.g.} Molecular Dynamics) and, more generally, to a generic description of the evolution of the system, where all atoms and molecules interact with each others---as in reality---and the elementary reactions take place. Provided that any changes in the spatial configuration of the system are monitored, then (i) the actual transition rates (= rate constants) can be determined, under the widely-accepted assumption in Transition State Theory and Chemical Kinetics that the rate constant from one configuration to another is independent of any previously visited configurations (Markov chain), and (ii) the actual changes in the multiplicities of the elementary reactions can be tracked, thus directly solving the variables of interest in Eq. \ref{Eq_LH_4}. 

Note that Eq. \ref{Eq_LH_4} is valid beyond the mean field approximation, since the multiplicities themselves have been defined for this purpose, directly carrying information about the presence of correlations. Within mean field, Eq. \ref{Eq_LH_4} decays naturally into typical rate equations for the coverages of the adsorbates, such as Eqs. \ref{Eq_LH_1_e2}-\ref{Eq_LH_1_e4}. In this manner, the proposed formalism provides a generalization of the traditional coverage-based approach, directly enabling the study of heterogeneous catalytic systems outside the mean field approach.

While traditionally one considers the adsorbate coverages as the natural variables required to describe the evolution of the system, here we have presented the idea that it is the collection of the multiplicities 
of a few local configurations
that provides a natural description of the configurational structure of the surface and, thus, its evolution. 

Finally, we stress that it is possible to identify the instantaneous multiplicity of a reaction with the instantaneous coverage for the corresponding local configuration. For this purpose, the  {\em instantaneous coverage of a local configuration} is defined as $\frac{\hat{p}_{\alpha}}{z_{\alpha}}/\hat{s}$, where $z_{\alpha}$ is the number of sites participating in the local configuration, and $\hat{p}_{\alpha}$ ($= \hat{m}_{\alpha} z_{\alpha}$) is the total number of sites participating in reaction $\alpha$, with $\hat{m}_{\alpha}$ and $\hat{s}$ as already defined. As an example, for dissociative adsorption of a triatomic molecule, the local configuration requires three neighbor empty sites and, thus, $z_{\alpha} = 3$. Similarly, $z_{\alpha}=2$ for bimolecular recombination reactions (since two neighbor sites participate in every elementary reaction) and also $z_{\alpha}=2$ for typical diffusion reactions (since the adsorbate hops between two sites). Considering Fig. \ref{Figure01} as a specific example, the instantaneous multiplicity for the desorption of A is equal to the instantaneous coverage for all sites occupied by molecules of type A, namely, $\frac{4}{1}/49=4/49$. Similarly, the multiplicity for the recombination of A and B is equal to the coverage by all pairs of nearest neighbor sites such that one site is occupied by A and the other by B ($\frac{4}{2}/49=2/49$). Since the relation between coverage and multiplicity is valid at any instant, it remains valid also between their averages.

\subsection{Rate constant for an elementary reaction}
\label{Rate constant for an elementary reaction}

For a typical rate law, $r_{\alpha} = k_{\alpha} \theta_A^{x'} \theta_B^{y'}$, the {\em specific reaction rate}, $k_{\alpha}$, also known as the {\em specific rate} or {\em rate constant}, refers to the part of the {\em rate}, $r_{\alpha}$, that does not depend on concentration/coverage, {\it i.e.} the part that does not depend on the number of locations where the reaction can be performed.
The statistical formulation of transition state theory (TST) \cite{Chorkendorff2003, Eyring1935, Laidler1983} describes the specific rate for an elementary  reaction as $k_{\alpha} = k_{\alpha}^{0} e^{-E_{\alpha}^{k} / k_B T}$, where $k_{\alpha}^{0} = \frac{k_B T}{h} \frac{q^{\ne}}{q}$ is the attempt frequency, with $q^{\ne}$ and $q$ the partition functions of the system in the transition and initial states of the reaction, respectively, and $h$ is Planck's constant. 
Determination of the partition functions leads to $k_{\alpha}^{0} = \frac{P A} {\sqrt{2 \pi m k_B T}}$ for nonactivated adsorption, where $m$ and $P$ are the mass and pressure of the adsorbed gas, respectively, and $A$ is the adsorption site area \cite{Chorkendorff2003}. Similarly, $\frac{q^{\ne}}{q} \approx 1$ and $k_{\alpha}^{0} \approx \frac{k_B T}{h}$ for diffusion, recombination and desorption \cite{Reuter2006, Farkas2012, Teschner2012, Shah2014}. See Eqs. \ref{eq_desorption}-\ref{eq_mu} in the Supporting information for a more complex treatment of the desorption case. 

Complementarily, the thermodynamic formulation of TST \cite{Chorkendorff2003, Laidler1983, Eyring1935b, Wynne-Jones1935} states that $k_{\alpha} = \frac{k_B T}{h}$ \\ $e^{ \Delta  S_{\alpha}^k / k_B} e^{- \Delta  H_{\alpha}^k / k_B T}$, where $ \Delta  S_{\alpha}^k$ and $ \Delta  H_{\alpha}^k$ are the entropy change  and enthalpy change, respectively, from the initial to the transition state. 
 Note the superindex $k$, which stresses the fact that both changes are contained in the value for the specific rate $k_{\alpha}$. 
The entropy barrier, $ \Delta  S_{\alpha}^k$, is usually assigned to the variation in the number of energy states that can be occupied at a given temperature, {\it i.e.} the difference in the partition functions of vibration, rotation and/or translation at the ground state of the reactants and at the transition state \cite{Chorkendorff2003}. 
 In fact, for elementary reactions at constant pressure for which the volume change is negligible ($\Delta V_{\alpha}^k \approx 0$ and, thus, $\Delta H_{\alpha}^k = E_{\alpha}^k + p\Delta V_{\alpha}^k \approx E_{\alpha}^k$), equating the statistical and thermodynamic formulations of $k_{\alpha}$ leads to $e^{ \Delta  S_{\alpha}^k / k_B} = \frac{q^{\ne}}{q}$.   
This results in negligible entropy barriers ($ \Delta  S_{\alpha}^k \approx 0$) for those reactions where $\frac{q^{\ne}}{q} \approx 1$, while noticeable barriers are expected for other descriptions of the partition function ratio.

Section \ref{Apparent activation energy of the $TOF$} shows that the 'rate' $r_{\alpha}$ (which contains both the specific rate, $k_{\alpha}$, and the number of locations where the elementary reaction can be performed per active site, $M_{\alpha}$) can be formulated similarly as $k_{\alpha}$ itself, simply by replacing $\Delta  S_{\alpha}^k$ with $\Delta  S_{\alpha}^k + S_{\alpha}^{ M  }$, where the configurational entropy $S_{\alpha}^{ M  }$ is directly related to the multiplicity $M_{\alpha}$.

\subsection{Total rate and the probability of an elementary reaction}
\label{Total rate and the probability of an elementary reaction}

\noindent
Let us define the {\em instantaneous total rate} as the sum of the specific rates (= rate costants) for all elementary reactions that can be performed at the current configuration: 
$ \hat{r}  = \Sigma_{\alpha \in \{e\} }  \hat{m}_{\alpha}  k_{\alpha}$. 
Here, the symbol $\in$ denotes 'in' so that $\alpha \in \{ e \}$ means that the sum is over any elementary reaction $\alpha$ contained in the {\em entire} collection of elementary reactions $\{ e \}$.
The corresponding average, referred to as the {\em total rate}, is:
\begin{equation}
 r = \langle \overline{ \hat{r} } \rangle = \sum_{\alpha \in \{e\} } m_{\alpha} k_{\alpha}. 
\end{equation}
The abundance of each reaction ($m_{\alpha}$) is useful to stress the dependence of the total rate on the configuration of the system, a feature that remains hidden if one uses the form $r = \Sigma_{i} k_{i}$ (no grouping of identical reactions).

Similarly, we consider another average quantity, the {\em total rate per active site}: 
\begin{eqnarray}
\hspace{-8mm}
R_{}     & = & r/s \label{Re_definition000} \\
            & = & \sum_{\alpha \in \{e\} } M_{\alpha} k_{\alpha} \label{Re_definition0} \\
            & = & \sum_{\alpha \in \{a\}}  {\scriptstyle M_{\alpha} k_{\alpha} +}  \sum_{\alpha \in \{d\}} {\scriptstyle M_{\alpha} k_{\alpha} +}  \sum_{\alpha \in \{h\}} {\scriptstyle M_{\alpha} k_{\alpha} +}  \sum_{\alpha \in \{r\}} {\scriptstyle M_{\alpha} k_{\alpha}} . \nonumber \\
            & & \label{Radhr_definitions} \\
           & = & R_{a} + R_{d} + R_{h} + R_{r} \label{Re_definition1} 
\end{eqnarray}
Here, we have explicitly separated all the elementary reactions ($\alpha \in \{e\}$) into adsorption reactions ($\alpha \in \{a\}$), desorption reactions ($\alpha \in \{d\}$), diffusion reactions ($\alpha \in \{h\}$) and recombination reactions ($\alpha \in \{r\}$). Additionally, we have defined $R_{g} = \sum_{\alpha \in \{g\}} M_{\alpha} k_{\alpha}$ with $g=a, d, h, r$ to denote (per active site): the total adsorption rate $R_{a}$, total desorption rate $R_{d}$, total hop rate $R_{h}$ (diffusion) and total recombination rate $R_{r}$.

Based on these definitions, we also define the {\em probability to observe reaction} $\alpha$: 
\vspace{-2mm}
\begin{equation}
\omega_{\alpha}^{R} =  \frac{ m_{\alpha} k_{\alpha} }{ r } =   \frac{ M_{\alpha} k_{\alpha} }{ R } = \frac{ M_{\alpha} k_{\alpha} }{ \Sigma_{\alpha' \in \{ e \} } M_{\alpha'}k_{\alpha'} }. \label{omega_definition}
\end{equation}
As shown in this study, the {\em reaction probabilities} of Eq. \ref{omega_definition} provide a complete and accurate picture of the undergoing competition between the different elementary reactions, for a fraction of the cost required to obtain similar insights based on the degrees of rate control and sensitivity  ($\chi_{\alpha_*}^{}$ and $\xi_{\alpha}^{}$). 

All averaged quantities defined above have corresponding instantaneous counterparts, which are well defined at any instant (during the transient or within the steady state). For instance, the {\em instantaneous total rate per active site} is $\hat{R} = \Sigma_{\alpha \in \{ e \} } \hat{M}_{\alpha}k_{\alpha}$, and the {\em instantaneous probability to observe an elementary reaction} is $\hat{\omega}_{\alpha}^{R} = \hat{M}_{\alpha} k_{\alpha} / \Sigma_{\alpha' \in \{ e \} } \hat{M}_{\alpha'}k_{\alpha'}$. The traditional 'rate' $r_{\alpha} = k_{\alpha} \theta_A^{x'} \theta_B^{y'} = k_{\alpha} M_{\alpha}$, which is an average quantity, is described as the {\it total rate per active site for reaction} $\alpha$ in our formalism. The corresponding instantaneous value is: $\hat{r}_{\alpha} = \hat{M}_{\alpha} k_{\alpha}$.

\subsection{Turnover frequency}
\label{Turnover frequency}

The turnover frequency ($TOF$) refers to the number of molecules of the product of interest in the gas phase, generated per active site per unit time \cite{Meskine2009, Hess2012, Farkas2012}.
It is the {\em rate} in 'degree of {\em rate} control' and '{\em rate} sensitivity'. Traditional mathematical formulations, such as $TOF = k_{\lambda} \theta_A^{x'}  \theta_B^{y'}$, are based on the assumption that the rate of one particular reaction ($\lambda$, in this case) is sufficiently low so that it acts as the RDS. Here, we follow previous theoretical studies, where it was recognized that the gaseous product of interest will typically be generated in different elementary reactions \cite{Meskine2009} and/or different products of interest will be generated \cite{Shah2014}.

As an example, let $AB$ refer to the product of interest and let us consider two different elementary reactions where $AB_{\rm (g)}$ is generated: (1) a recombination reaction with direct desorption: $A_{\rm X} + B_{\rm X} \rightarrow 2 V + AB_{\rm (g)}$, and (2) a desorption reaction: $AB_{\rm Y} \rightarrow V + AB_{\rm (g)}$. Here, $V$ refers to a vacant site, while $X$ and $Y$ denote different site types populated by species $A$, $B$ and $AB$. Note that, in this example, the way $A_{\rm X}$, $B_{\rm X}$ and $AB_{\rm Y}$ were formed in previous elementary reactions is irrelevant in order to determine the TOF, since the production of $AB_{\rm (g)}$ occurs through reactions (1) and (2) only. If $k_{1}$ and $k_{2}$ are the specific rates (or rate constants) for both reactions, respectively, and the two reactions are present $m_{1}$ and $m_{2}$ times on the surface with a total of $s$ active sites, then the $TOF$ is simply formulated as: $TOF  = (m_{1} k_{1} + m_{2} k_{2} ) / s$. This can be re-written as:  $TOF  = \sum_{\alpha \in \{ 1,2 \}} M_{\alpha} k_{\alpha}$, where $M_{\alpha}=\frac{ m_{\alpha} }{ s }$ is the multiplicity for reaction $\alpha$. Note that $ M_{\alpha} k_{\alpha} = \frac{ m_{\alpha} k_{\alpha} }{ s }$ describes how many molecules of $AB_{\rm (g)}$ are generated per unit time per active site due to reaction $\alpha$. 

If more than two reactions explicitly contribute to the generation of the gaseous product of interest, the $TOF$ is generalized as:
\begin{equation}
TOF = \sum_{\alpha \in \{ x \}} M_{\alpha} k_{\alpha} , \label{TOF_definition}
\end{equation}
where $\{ x \}$ denotes the collection of elementary reactions where the target product {\em exits} the catalyst surface ({\it i.e.} those reactions whose final state contains the target product in the gas phase). 
The use of the multiplicities in Eq. \ref{TOF_definition} (instead of traditional products/powers of the adsorbate coverages) is justified by the master equation (Eq. \ref{Eq_LH_4}), which shows that the multiplicities are the natural variables describing the evolution of the system.

If the target gaseous product is generated in reversible elementary reactions ({\it e.g.} $A_{\rm X} + B_{\rm X}  \rightleftarrows 2 V + AB_{\rm (g)}$ and/or $AB_{\rm Y}  \rightleftarrows V + AB_{\rm (g)}$) with $k_{\alpha}^+$ ($k_{\alpha}^-$) denoting the corresponding forward (backward) rate constant, the $TOF$ is defined as:
\begin{equation}
TOF = \sum_{\alpha \in \{ x \}} (M_{\alpha}^+ k_{\alpha}^+ - M_{\alpha}^- k_{\alpha}^-) . \label{TOF_definition_reversible}
\end{equation}
If we are interested in more than one product, the $TOF$ is simply the sum of several expressions, one for each product $P$: 
\begin{equation}
TOF =  \sum_{P} \sum_{\alpha \in \{ x_P \}} (M_{\alpha}^{P,+} k_{\alpha}^{P,+} - M_{\alpha}^{P,-} k_{\alpha}^{P,-}). \label{TOF_definition_SeveralProducts}
\end{equation}
In Section \ref{Application} we consider a system with one product of interest (CO$_2$) and another system with two products of interest (NO and N$_2$).
Note that Eq. \ref{TOF_definition_SeveralProducts} transforms into Eq. \ref{TOF_definition_reversible} by simply summing over $\alpha \in \{ x_1 \}, \{ x_2 \}, ...$ in Eq. \ref{TOF_definition_reversible}. In turn, Eq. \ref{TOF_definition_reversible} can be formulated as Eq. \ref{TOF_definition} by simply using negative multiplicities for the reverse reactions. Thus, without loss of generality, we focus on using Eq. \ref{TOF_definition} as a general description for the $TOF$.

As with other variables in previous sections, we have defined the $TOF$ as an average quantity, determined in the steady state: $TOF = \sum_{\alpha \in \{ x \}} M_{\alpha} k_{\alpha}$. However, our formalism allows considering also the instantaneous value, $\widehat{TOF} = \sum_{\alpha \in \{ x \}} \hat{M}_{\alpha} k_{\alpha}$, which is well defined at any instant, during the transient and within the steady state.

\vspace{-2mm}
\subsection{Apparent activation energy of the $TOF$ }
\label{Apparent activation energy of the $TOF$}

As shown in Section \ref{Rate equations and master equation}, the values of the multiplicities, $M_{\alpha}$, are functions of the actual values of the rate constants, $k_{\alpha}$. In this manner, the $M_{\alpha}$'s are functions of temperature.
Thus, for an Arrhenius plot of $\log(TOF)$ {\it vs} inverse temperature, $\beta = 1 / k_B T$, the apparent activation energy, $E_{app}^{TOF}  = - \frac{ \partial   \log(TOF) }{ \partial   \beta } = - \frac{ 1 } {TOF} \frac{ \partial   (TOF) }{ \partial   \beta }$, is given by:
$E_{app}^{TOF}  = - \frac{ 1 } { \sum_{\alpha \in \{ x \}} M_{\alpha} k_{\alpha} } \frac{ \partial   \sum_{\alpha \in \{ x \}} M_{\alpha} k_{\alpha} }{ \partial   \beta }$. Using $k_{\alpha} = k_{\alpha}^{0} e^{-E_{\alpha}^{k} \beta}$ and $E_{\alpha}^{M} =  - \frac{ \partial   \log(M_{\alpha}) }{ \partial \beta }$, and applying the chain rule to $\sum_{\alpha \in \{ x \}} M_{\alpha} k_{\alpha}$ easily leads to:
\vspace{-5mm}
\begin{equation}
\hspace{-1mm}
\begin{tabular}{cc}
& $\scriptstyle \epsilon_{\alpha}^{TOF}$ \\
$\displaystyle E_{app}^{TOF} = \sum_{\alpha \in \{ x \}}$ &  $\overbrace{ \scriptstyle \omega_{\alpha}^{TOF} (E_{\alpha}^{k} + E_{\alpha}^{k^{0}} + E_{\alpha}^{M} ) } $ , \label{EappTOF_5} \\
$\omega_{\alpha}^{TOF} = \frac{  M_{\alpha} k_{\alpha} }{ TOF }$ & $= \frac{  M_{\alpha} k_{\alpha} }{ \Sigma_{\alpha' \in \{ x \}} M_{\alpha'} k_{\alpha'} } \; , \alpha \in \{ x \} \; \;$
\end{tabular}
\end{equation}
where $E_{\alpha}^{k^{0}} =  - \frac{ \partial   \log(k_{\alpha}^{0}) }{ \partial   \beta }$ 
and the weight $\omega_{\alpha}^{TOF}$
for $\alpha \in \{ x \}$ is the probability of observing reaction $\alpha$ amongst all reactions explicitly contributing to the TOF. Since these weights are normalized between 0 and 1, Eq. \ref{EappTOF_5} describes the apparent activation energy as a proper weighted average.

If $k_{\alpha}^{0}$ depends on temperature, its energy contribution ($E_{\alpha}^{k^{0}}$) needs to be added, as indicated in Eq. \ref{EappTOF_5}. Assuming momentarily that $k_{\alpha}^{0}$ is temperature-independent, then Eq. \ref{EappTOF_5} is a weighted-average over the elementary activation energies ($E_{\alpha}^{k}$), each one modified by an effective energy ($E_{\alpha}^{M}$), which originates from the temperature dependence of the corresponding multiplicity. From a traditional perspective, this can be understood as an underlying change in configurational entropy, since modifying the temperature alters the morphology (and the configuration) of the system. 

Recalling { Boltzmann's exact formulation  of entropy ($S$) as the natural logarithm of the number of possible  microscopic configurations ($\Omega$) multiplied by the Boltzmann constant ($k_{B}$),
$S = k_B \log \Omega$, in our case $\Omega$ can be directly identified as $M_{\alpha}$, {\it i.e.} the number of local microscopic configurations where reaction $\alpha$ can be performed on the surface per active site. Thus, we simply define the configurational entropy $S_{\alpha}^{ M  }$ for reaction $\alpha$ as:
\begin{eqnarray}
\hspace{-5mm}
S_{\alpha}^{ M  } = k_{B} \log M_{\alpha}  \; \; \Leftrightarrow \; \; M_{\alpha} = e^{S_{\alpha}^{ M  } / k_B} . \label{Eq_ConfigurationalEntropy}
\end{eqnarray}
Then, the total rate per active site for reaction $\alpha$ becomes 
$ r_{\alpha} =  M_{\alpha}k_{\alpha} = \frac{k_B T}{h} e^{(S_{\alpha}^{ M  } + \Delta  S_{\alpha}^k) / k_B} e^{- \Delta  H_{\alpha}^{k} / k_B T}$.  
Thus, the 'rate' ($r_{\alpha}$) can be formulated in a similar manner as the 'rate constant' ($k_{\alpha}$) by simply considering the entropy sum $S_{\alpha}^{ M  } + \Delta  S_{\alpha}^k$, where the configurational entropy, $S_{\alpha}^{ M  }$, is directly related to the multiplicity of reaction $\alpha$, and the entropy barrier for the reaction itself, $ \Delta  S_{\alpha}^k$, is related to the change in the number of molecular energy levels due to vibration, rotation and/or translation from the initial to the transition state. 
While traditionally the latter is contained in the value of the rate constant $k_{\alpha}$, in this study we explicitly consider the presence of the configurational part $M_{\alpha}=e^{S_{\alpha}^{ M  }/k_B}$ in $r_{\alpha}$. This enables a direct analysis of the role of the relative abundance of each elementary reaction in describing the apparent activation energy. 

The equation $S = k_B \log \Omega$ (and, correspondingly, Eq. \ref{Eq_ConfigurationalEntropy}) is valid under the fundamental assumption of equiprobable microscopic configurations in Statistical Mechanics (all microscopic configurations are equally probable). The number of possible microscopic configurations ($\Omega$) should not be confused with the partition function ($Q$), typically used to derive expressions for all thermodynamic variables (including the entropy) in the canonical ensemble (see {\it e.g.} Section 3.3.3 in Ref. \cite{Chorkendorff2003}): $S = \left[ \frac{\partial (k_B T \log Q) } { \partial T } \right]_{N,V} = k_B \log Q + k_B T \left[ \frac{ \partial \log Q }{ \partial T} \right]_{N,V}$, where the derivatives are taken at constant particle number ($N$) and volume ($V$). 

Based on Eq. \ref{Eq_ConfigurationalEntropy}, the change in configurational entropy with inverse temperature is:
\begin{eqnarray}
\hspace{-5mm}
\frac{\partial S_{\alpha}^{ M  } }{\partial \beta} = k_B \frac{\partial \log M_{\alpha} }{\partial \beta} =  - k_B E_{\alpha}^{M}. & \label{eq_entropy_4_0} 
\end{eqnarray}
Thus, $E_{\alpha}^{M}$ is essentially the negative of the change in configurational entropy with inverse temperature and we refer to it as the {\em configurational contribution} to the apparent activation energy.

This perspective agrees well with recent reports, where the configuration and energy dependence of the $TOF$ has been discussed \cite{Hess2012, Teschner2012, Hess2017}. As an example, modifications in the coverage of the empty sites give rise to configurational entropy contributions to the apparent activation energy \cite{Teschner2012}. In our case, however, a more general scenario is considered. Some elementary reactions may involve several sites/species and, thus, cannot be simply described in terms of the coverage of the intermediates under all possible circumstances. Instead, the multiplicities, which characterize the coverage for rather complex collections of sites, appear as the natural variables to describe the relative presence of the various reactions on the surface. 
Note that our formalism places the emphasis on the determination of the multiplicities and their variation with temperature in order to describe the apparent activation energy. The configurational entropy is not really needed and has been provided here as a link to traditional thinking.

\subsection{ Apparent activation energy of $R$ }
\label{Apparent activation energy of $R$}

From the resemblance of Eq. \ref{Re_definition0} to Eq. \ref{TOF_definition}, also the apparent activation energy of the total rate per site $R_{}$ is easily obtained:
\vspace{-5mm}
\begin{equation}
\begin{tabular}{cc}
& $\epsilon_{\alpha}^{R}$ \\
$\displaystyle E_{app}^{R_{}} = \sum_{\alpha \in \{ e \}}$ &  $\overbrace{ \scriptstyle \omega_{\alpha}^{R} (E_{\alpha}^{k} + E_{\alpha}^{k^{0}} + E_{\alpha}^{M} ) } \; , $ \label{Eapp_e_1} \\
$\omega_{\alpha}^{R_{ }} = \frac{ M_{\alpha} k_{\alpha} }{ R }$ & $= \frac{ M_{\alpha} k_{\alpha} }{ \Sigma_{\alpha' \in \{e\}} M_{\alpha'}k_{\alpha'} } \; , \alpha \in \{ e \} $
\end{tabular}
\end{equation}
where $\omega_{\alpha}^{R_{ }}$
for $\alpha \in \{ e \}$ is the probability of observing elementary reaction $\alpha$ amongst all elementary reactions. Thus, the probabilities of Eq. \ref{omega_definition} appear naturally within our formalism (Eq. \ref{Eapp_e_1}), regulating the contribution of every reaction to the apparent activation energy of $R$. Since $M_{\alpha}$ may increase, decrease or remain constant with temperature, $E_{\alpha}^{M}$ can be positive, negative or zero. Thus, Eqs. \ref{EappTOF_5} and \ref{Eapp_e_1} may lead to positive, negative or zero apparent activation energy, just as Eqs. \ref{E_app_temkin} and \ref{E_app_using_sensitivities}.

\subsection{Rate Determining Step and Rate Controlling Steps}
\label{Rate Determining Step and Rate Controlling Steps}

If a particular reaction (say $\lambda$) can be assigned as the RDS, then, by definition, the $TOF$ can be written solely in terms of that  reaction:
\vspace{-2mm}
\begin{eqnarray}
TOF & = & M_{\lambda} k_{\lambda}  \; \; ({\rm for}\; \; \lambda = {\rm RDS}) . \label{TOF_definition_RDS} 
\end{eqnarray}
This means that the apparent activation energy is:
\vspace{-2mm}
\begin{eqnarray}
E_{app}^{TOF} & = &  E_{\lambda}^{k} + E_{\lambda}^{k^{0}} + E_{\lambda}^{M} \; \; ({\rm for}\; \; \lambda = {\rm RDS}). \nonumber \\
 & & \label{EappTOF_6}
\end{eqnarray}
This is a very simple, yet meaningful result. Even if $E_{app}^{TOF}$ is dominated by a single reaction ($\lambda$), in general, $E_{app}^{TOF}$ should not be identified with the corresponding elementary activation energy alone, $E_{\lambda}^{k}$, as still accepted by some researchers (see the Discussion below). This will neglect the configurational contribution, $E_{\lambda}^{M}$, as well as the temperature dependence of the rate prefactor, $E_{\lambda}^{k^0}$, should it be relevant.

In general, the RDS may change as the temperature and/or partial pressures are modified. To assign the RDS to a particular reaction, we consider Eq. \ref{TOF_definition_RDS} and define the {\em relative error in representing the $TOF$ using reaction} $\alpha$:
\vspace{-2mm}
\begin{eqnarray}
\delta_{\alpha}^{TOF} & = & \left| 1 - \frac{ M_{\alpha} k_{\alpha} } { TOF } \right| \;\; (\alpha \in \{ e \}) ,  \label{delta_definition}
\end{eqnarray}
which is 0 if $\alpha = $ RDS, while it may take unbound, positive values if $M_{\alpha}k_{\alpha}$ deviates largely from the TOF. Then, we define the {\em proximity to the TOF} as:
\vspace{-2mm}
\begin{eqnarray}
\hspace{-5mm}
\sigma_{\alpha}^{TOF} & = & 1 - min( 1 , \delta_{\alpha}^{TOF}) \;\; (\alpha \in \{ e \}) , \label{sigma_definition}
\end{eqnarray}
so that $0 \le \sigma_{\alpha}^{TOF} \le 1$, taking 1 if $\alpha = $ RDS and 0 if $M_{\alpha}k_{\alpha}$ deviates significantly from the TOF.
By definition, the proximity $\sigma_{\alpha}^{TOF}$ is comparable to $\chi_{\alpha}$ (the degree of rate control), both taking the value 1 when reaction $\alpha$ is the RDS. In addition, similarly to $\xi_{\alpha}$ (the rate sensitivity), also the proximity $\sigma_{\alpha}^{TOF}$ provides crucial information about the sensitivity of the $TOF$ to the different reactions. 

Considering Eq. \ref{Eapp_e_1}, we note that the probability of observing any reaction explicitly contributing to the $TOF$ is given by $\omega_{TOF}^{R}= TOF / R $. Thus, those reactions with probability  $\omega_{\alpha}^{R}  >> \omega_{TOF}^{R}$ ({\it i.e.} $ M_{\alpha} k_{\alpha} >> TOF $) will occur much more frequently than any  reaction explicitly contributing to the $TOF$ and, thus, a small variation in their rate constants, $k_{\alpha}$, will essentially leave the $TOF$ unchanged (see below one exception, due to time scaling).
The same applies to the reactions with $\omega_{\alpha}^{R}  << \omega_{TOF}^{R}$ ({\it i.e.} $ M_{\alpha} k_{\alpha} << TOF $). Only those reactions with $M_{\alpha} k_{\alpha} / R$ around $TOF / R$ may noticeably affect the $TOF$. In this manner, in probability space, proximity to the $TOF$ means sensitivity by the TOF. This provides a procedure to identify any RCS, in addition to the RDS (if it exists).

An advantage of our formulation is the direct use of  the values of the $TOF$ and $M_{\alpha}k_{\alpha}$ in the definition of $\sigma_{\alpha}^{TOF}$, instead of the derivatives of the $TOF$ with respect to the rate constants $k_{\alpha}$, as required in the determination of $\chi_{\alpha}$ and $\xi_{\alpha}$ (see text after Eq. \ref{E_app_using_sensitivities}). Thus, our approach avoids (i) the need of performing a large amount of simulations (as required by previous methods, in order to characterize the dependence of the $TOF$ on every rate constant) as well as (ii) the emergence of potential inaccuracies from the additional processing (as required by previous methods, in order to determine the derivatives of the $TOF$ and, thus, $\chi_{\alpha}$ and $\xi_{\alpha}$). 

Finally, we note that any reaction with $\omega_{\alpha}^{R} \sim 1$ is exceptional, affecting the $TOF$ by {\em scaling} the time increment, even if its proximity to the $TOF$ is very low. This results from the fact that the inverse of the total rate $r_{}$ (= $R s$) has dimensions of time and, in fact, it provides a natural variable to determine the time increment. For instance, in a KMC simulation the instantaneous time increment is calculated as $  \hat{\Delta t}  = -\log(u)/\hat{r_{}} $, where $u \in (0,1]$ is a uniform random number \cite{Reuter2006, Voter2007, Chatterjee2007, Reuter2011, Jansen2012, Temel2007, Gosalvez2017}. Thus, considering the average values in the steady state, those reactions with large $\omega_{\alpha}^{R}$ essentially control the value of $R_{}$ and, accordingly, the value of $\Delta t$. In this manner, variations in their rates end up affecting the value of $TOF$ by scaling $\Delta t$. It is not the same generating $n$ molecules per site per minute than generating the same $n$ molecules per site every five minutes. In this manner, we distinguish between two sources for variations in the $TOF$ in general, proximity and scaling. Proximity is signaled by $\sigma_{\alpha}^{TOF} \sim 1$ or, equivalently, $\omega_{\alpha}^{R} \sim \omega_{TOF}^{R}$. Scaling is indicated by $\omega_{\alpha}^{R} \sim 1$.

\section{Application}
\label{Application}

To illustrate the validity of the proposed multiplicity analysis we consider a reaction mechanism containing a total of 21 elementary reactions for the oxidation of CO on RuO$_2$(110) \cite{Meskine2009, Reuter2006, Reuter2011, Hess2012, Farkas2012, Temel2007, Hoffmann2017, Over2000}. 
We also consider a distinctively different reaction mechanism containing a total of 18 elementary reactions for the selective oxidation of NH$_3$ on RuO$_2$(110) as well \cite{Hong2010}. The two reaction mechanisms are schematically shown on Figures \ref{Figure02}(a) and \ref{Figure02}(b). Note that in general the surface dictates the actual symmetry of the neighborhood around each surface site as well as the particular collection of elementary reactions that may take place. Since the collection of elementary reactions (and elementary activation energies) is very different for the two selected application examples and symmetry is incorporated through the actual values of the multiplicities, we believe the two cases are sufficient to illustrate the general applicability of our formalism to different catalytic reactions. 
In fact,} we emphasize that similar multiplicity formulations have already been successfully applied to other surface processes, such as two-dimensional epitaxial growth (with triangular and rectangular lattices) and anisotropic etching (in 3 dimensions) \cite{Gosalvez2017, Gosalvez_Ea_NJP}. 

The oxidation of NH$_3$ provides an example of a highly sequential catalytic reaction, taking place as a cascade of elementary abstraction reactions (between adsorbed NH$_3$/NH$_2$/NH and adsorbed O/OH), progressively stripping the H atoms until bare N is present at the surface, where it recombines with either adsorbed O (to form NO, which is desorbed later) or with itself (to form N$_2$, which is desorbed immediately). On the contrary, the oxidation of CO is an example of a highly parallel reaction mechanism, where basically all elementary reactions are enabled on all active sites. 
Below, we concentrate on the presentation of the case for the oxidation of CO, leaving the corresponding information for NH$_3$ to the Supporting Information.
\begin{figure*}[htb!]
  \begin{center}
    \includegraphics[width=2.0\columnwidth]{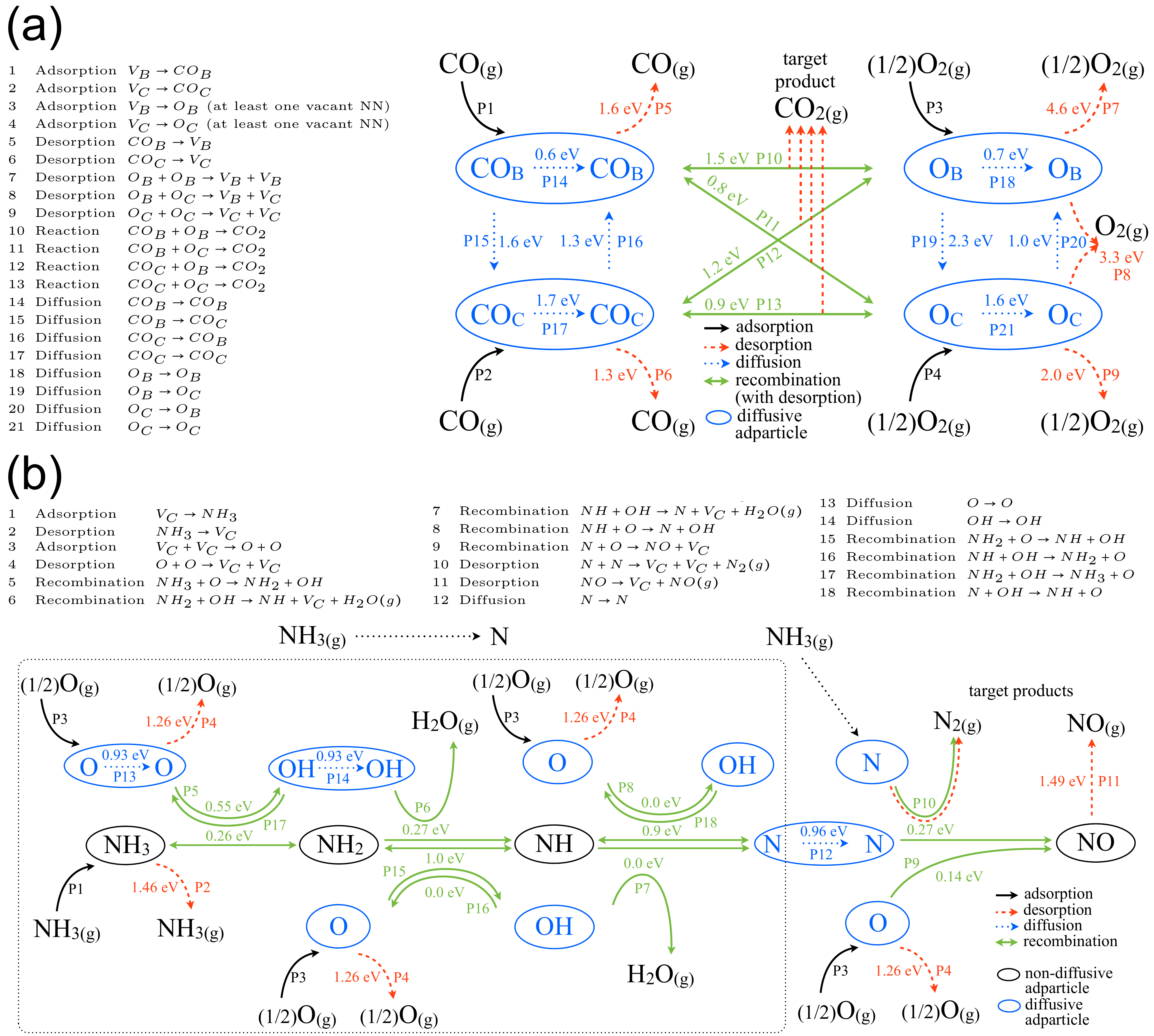}
  \end{center}
  \caption{
  (a) Reaction mechanism and corresponding graphical representation for the oxidation of CO on RuO$_2$\{110\} according to Ref. \cite{Reuter2006} (elementary barriers, in eV, are for model I, see main text).
  (b) Same as part (a), now for the selective oxidation of NH$_3$ on RuO$_2$\{110\} according to Ref. \cite{Hong2010}.  In this case, all elementary reactions occur only at C sites (see main text).
}
  \label{Figure02}
\end{figure*}

Until occurrence of desorption, diffusion or recombination, the adsorbed CO and O adspecies remain chemisorbed on the RuO$_2$(110) surface on both bridge (B) and cus (C) sites, which form alternating morphological rows (B-C-B-C-...), with every row parallel to the [110] crystallographic direction \cite{Temel2007}. Such an array of adsorption sites can be described using a rectangular unit cell, with lattice parameters $a_{x}$ = 6.43 {\AA} and $a_{y}$ = 3.12 {\AA} along the [$\overline{1}10$] and [110] directions, respectively, with two sites per unit cell \cite{Temel2007}: one B site located at $(0,0)$ and one C site located at $(\frac{1}{2},0)$. Thus, the area per site $A_{s}$ is half the unit cell area $A_{u.c.}$: $A_s = \frac{1}{2} A_{u.c.} = \frac{1}{2} a_{x} a_{y} = 10.03$ {\AA}$^2$. For the purposes of this report, the system can be treated as a two-dimensional array of $L \times L$ total sites with periodic boundary conditions.

As shown in Figure \ref{Figure02}(a) and extensively described in Section \ref{Oxidation_of_CO} of the Supporting Information, the currently accepted reaction mechanism for the oxidation of CO on RuO$_2$(110) contains a total of 21 elementary reactions, including dissociative adsorption of O$_2$ on two neighbor vacant sites ($V_{\rm X} + V_{\rm Y} \rightarrow O_{\rm X} + O_{\rm Y}$, where X and Y stand for either B or C sites), non-dissociative adsorption of CO on vacant sites ($V_{\rm X} \rightarrow CO_{\rm X}$), associative desorption of O$_2$ from two neighbor O atoms ($O_{\rm X} + O_{\rm Y} \rightarrow V_{\rm X} + V_{\rm Y}$), direct desorption of CO ($CO_{\rm X} \rightarrow V_{\rm X}$), surface diffusion of CO and O from B or C sites to B or C sites ($CO_{\rm X} \rightarrow CO_{\rm Y}$ and $O_{\rm X} \rightarrow O_{\rm Y}$), and recombination of CO on B or C sites with neighboring O on B or C sites ($CO_{\rm X} + O_{\rm Y} \rightarrow CO_{2}$). The reaction mechanism assumes that CO$_2$ is immediately desorbed after recombination. Thus, potential diffusion and/or decomposition of CO$_2$ admolecules on the surface is disregarded. As a result, the $TOF$ in this system corresponds to the total recombination rate: $TOF = R_{r} = \sum_{\alpha \in \{r\}} M_{\alpha} k_{\alpha}$. 

A more complete description of the adsorption of oxygen can be obtained by considering both adsorbed $O_2^*$ (mono) and $O_2^{**}$ (dihapto) adsorbates on the C sites, leading to a two-step adsorption-desorption reaction $O_2 \leftrightarrows O_2^{**} \leftrightarrows 2O^*$ at moderate coverages \cite{Pogodin2016}, rather than the one-step reaction $O_2 \leftrightarrows 2O^*$ assumed in the traditional reaction mechanism \cite{Meskine2009, Reuter2006, Reuter2011, Hess2012, Farkas2012, Temel2007, Hoffmann2017}. To directly compare our results to the traditional mechanism, the one-step  route is considered. The proposed multiplicity analysis can also be applied to the two-step route.

Table \ref{Table_Processes} in Section \ref{Oxidation_of_CO} of the Supporting Information provides the values for the attempt frequencies ($k_{\alpha}^{0}$) and activation energies ($E_{\alpha}^{k}$) used in four different models for the same reaction mechanism, here referred to as: I. Reuter, II. Seitsonen, III. Kiejna, and IV. Farkas. 
The four models differ in the actual values for the atomistic activation energies $E_{\alpha}^{k}$, which where obtained using different implementations of Density Functional Theory (models I-III) and experiment (model IV).
Moreover, model IV considers explicitly the presence of repulsion between nearest neighbor COs located at C sites, thus allowing to test the validity of the proposed formalism when adsorbate-adsorbate interactions are included beyond the adsorbate correlations already occurring at high coverages in the other models.
\begin{figure*}[htb!]
  \begin{center}
    \includegraphics[width=1.8\columnwidth]{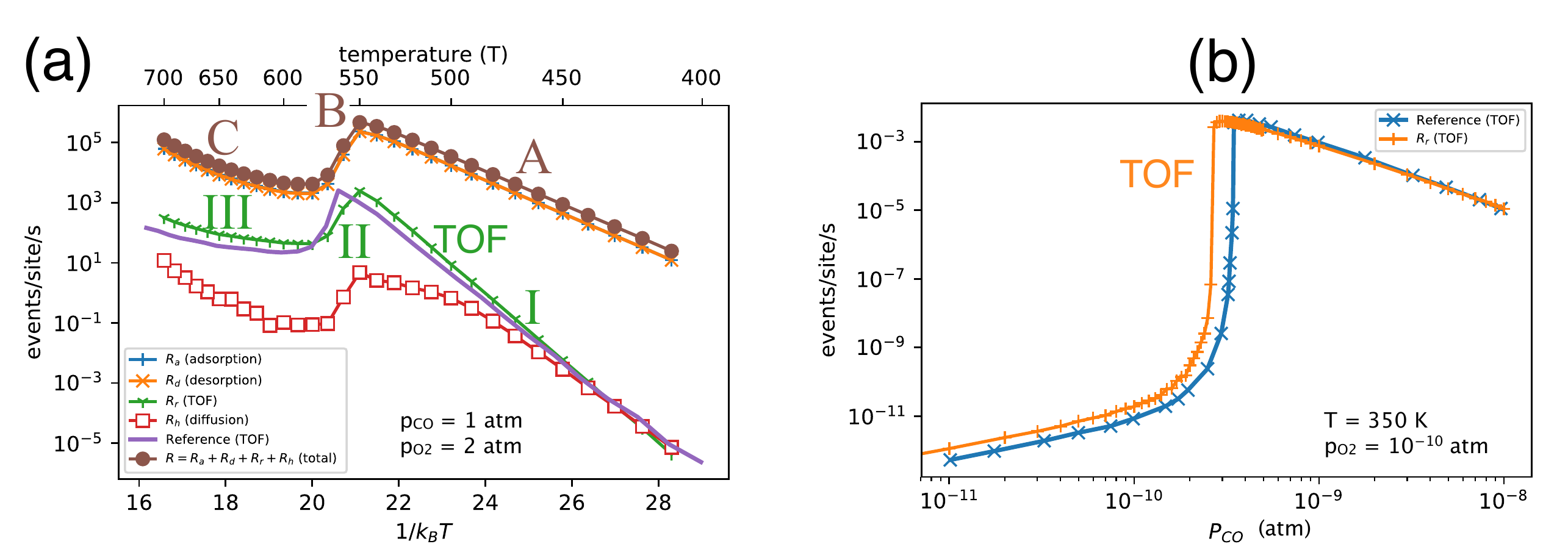}
  \end{center}
  \caption{Results for model I: 
    (a) Arrhenius plot for the total rates per site $R_{a}$, $R_{d}$, $R_{h}$, $R_{r}$ ($= TOF$), and $R_{}=R_{a}+R_{d}+R_{h}+R_{r}$ {\it vs} inverse temperature $\beta=1/k_B T$. (b) CO pressure dependence of $R_{r}$ ($= TOF$). Reference $TOF$ data from Ref. \cite{Meskine2009}.
}
  \label{Figure03}
\end{figure*}

Since the adsorption rate of CO on both B and C sites is the same in models I through IV, both  reactions $V_{\rm X} \rightarrow CO_{\rm X}$ (with X = B and C) have the same rate and, thus, are jointly referred to as $V \rightarrow CO$ in the rest of the report. Similarly, provided there is at least one vacant nearest neighbor (NN) to accommodate another O atom, the adsorption rate of an O atom is the same for B and C sites and, thus, simply referred to as $V \rightarrow O$ below.

Further details about the reaction mechanism for the oxidation of NH$_3$ on RuO$_2$(110) are provided in Section \ref{Selective_oxidation_of_NH_3} of the Supporting Information. In either case, oxidation of CO or NH$_3$, the catalytic process is simulated using a typical lattice-gas model and the rejection-free, time-dependent implementation of the KMC method \cite{Reuter2006, Voter2007, Chatterjee2007, Reuter2011, Jansen2012, Gosalvez2017}. See Section \ref{computational_method} of the Supporting Information for details.

\section{Results}
\label{Results}

Here we concentrate on the presentation of the results for the oxidation of CO, leaving the case of the oxidation of NH$_3$ to the Supporting Information (see Section \ref{Multiplicity_Selective_oxidation_of_NH3}). Fig. \ref{Figure03}(a) shows the temperature dependence of the total rate per site ($R_{}$) for model I at $p_{\rm CO} = 1$ atm and  $p_{\rm O_{2}} = 2$ atm. The plot also displays $R_{a}$, $R_{d}$, $R_{h}$, and $R_{r}$ ($= TOF$), as well as the corresponding $TOF$ data from Meskine {\it et al.} \cite{Meskine2009}. This demonstrates that our implementation is similar to that by Reuter and Scheffler \cite{Reuter2006, Temel2007, Meskine2009}. This is further confirmed in Fig. \ref{Figure03}(b), where basically the same pressure dependence is observed for our $TOF$ and that in Ref. \cite{Meskine2009}. The small, horizontal shift in the data for both temperature (Fig. \ref{Figure03}(a)) and pressure (Fig. \ref{Figure03}(b)) is assigned to (i) minor differences in some parameters used in the attempt frequencies (we carefully tried following every detail of their implementation) and, perhaps, (ii) differences in the detection of the onset of the steady state, which in our case is carried out automatically (see Section \ref{computational_method} of the Supporting Information). The validity of our implementation is further confirmed in Fig. \ref{Figure1S} for models II, III and IV, as shown in Section \ref{Oxidation_of_CO_using_models_II_III_and_IV} of the Supporting Information.

\begin{figure*}[htb!]
  \begin{center}
    \hspace{-4 mm}
    \includegraphics[width=1.8\columnwidth]{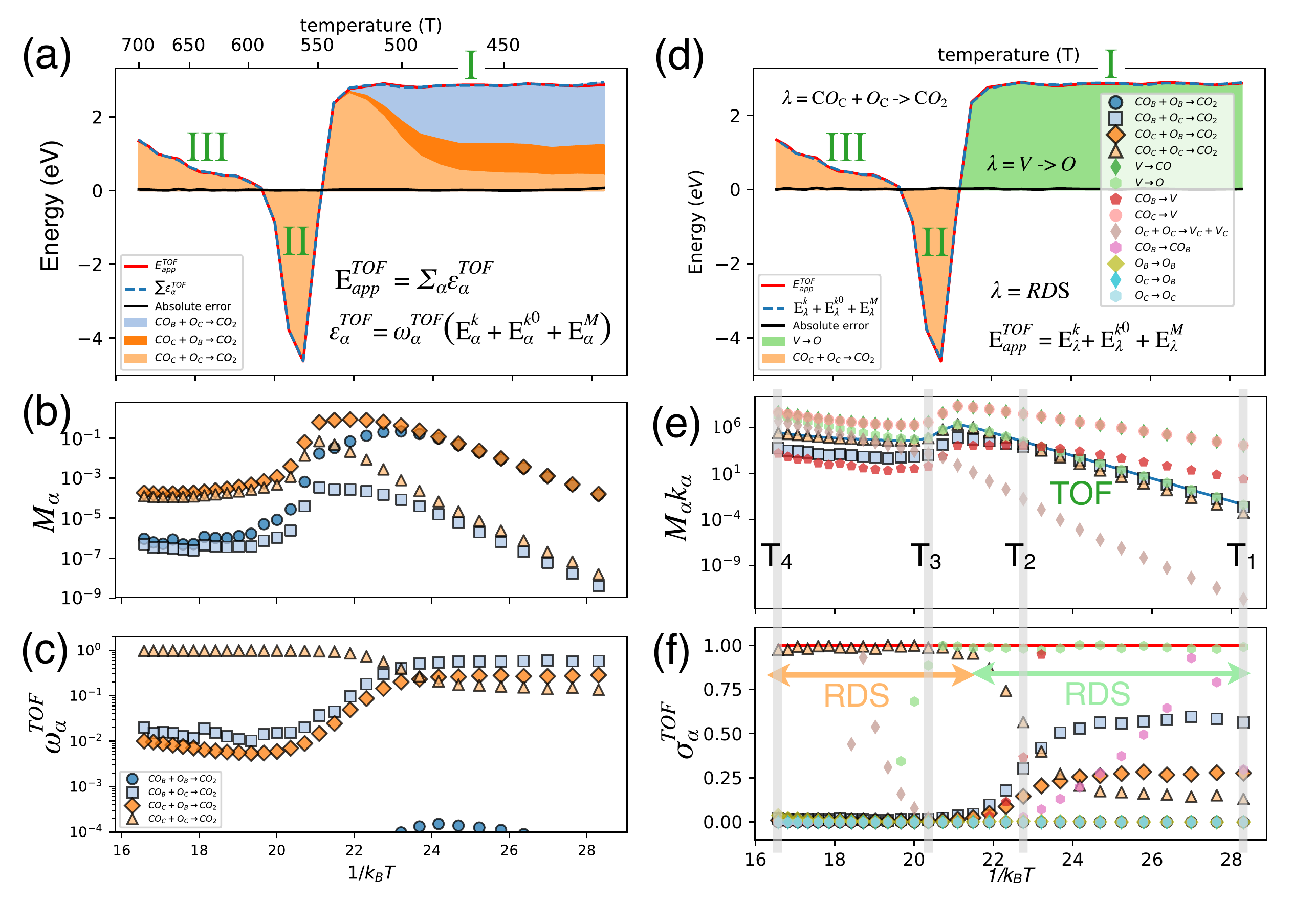}
  \end{center}
  \caption{Temperature dependence for model I (oxidation of CO on RuO$_2$\{110\}): 
  (a) Apparent activation energy ($E_{app}^{TOF}$) for the $TOF$ in Fig. \ref{Figure03}(a). $E_{app}^{TOF}$ is described well by $\sum_{\alpha \in \{ x \}} \epsilon_{\alpha}^{TOF}$, where $\epsilon_{\alpha}^{TOF} = \omega_{\alpha}^{TOF} (E_{\alpha}^{k} + E_{\alpha}^{k^{0}} + E_{\alpha}^{M})$. The absolute error $|E_{app}^{TOF} -  \sum_{\alpha \in \{ x \}} \epsilon_{\alpha}^{TOF} |$ is also plotted. 
  (b), (c) Multiplicities ($M_{\alpha}$) and probabilities ($\omega_{\alpha}^{TOF}$) for those elementary reactions explicitly contributing to the TOF, respectively. 
  (d) Same as (a), now describing $E_{app}^{TOF}$ as $E_{\lambda}^{k} + E_{\lambda}^{k^{0}} + E_{\lambda}^{M}$ for the RDS. The absolute error $|E_{app}^{TOF} -  (E_{\lambda}^{k} + E_{\lambda}^{k^{0}} + E_{\lambda}^{M}) |$ is also plotted.  
  (e) $M_{\alpha} k_{\alpha}$ ($=r_{\alpha}$) for any elementary reaction with probability $\omega_{\alpha}^{R} \ge 10^{-8}$ at any temperature. The $TOF$ is matched by $M_{\lambda}k_{\lambda}$ ($=r_{\lambda}$) for some $\lambda$ within some range of temperature .
  (f) Proximity to the $TOF$ ($\sigma_{\alpha}^{TOF}$), enabling the assignment of the RDS at every temperature.
  }
  \label{Figure04}
\end{figure*}

\begin{figure*}[htb!]
  \begin{center}
    \hspace{-4 mm}
    \includegraphics[width=1.8\columnwidth]{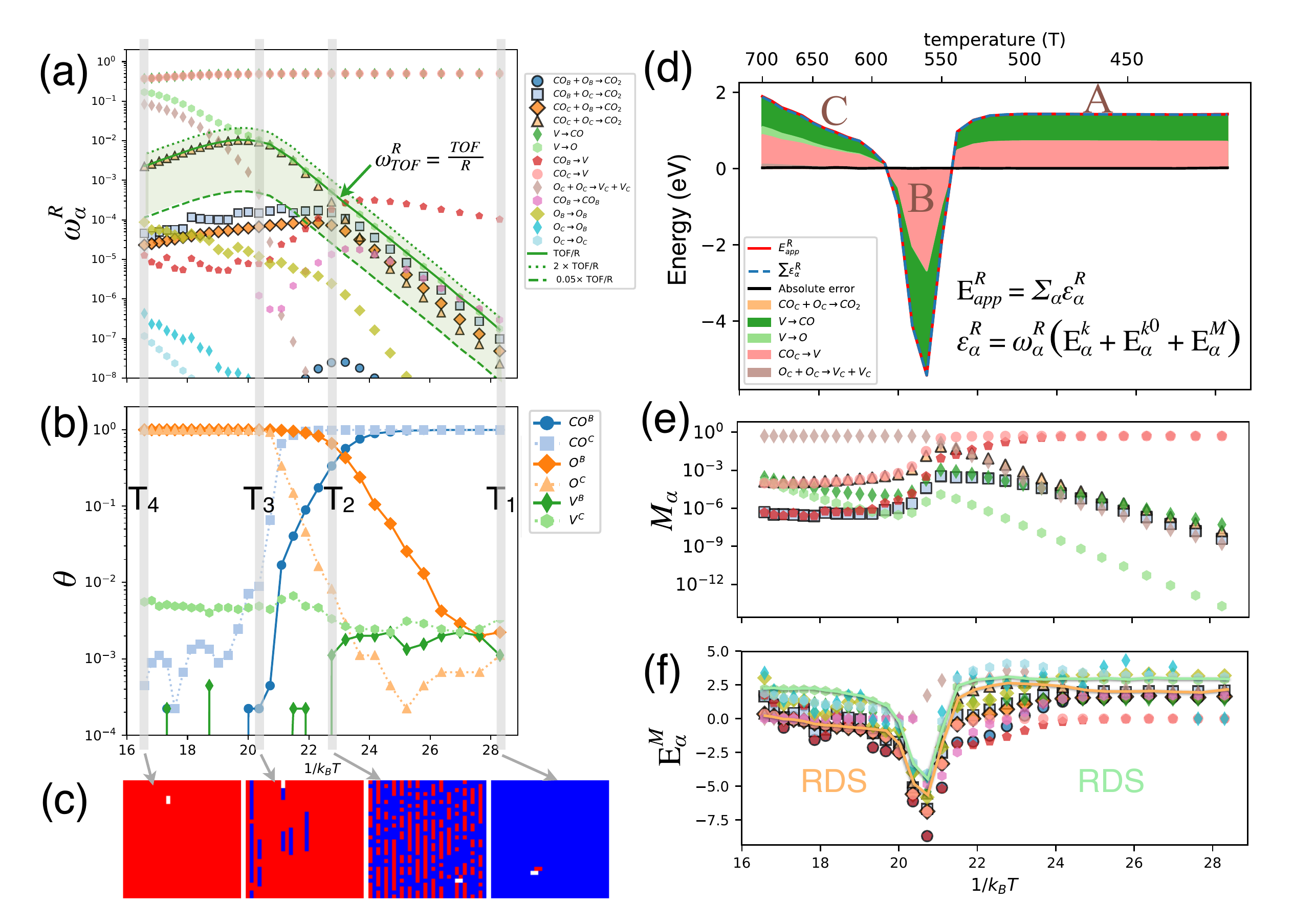}
  \end{center}
  \caption{Temperature dependence for model I (continued, {\it cf} Fig. \ref{Figure04}): 
     (a) Reaction probabilities $\omega_{\alpha}^{R}$ ($\ge 10^{-8}$). 
     (b) Coverage for all adspecies ($\theta_{\rm CO_{B}}$, $\theta_{\rm CO_{C}}$, $\theta_{\rm O_{B}}$, $\theta_{\rm O_{C}}$, $\theta_{\rm V_{B}}$ and $\theta_{\rm V_{C}}$ $\ge 10^{-4}$). 
     (c) Morphology snapshots at various temperatures.
  (d) Apparent activation energy ($E_{app}^{R}$) for the total rate per site $R$ in Fig. \ref{Figure03}(a). $E_{app}^{R}$ is described well by $\sum_{\alpha \in \{ e \}} \epsilon_{\alpha}^{R}$, where $\epsilon_{\alpha}^{R} = \omega_{\alpha}^{R} (E_{\alpha}^{k} + E_{\alpha}^{k^{0}} + E_{\alpha}^{M})$. The absolute error $|E_{app}^{R} -  \sum_{\alpha \in \{ e \}} \epsilon_{\alpha}^{R} |$ is also plotted. 
  (e), (f) Multiplicities ($M_{\alpha}$) and effective configurational energies ($E_{\alpha}^{M}$) for any elementary reaction with probability $\omega_{\alpha}^{R} \ge 10^{-8}$ at any temperature, respectively. $E_{\alpha}^{M}$ applies to frame (e) of the current figure and frame (b) of Fig. \ref{Figure04}.
  }
  \label{Figure05}
\end{figure*}

Without crossing each other, the curves in Fig. \ref{Figure03}(a) display three regions, labelled $A$, $B$ and $C$ for the total rate per site $R$, and I, II and III for the TOF. Accordingly, this model is dominated by adsorption and desorption reactions in the complete range of temperature, with both recombination and diffusion occurring much less frequently. The derivative of the $TOF$ of Fig. \ref{Figure03}(a), {\it i.e.} the apparent activation energy $E_{app}^{TOF}=- \frac{  \partial   \log(TOF) }{  \partial   \beta }$, is displayed in Fig. \ref{Figure04}(a). Beyond the constant value in region I, an excursion through negative values is observed in region II and a positive, roughly linear increase occurs in region III. The temperature dependence of $E_{app}^{TOF}$ in all three regions is accurately described by Eq. \ref{EappTOF_5} (absolute error $ | E_{app}^{TOF} - \sum_{\alpha \in \{ x \}} \epsilon_{\alpha}^{TOF} | < 0.07$ eV). In region I, where $E_{app}^{TOF}$ remains constant at $\sim 2.87$ eV (2.85 eV was reported in Ref. \cite{Meskine2009}), three recombination reactions participate. The dominating reaction changes from $CO_{\rm B} + O_{\rm C} \rightarrow CO_{2}$ (at the lowest temperatures) to $CO_{\rm C} + O_{\rm C} \rightarrow CO_{2}$ (near the onset of region II), with the third reaction, $CO_{\rm C} + O_{\rm B} \rightarrow CO_{2}$, losing importance with increasing temperature. Since $E_{\alpha}^{k}$ is constant for each reaction and $E_{\alpha}^{k^0}$ is weakly dependent on temperature (see Eqs. \ref{E_alpha_ads_k0}-\ref{E_alpha_recNdif_k0} in Section \ref{Oxidation_of_CO} of the Supporting Information), the overall temperature dependence of each contribution $\epsilon_{\alpha}^{TOF}$ in Eq. \ref{EappTOF_5} is mainly due to (i) the slope $E_{\alpha}^{M}$ of the multiplicity $M_{\alpha}$, as shown in Figs. \ref{Figure04}(b) and \ref{Figure05}(e-f), and (ii) the actual recombination probability, $\omega_{\alpha}^{TOF}$, as shown in Fig. \ref{Figure04}(c). 

Alternatively, considering the presence of a Rate Determining Step (RDS), Fig. \ref{Figure04}(d) shows that the temperature dependence of $E_{app}^{TOF}$ in all three regions is accurately described also by Eq. \ref{EappTOF_6} (absolute error $ | E_{app}^{TOF} - ( E_{\lambda}^{k} + E_{\lambda}^{k^{0}} + E_{\lambda}^{M} ) |  < 0.05$ eV). While at higher temperatures ($\beta < 21$) the RDS is one recombination reaction ($\lambda = CO_{\rm C} + O_{\rm C} \rightarrow CO_{2}$) at lower temperatures ($\beta > 21$) it corresponds to O adsorption ($\lambda = V \rightarrow O$). Based on the similarity of $M_{\alpha}k_{\alpha}$ with respect to the TOF, as shown in Fig. \ref{Figure04}(e), the actual proximity to the $TOF$ ($\sigma_{\alpha}^{TOF}$) is presented in Fig. \ref{Figure04}(f). This allows assigning the RDS, since $\sigma_{\alpha}^{TOF} \approx 1$ for $\alpha = CO_{\rm C} + O_{\rm C} \rightarrow CO_{2}$ and $\alpha = V \rightarrow O$ at high and low temperatures, respectively. 

Fig. \ref{Figure04}(d) shows that merely observing a linear Arrhenius behavior within some range of temperatures (region I) does not guarantee that $E_{app}^{TOF}$ ($\approx 2.87$ eV) corresponds to the highest elementary activation energy in the system $E_{\alpha, max}^{k}$ (=4.6 eV, for the desorption $O_{\rm B} + O_{\rm B} \rightarrow V_{\rm B} + V_{\rm B}$). The RDS depends on the whole reaction mechanism and does not necessarily correspond to the reaction with the highest elementary activation energy. In addition, even if $E_{app}^{TOF}$ can be assigned to one  reaction ($\lambda$), $E_{app}^{TOF}$ should not be identified with the elementary activation energy alone, $E_{\lambda}^{k}$ (= 0 eV, for $\lambda = V \rightarrow O$), since this disregards the configurational contribution, $E_{\lambda}^{M}$ ($\approx$ 2.87 eV, for $\lambda = V \rightarrow O$) and the term $E_{\lambda}^{k^0}$ (negligible here). We conclude that Eqs. \ref{EappTOF_5} and \ref{EappTOF_6} are more accurate than Eq. \ref{E_app_using_sensitivities}, previously applied to region I only, resulting in an error of $0.25$ eV \cite{Meskine2009}. Similarly, Eqs. \ref{EappTOF_5} and \ref{EappTOF_6} are more accurate than the traditional Temkin formula (Eq. \ref{E_app_temkin}), as shown in Section \ref{Wrong_apparent_activation_energies_based_on_the_Temkin_formulation} of the Supporting Information.

In addition to enabling the determination of the RDS, Fig. \ref{Figure04}(f) provides crucial information by showing which reactions affect the $TOF$ significantly, {\it i.e.} the Rate Controlling Steps (RCSs). For $\beta > 21$, in addition to the RDS ($\lambda = V \rightarrow O$), the $TOF$ is sensitive to the three recombination reactions discussed in Figs. \ref{Figure04}(a)-(c), as well as one diffusion type ($CO_{\rm B} \rightarrow CO_{\rm B}$, especially for $\beta \approx 26-28$) and one desorption reaction ($CO_{\rm B} \rightarrow V$, for $\beta \approx 23-24$). For $\beta < 21$, the $TOF$ is sensitive only to the RDS ($CO_{\rm C} + O_{\rm C} \rightarrow CO_{2}$), with a sensitivity spike for one desorption reaction ($O_{\rm C} + O_{\rm C} \rightarrow V_{\rm C} + V_{\rm C}$, at $\beta \sim 19$, approaching 1 sharply from both left and right). 

Although these proximity curves might look whimsical--especially the spikes--they can be easily understood from the actual reaction probabilities shown in Fig. \ref{Figure05}(a). The figure also displays the probability to observe any reaction explicitly contributing to the TOF, $\omega_{TOF}^{R} = TOF / R$, as well as two additional curves, namely, $2\omega_{TOF}^{R}$ and $0.05\omega_{TOF}^{R}$. Any elementary reaction with probability $\omega_{\alpha}^{R}$ between $\omega_{TOF}^{R}$ and $2\omega_{TOF}^{R}$ will lead to proximity values $\sigma_{\alpha}^{TOF}$ between 1 and 0. Likewise, if $\omega_{\alpha}^{R}$ falls between $\omega_{TOF}^{R}$ and $0.05\omega_{TOF}^{R}$ the proximity will lie between 1 and 0.05. 
[See Section \ref{Cut_offs_in_the_proximity} of the Supporting Information for further details about the cut-offs $2\omega_{TOF}^{R}$ and $0.05\omega_{TOF}^{R}$.] 
Thus, a spike in $\sigma_{\alpha}^{TOF}$ (approaching value 1 from left and right) will appear when $\omega_{\alpha}^{R}$ crosses $\omega_{TOF}^{R}$ within a small range of temperature. 

Similarly, any curve for $\sigma_{\alpha}^{TOF}$ in Fig. \ref{Figure04}(f) can be easily interpreted from the actual behavior of the corresponding reaction probability within the band displayed in Fig. \ref{Figure05}(a). Most importantly, Fig. \ref{Figure05}(a) stresses that, in probability space, proximity to the $TOF$ means sensitivity by the TOF. As explained in the last paragraph of Section \ref{Rate Determining Step and Rate Controlling Steps}, the $TOF$ is also sensitive to variations in the rates of those reactions with $\omega_{\alpha}^{R} \sim 1$ through their scaling of time. Such reactions essentially control the total rate $r$ (= $R s$) and, thus, the time increment $\Delta t \propto 1/ r $. In this manner, according to Fig. \ref{Figure05}(a), the $TOF$ will also be sensitive to the adsorption and desorption of CO ($V \rightarrow CO$ and $CO_C \rightarrow V$, respectively), in agreement with Fig. 5 of Ref. \cite{Meskine2009}.

\begin{figure*}[htb!]
  \begin{center}
    \includegraphics[width=1.8\columnwidth]{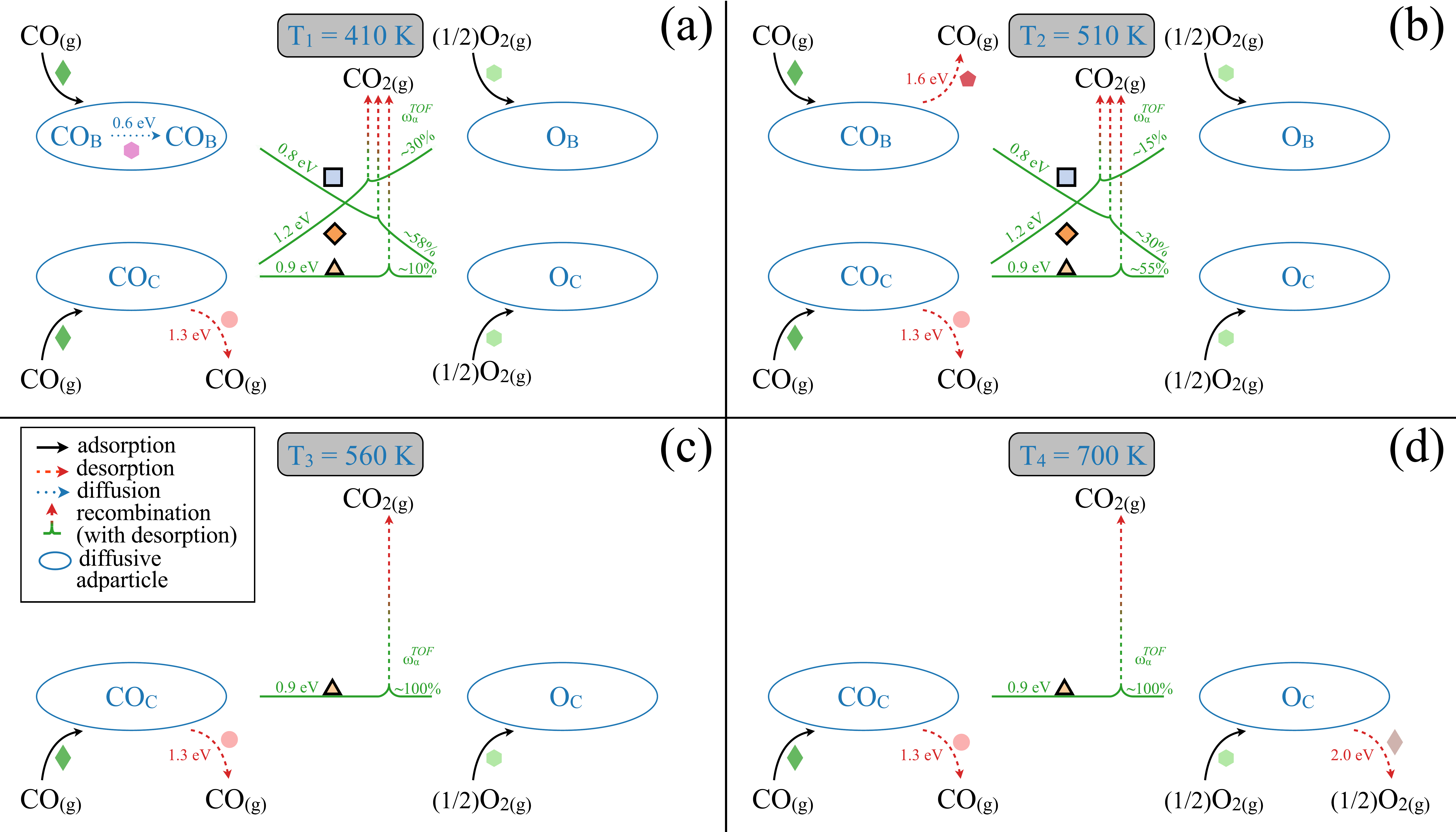}
  \end{center}
  \caption{Temperature dependence for model I ({\it cf} Figs. \ref{Figure04} and \ref{Figure05}): Elementary reactions having a leading role (Rate Controlling Steps, RCSs) according to the multiplicity analysis proposed in this study ($\omega_{\alpha}^{R} \sim \omega_{TOF}^{R}$ and $\omega_{\alpha}^{R} \sim 1$ in Fig. \ref{Figure05}(a)): (a) 410 K, (b) 510 K, (c) 560 K, (d) 700 K.
}
  \label{Figure06}
\end{figure*}

Furthermore, the reaction probabilities of Fig. \ref{Figure05}(a) are useful to directly extract meaningful information about the catalytic process. For this purpose, Fig. \ref{Figure05}(a) is best analyzed jointly with Fig. \ref{Figure05}(b), which shows the temperature dependence of the coverage by all adspecies ($\theta_{\rm CO_{B}}$, $\theta_{\rm CO_{C}}$, $\theta_{\rm O_{B}}$, $\theta_{\rm O_{C}}$, $\theta_{\rm V_{B}}$ and $\theta_{\rm V_{C}}$). For completeness, Fig. \ref{Figure05}(c) additionally shows typical surface morphologies (configurations) for the system at four characteristic temperatures $T_1 < T_2 < T_3 < T_4$ (410, 510, 560 and 700 K). At any temperature, the adsorption of CO and the desorption of CO from C sites ($V \rightarrow CO$ and $CO_{\rm C} \rightarrow V$, respectively) are so overwhelmingly probable ($\omega_{\alpha}^{R} \sim 0.5$) with respect to the $TOF$ ($\omega_{TOF}^{R} \sim 10^{-2}-10^{-7}$) that the two reactions can be regarded as completely equilibrated (one to one), thus minimally interfering with any $TOF$ event. At $T_1$, the next most probable reaction is the desorption of CO from B sites ($CO_{\rm B} \rightarrow V$, with $\omega_{\alpha}^{R} \sim 10^{-4}$ ), which is also equilibrated with the corresponding adsorption of CO at B sites (included in the $V \rightarrow CO$ curve). With probabilities between $3 \times 10^{-7}$ and $2 \times 10^{-8}$, we then find a diffusion reaction ($CO_{\rm B} \rightarrow CO_{\rm B}$), an adsorption reaction ($V \rightarrow O$) and the three recombination reactions already discussed in relation to Figs. \ref{Figure04}(a)-(c). Since the surface is essentially CO-terminated (Fig. \ref{Figure05}(b)), for these recombinations to occur the adsorption of O must take place. In other words, $V \rightarrow O$ is the RDS, in agreement with Fig. \ref{Figure04}(f). The corresponding Rate Controlling Steps (RCSs) at $T_1$ are summarized in Fig. \ref{Figure06}(a).

At $T_2$, recombination now occurs mostly due to the $CO_{\rm C} + O_{\rm C} \rightarrow CO_{2}$ route, rather than $CO_{\rm B} + O_{\rm C} \rightarrow CO_{2}$ (which dominated at $T_1$), while the $CO_{\rm C} + O_{\rm B} \rightarrow CO_{2}$ channel becomes gradually less relevant with increasing temperature. Another difference with respect to $T_1$ is that there is plenty of O on the B sites at $T_2$ (Fig. \ref{Figure05}(b)), but the previous sentence concluded that CO$_{\rm C}$ and CO$_{\rm B}$ typically react with O$_{\rm C}$. Thus, the system is ready to generate CO$_2$ as soon as O is adsorbed on the C sites. In this manner, $V \rightarrow O$ remains as the RDS, as shown in Fig. \ref{Figure04}(f). The corresponding RCSs at $T_2$ are summarized in Fig. \ref{Figure06}(b). At $T_3$, there is plenty of O on both B and C sites (see Fig. \ref{Figure05}(b)) while the small coverage of the C sites by CO is large enough to enable recombination through the $CO_{\rm C} + O_{\rm C} \rightarrow CO_{2}$ route, with probability $\omega_{\alpha}^{R} \sim 9 \times 10^{-3}$ comparable to that for O adsorption ($V \rightarrow O$, with $\omega_{\alpha}^{R} \sim 1 \times 10^{-2}$). Although $CO_{\rm C}$ and $O_{\rm C}$ units are constantly in contact, their recombination takes some time. Thus, the recombination itself is the RDS, in agreement with Fig. \ref{Figure04}(f). Finally, at $T_4$, not only the adsorption and desorption of CO are equilibrated ($\omega_{\alpha}^{R} \sim 0.4$) but also the adsorption and desorption of molecular O$_2$ ($\omega_{\alpha}^{R} \sim 0.2/2=0.1$ and $\sim 0.1$, respectively). Thus, on a mostly O-covered surface (see Fig. \ref{Figure05}(b)), adsorption and desorption of CO at C sites occurs frequently, but hardly ever this leads to a recombination ($CO_{\rm C} + O_{\rm C} \rightarrow CO_{2}$, with $\omega_{\alpha}^{R} \sim 2 \times 10^{-3}$). Thus, the recombination itself is the RDS, in agreement with Fig. \ref{Figure04}(f). The RCSs at $T_3$ and $T_4$ are summarized in Figs. \ref{Figure06}(c)-(d).

Finally, Figs. \ref{Figure05}(d)-(f) show the corresponding contributions to the apparent activation energy for the total rate per site $R_{}$ according to Eq. \ref{Eapp_e_1} (absolute error $ | E_{app}^{R} - \sum_{\alpha \in \{ e \}} \epsilon_{\alpha}^{R} | < 0.03$ eV). This demonstrates that monitoring the multiplicities enables describing both easily and accurately any of the total rates per site ($R_{a}$, $R_{d}$, $R_{h}$, $R_{r}$ and $R_{}$). As in Fig. \ref{Figure03}(a), Fig. \ref{Figure05}(d) confirms that the total rate is dominated by adsorption and desorption reactions, in particular, the adsorption and desorption of CO ($V \rightarrow CO$ and $CO_{\rm C} \rightarrow V$), while the adsorption of O ($V \rightarrow O$) becomes relevant in region C. As indicated above, the $TOF$ is sensitive to variations in the rates of these reactions through their ability to scale the time increment  $\Delta t \propto 1/r$ with $r = R s$.

We stress that the temperature dependencies of $E_{app}^{TOF}$ and $E_{app}^{R}$ are well explained by Eqs. \ref{EappTOF_5} and \ref{EappTOF_6} also for models II-IV (for the oxidation of CO) as well as for a distinctively different model that describes the selective oxidation of NH$_3$ on RuO$_2$(110) (see Sections \ref{Multiplicity_Oxidation_of_CO_using_models_II_III_and_IV} and \ref{Multiplicity_Selective_oxidation_of_NH3} of the Supporting Information, respectively). This is valid even in the case of model IV, which explicitly considers adsorbate-adsorbate interactions. Similarly, based on directly inspecting the corresponding reaction probabilities, essential understanding is obtained about the overall catalytic reaction for each model, including the assignment of the RDS to one or more elementary reactions. These results strongly indicate that the proposed multiplicity analysis can be used to obtain a deep understanding for any reaction mechanism / catalytic model.

\section{Discussion}
\label{Discussion}

\subsection{Novelty}
\label{Novelty}

This study presents the use of the multiplicities to formulate novel expressions for the $TOF$ (Eq.  \ref{TOF_definition}) and its apparent activation energy (Eqs. \ref{EappTOF_5} and \ref{EappTOF_6}), as well as to describe the relative importance of every elementary reaction via the reaction probabilities (Eq. \ref{omega_definition}). The application to two model catalytic reactions (the oxidation of CO on RuO$_2$(110) and the selective oxidation of NH$_3$ on the same surface) and the computational aspects (the Kinetic Monte Carlo simulations) are secondary features, used to confirm the validity of the proposed equations.

The primary result is Eq. \ref{TOF_definition}. This formulation of the $TOF$ follows from the observation that every elementary reaction occurring on a catalyst surface is available at different locations. 
Thus, in addition to a characteristic rate constant, $k_{\alpha}$, each elementary reaction has an associated multiplicity, $M_{\alpha}$, which is directly linked to configurational entropy (Eq. \ref{Eq_ConfigurationalEntropy}). 
While traditionally one considers the adsorbate coverages as the natural variables to describe the system (and, thus, the $TOF$), the proposed master equation (Eq. \ref{Eq_LH_4}) shows that, instead, one may consider the multiplicities of the local configurations as the irreducible variables. 
For spatially extended systems whose morphology (spatial configuration) can be monitored, the multiplicities of the elementary reactions can then be tracked and the proposed expression for the $TOF$ is fully justified.

Considering all elementary reactions, $\{ e \}$, the proposed expression, $TOF = \Sigma_{ \alpha \in \{x\} } M_{\alpha} k_{\alpha}$, focuses on the particular subset of reactions, $\{ x \}$, whose reaction products explicitly contain the desired target molecule (or molecules) in the gas phase. 
If there happens to be an elementary reaction, $\lambda \in \{ e \}$, so that $M_{\lambda} k_{\lambda} = \Sigma_{ \alpha \in \{x\} } M_{\alpha} k_{\alpha}$, then that reaction is the RDS. In this particular case, our expression ($TOF = k_{\lambda} M_{\lambda}$) can be directly compared with traditional formulations ({\it e.g.} $TOF = k_{\lambda} \theta_{A} \theta_{B}$, if the RDS is the recombination of two adsorbates, $A$ and $B$, in the mean field approximation, or $TOF = k_{\lambda} \theta_{A}^{x'} \theta_{B}^{y'}$, considering the two adsorbates have partial reaction orders $x'$ and $y'$, which describe phenomenologically the presence of correlated configurations beyond the mean field approach). Thus, the traditional coverage dependence is replaced with the multiplicity, $M_{\lambda}$, which is an exact measure of the 'concentration of the reaction', {\it i.e.} the reaction abundance per active site, valid within and beyond mean field.

In spite of the simplicity of Eq. \ref{TOF_definition}, we are not aware of any previous, similar approach. Direct formulation of the $TOF$ in terms of the multiplicities (or their traditional counterparts, the coverages by the reaction intermediates) was explicitly disregarded in Ref. \cite{Meskine2009} (see the text after Eq. (9) in that study). However, formulations of the $TOF$ in terms of the coverage of one or several intermediates are a standard procedure in chemical kinetics \cite{Campbell2017, Chorkendorff2003, Lynggaard2004, Bond2005, Bond2006, Teschner2012} (see several examples in Section \ref{Apparent_activation_energy_in_the_Langmuir_Hinshelwood_model} of the Supporting Information).  Furthermore, the present study strongly supports the idea that the $TOF$ is described naturally by using the multiplicities.

Regarding Eq. \ref{EappTOF_5} (Eq. \ref{EappTOF_6}), every configurational contribution $E_{\alpha}^{M}$ ($E_{\lambda}^{M}$) to the apparent activation energy $E_{app}^{TOF}$ reflects the temperature dependence of the coverage for a particular collection of sites. As shown in Section \ref{Apparent_activation_energy_in_the_Langmuir_Hinshelwood_model} of the Supporting Information for the Langmuir-Hinshelwood model with recombination as the RDS ($\lambda = CO_{\rm C} + O_{\rm C} \rightarrow CO_2$), the configurational contribution $E_{\lambda}^{M}$ contains the temperature dependence of $M_{\lambda}$ in the same manner as the Temkin contribution $-x \Delta H_{CO} -y \Delta H_{O}$ carries the temperature dependence for the approximation $M_{\lambda} \approx \theta_{CO} \theta_{O}$. Since here $M_{\lambda}$ characterizes the coverage of all neighbor site pairs occupied by $CO$ and $O$, replacing $M_{\lambda}$ by $\theta_{CO}\theta_{O}$ becomes a poor approximation when the interplay of all reactions leads to structured morphologies ({\it i.e.} non-random configurations). 

Regarding Eq. \ref{omega_definition}, the probability of observing any particular elementary reaction, $\omega_{\alpha}^{R}$, provides a precise measure of the relative importance of every reaction. In addition to enabling a deep understanding of the way the overall reaction is conducted, $\omega_{\alpha}^{R}$ allows easy identification of the Rate Determining Step (RDS), if it exists, as well as the Rate Controlling Steps (RCSs). Overall, this provides a straightforward alternative to computationally-expensive approaches based on the degree of rate control ($\chi_{\alpha}$) and/or the rate sensitivity  ($\xi_{\alpha}$).

\subsection{ Sensitivity analysis }
\label{Sensitivity analysis}

Regarding the analysis of the promotion or hindering of the $TOF$, traditionally $\xi_{\alpha}$ and $\chi_{\alpha_*}$ provide this information by construction, directly measuring the changes in the $TOF$ by varying one rate constant ($\xi_{\alpha}$) or two rate constants ($\chi_{\alpha_*}$) while keeping all other rate constants fixed. In this context, the proposed multiplicity approach should become very useful, substantially reducing the overall cost of the traditional sensitivity analysis. By designating which elementary reactions significantly modify the $TOF$, the sensitivity analysis for all other elementary reactions can be directly discarded, with the corresponding enormous saving in computational effort. 

This is summarized in various plots, such as Fig. \ref{Figure05}(a), where the probability of any elementary reaction--or any desired combination of reactions, such as the $TOF$--is shown as a function of inverse temperature. Similar plots are possible as a function of the partial pressure for any desired gas species. By considering such plots, the sensitivity analysis can be reliably restricted to only those elementary reactions whose probability is either (i) larger than about 0.01 (thus affecting the $TOF$ by scaling the time increment), or (ii) lies within the indicated band around the $TOF$ (thus affecting the $TOF$ by proximity). In other words, the proposed multiplicity analysis performed at fixed conditions directly indicates which elementary rate constants will affect the $TOF$ and which ones will not. The actual promotion or hindering of the $TOF$ can then be determined by performing the sensitivity analysis only on the affecting rate constants.

Regarding the RDS for model I (Fig. \ref{Figure04}(f)), our results agree with (and clarify) the data presented in Fig. 5 of Ref. \cite{Meskine2009} (see Section \ref{Rate_Determining_Step_for_model_I} of the Supporting Information for a deeper comparison). In fact, some of the values shown for the rate sensitivity  $\xi_{\alpha}$ in Fig. 5 of Ref. \cite{Meskine2009} have the same qualitative shape as $E_{app}^{TOF}$ in Fig. \ref{Figure04}(d) and various $E_{\alpha}^{M}$ curves in Fig. \ref{Figure05}(f) of this study. This shows that their sensitivity analysis and our multiplicity approach contain similar information. However, according to Fig. \ref{Figure04}(f) at low temperature, we expect the $TOF$ to be rather sensitive to the same three recombination  reactions that describe $E_{app}^{TOF}$ accurately in Fig. \ref{Figure04}(a). We find it puzzling that no sizable values for $\chi_{\alpha}$ and/or $\xi_{\alpha}$ were found in region I in Ref. \cite{Meskine2009} for any of the three recombination reactions. 

This suggests that, in addition to the large computational effort, the actual numerical determination of some $\xi_{\alpha}$ might be quite difficult in practice, presumably due to the inherent noise in the KMC simulations. As evidenced by the ongoing search for 'practical approaches' \cite{Hoffmann2017}, there is a need to reduce the computational cost of the $\xi_{\alpha}$ analysis. Our method provides an alternative, only requiring the monitoring of the multiplicities of the different reactions, thus reducing the computational burden to a minimum. In particular, our approach avoids the determination of noisy derivatives, thus resulting in clearer trends, and it includes detailed information about the relative competition between the different reactions, simply by plotting the reaction probabilities, as in Fig. \ref{Figure05}(a). Furthermore, our approach distinguishes between two different sources for variations in the $TOF$ (proximity: $\omega_{\alpha}^{R} \sim \omega_{TOF}^{R}$, and scaling: $\omega_{\alpha}^{R} \sim 1$).

\subsection{Comparison to traditional descriptions of $E_{app}^{TOF}$}
\label{Comparison to traditional descriptions of EappTOF}

According to one line of traditional thinking, when there is only one dominating reaction, the apparent activation energy $E_{app}^{TOF}$ coincides with the elementary activation energy $E_{\lambda}^{k}$ of that particular reaction (the RDS or bottleneck). An example is Eq. \ref{E_app_using_sensitivities}, which exactly gives $E_{app}^{TOF} = E_{\lambda}^{k}$ when a single RDS exists. This was seen as a positive feature in Ref. \cite{Meskine2009} (see text after Eq. (12) in that study). However, according to Eq. \ref{EappTOF_6} of this report, a better description when a RDS exists is $E_{app}^{TOF} = E_{\lambda}^{k} + E_{\lambda}^{k^0} + E_{\lambda}^{M}$. Since  $E_{\lambda}^{k^0}$ is typically small, the important difference with respect to such traditional view is that $E_{app}^{TOF}$ differs from $E_{\lambda}^{k}$ due to the presence of an important configurational entropy contribution, $E_{\lambda}^{M}$, which contains the actual changes experienced by the coverage of the collection of sites where the RDS takes place.

Another line of traditional thinking, represented by Eq. \ref{E_app_temkin}, correctly considers the presence of an additional contribution to $E_{app}^{TOF}$, but describes it as a weighted sum of formation enthalpies (or adsorption heats) with phenomenological reaction orders as weights. Although this formulation remains valid beyond the mean field approximation, the reasoning behind is based on general arguments about the mathematical dependence on real-valued powers of the adsorbate coverages in the presence of correlated configurations. Effectively, this transfers the dependence on the spatial configuration (including any possible correlations) into a dependence on gas properties (the partial pressures), thus shifting the focus from the surface to the gas phase and masking the actual microscopic origin, which ultimately lies on the multiplicities themselves, as stressed in the present study. The introduction of the multiplicities in the present work directly enables placing the focus back on the actual structure of the surface.

A recent attempt to explain the apparent activation energy uses a generalized version of Eq. \ref{E_app_using_sensitivities} based on $\chi_{\alpha}$ (instead of $\xi_{\alpha}$) \cite{Jorgensen2017}:
\vspace{-1mm}
\begin{eqnarray}
\hspace{-5mm}
E_{app}^{TOF} & = & \sum_{\alpha \in \{d_{*}, h_{*}, r_{*} \}} \chi_{\alpha} \left( E_{\alpha}^k + k_{B}T + T^2\frac{\partial  ( \Delta   S_{\alpha}^{  k   }  )    }{\partial T}  \right)  \nonumber\label{E_app_jorgensen_1} \\
      &  & + \sum_{\alpha \in \{ a_{*} \}} \chi_{\alpha} \left( E_{\alpha}^k - \frac{k_{B}T}{2}  + k_B T^2\frac{\partial \log  s_{\alpha}    }{\partial T}   \right)  \nonumber \label{E_app_jorgensen_2} \\
      &  & - k_B T^2 \sum_{X} \frac{\partial n_X}{\partial T} \log p_X , \label{E_app_jorgensen_3}
\end{eqnarray}
where $  s_{\alpha}   $ is the sticking probability for adsorption reaction $\alpha$,
$p_X$ is the partial pressure for species $X$ and $n_X$ is the corresponding reaction order, which stems from the assumption of a power-law dependence on pressure \cite{Jorgensen2017}: 
\vspace{-1mm}
\begin{eqnarray}
TOF & = & A e^{-E_{app}^{TOF}/k_B T} \displaystyle\prod_{X} p_X^{n_X} . \label{TOF_PowerLaw}
\end{eqnarray}
Since each $\chi_{\alpha}$ considers simultaneously the forward and backward rates, the summations in $\alpha$ run over the forward reactions only ($\alpha \in \{d_{*}, h_{*}, r_{*} \}$ for desorption, diffusion and recombination, and  $\alpha \in \{a_{*} \}$ for adsorption). 
For diffusion, recombination and desorption, Ref. \cite{Jorgensen2017} assumes the rate constants to be: $k_{\alpha} = k_{\alpha}^0 e^{-E_{\alpha}^k / k_B T}$, where $k_{\alpha}^0 = \frac{ k_B T }{ h } e^{\Delta S_{\alpha}^k / k_B}$. 
Considering the thermodynamic formulation of the reaction rate in TST (see Section \ref{Rate constant for an elementary reaction}), Ref. \cite{Jorgensen2017} effectively approximates the enthalpy change by using the energy barrier ($\Delta H_{\alpha}^k \approx E_{\alpha}^k$). In turn, the rate constants for adsorption in Ref. \cite{Jorgensen2017} are: $k_{\alpha} = s_{\alpha} \cdot \frac{ A_s }{ \sqrt{ 2 \pi m_X k_B T} }$, where $A_s$ is the adsorption site area,  $s_{\alpha}$ is the sticking probability, $m_X$ is the mass of the adsorbed molecule and the typical dependence on pressure $p_X$ (as in $\frac{ A_s p_X}{ \sqrt{ 2 \pi m_X k_B T} }$) is modeled outside $k_{\alpha}$ (see Eq. \ref{TOF_PowerLaw}).  

Using $ k_B T^2 \frac{\partial}{\partial T} = -\frac{\partial}{\partial \beta}$, 
we re-write Eq. \ref{E_app_jorgensen_3} simply as:
\vspace{-1mm}
\begin{eqnarray}
\hspace{-5mm}
E_{app}^{TOF} & = & \sum_{\alpha \in \{ e_{*} \}} \scriptstyle \chi_{\alpha} \left( E_{\alpha}^k + E_{\alpha}^{k^0}  \right) + \displaystyle \sum_{X} \textstyle \frac{\partial n_X}{\partial \beta} \log p_X , \nonumber \\
& & \label{E_app_jorgensen_123b}
\end{eqnarray}
where we have used   
the definition in Eq. \ref{EappTOF_5} for 
$E_{\alpha}^{k^0} = - \frac{ \partial \log k_{\alpha}^0}{ \partial \beta } = k_B T^2 \frac{ \partial \log k_{\alpha}^0} { \partial T} = \frac{ k_B T^2 } { k_{\alpha}^0 } \frac{ \partial k_{\alpha}^0 }{ \partial T}$,
resulting in 
$E_{\alpha}^{k^0} = k_B T + T^2 \frac{ \partial ( \Delta S_{\alpha}^k ) } { \partial T } $
for diffusion/recombination/desorption, and  
$ E_{\alpha}^{k^0} = - \frac{ k_B T }{ 2 } + k_B T^2\frac{\partial \log s_{\alpha}}{\partial T} $
for adsorption.

In this study, we consider various expressions for $E_{\alpha}^{k^0}$ (see Eqs. \ref{E_alpha_ads_k0}-\ref{E_alpha_recNdif_k0} in Section \ref{Oxidation_of_CO} of the Supporting Information). For desorption, as an example, equating the value of $E_{\alpha}^{k^0}$ in Ref. \cite{Jorgensen2017}  ($E_{\alpha}^{k^0}= k_B T + T^2 \frac{ \partial (\Delta S_{\alpha}^k ) } { \partial T } $) and that in Eq. \ref{E_alpha_des_k0} of the Supporting Information gives: $T^2 \frac{ \partial ( \Delta S_{\alpha}^k ) } { \partial T } = 2 k_B T + \frac{h \nu_{\rm X} e^{-h \nu_{\rm X} / k_BT}}{1-e^{-h \nu_{\rm X} / k_BT}}$. Thus, the present study considers the temperature dependence of the entropy barrier ($\Delta S_{\alpha}^k$) for some reactions. 

The use of the standard expression for non-activated adsorption ($k_{\alpha} = s_{\alpha} \cdot \frac{p_X A_s} {\sqrt{2 \pi m_X k_B T}}$, see Section \ref{Oxidation_of_CO} of the Supporting Information) does not limit the conclusions of the present report. Although we may complicate the study by including more complex adsorption rate constants involving entropy barriers and/or energy barriers, this will only affect the actual value of $k_{\alpha}$ for the modified reactions and, accordingly, the value of $E_{\alpha}^k + E_{\alpha}^{k^0}$. However, the important configurational term emphasized in this report, $E_{\alpha}^{M}$, will still be needed in order to describe $E_{app}^{TOF}$ properly according to Eq. \ref{EappTOF_5}.

Although Eq. \ref{E_app_jorgensen_123b} shares two energy contributions with  Eq. \ref{EappTOF_5}, namely, $E_{\alpha}^k + E_{\alpha}^{k^0}$, there are marked differences between the two formulations. In Eq. \ref{E_app_jorgensen_123b}, the first summation is over all forward reactions ($\alpha \in \{ e_{*} \}$) while the corresponding summation in Eq. \ref{EappTOF_5} is over those reactions explicitly contributing to the $TOF$. Similarly, the first summation in Eq. \ref{E_app_jorgensen_123b} uses $\chi_{\alpha}$ as the weight, thus making it difficult to apply this formula to systems outside a mean field formulation (due to the huge computational effort as well as the impact on accuracy due to the numerical derivatives for noisy variables). On the other hand, the weights appearing in Eq. \ref{EappTOF_5} are reaction probabilities,
which can be effortlessly determined and easily interpreted within the range [0,1]. 

In addition, Eq. \ref{E_app_jorgensen_123b} contains a second summation over the partial pressures of the gas species, directly resulting from the power-law approximation for the overall prefactor of the $TOF$ (Eq. \ref{TOF_PowerLaw}). In comparison, our formulation avoids any such approximation, not even including an overall prefactor (see Eq. \ref{TOF_definition}), simply recognizing that every elementary reaction is present on the surface with a relative abundance ($M_{\alpha}$).
The use of the multiplicities and the lack of an overall prefactor makes a key difference, leading to a single summation with probabilities as weights (Eq. \ref{EappTOF_5}) instead of splitting the dependence into two complex summations (Eq. \ref{E_app_jorgensen_123b}).

\subsection{ Eley-Rideal mechanism }
\label{Eley-Rideal mechanism}

For reactions between an adsorbed molecule and a gas molecule, the Eley-Rideal mechanism can be formulated as:
\begin{equation}
\begin{tabular}{ccccc}
$A_{(g)}$ & & & & $C_{(g)}$ \\
${\scriptstyle \colvec{2}{k_{d}^{A}}{[R1]} } \updownarrows {\scriptstyle  \colvec{2}{k_{a}^{A}p_{A}}{[R2]} }$ & & & & ${\scriptstyle \colvec{2}{k_{d}^{C}}{[R5]} } \updownarrows {\scriptstyle \colvec{2}{k_{a}^{C}p_{C}}{[R6]} }$ \\
$A$ & $+$ & $B_{(g)}$ & $\xrightarrow[RDS]{k_{r} [R7]}$ & $C$ , \\
 & & ${\scriptstyle \colvec{2}{k_{d}^{B}}{[R3]} } \updownarrows {\scriptstyle \colvec{2}{k_{a}^{B}p_{B}}{[R4]} }$ & &  \\
 & & $B$ & &  
\label{ER_eq}
\end{tabular}
\end{equation}
where typically the irreversible reaction between $A$ and $B_{(g)}$ is considered as the Rate Determining Step (RDS). Thus, traditionally one writes: $TOF = r_{7} \approx k_{r} \theta_A p_B $ (mean-field approximation). 
Further assuming Langmuir adsorption equilibria one obtains: $\theta_{A} = K_{A}p_{A} / D $, with $D = 1 + K_{A}p_{A} + K_{B}p_{B} + K_{C}p_{C}$ and $K_{X} \propto e^{\Delta H_{X} /k_BT}$, with $\Delta H_{X}$ the heat of adsorption of $X$, as described in the Introduction, before Eq. \ref{E_app_temkin}.
This directly leads to the traditional expression: $TOF \approx k_{r} \theta_A p_B = \frac{ k_{r} }{ K_{B} } (K_{A}p_{A})  (K_{B}p_B) / D = k_{r} K_{B}^{-1} (K_{A}p_{A})^x (K_{B}p_{B})^{y} (K_{C}p_{C})^z $, where $x$, $y$ and $z$ are the partial reaction orders. 
Thus, the general expression in Eq. \ref{E_app_temkin} for $E_{app}^{TOF}$ remains valid for the Eley-Rideal mechanism. 
Even if the adsorbates are not well-mixed on the catalyst surface ({\it e.g.} forming islands, so that $B_{(g)}$ may react with $A$ only if $A$ is located at specific sites, {\it e.g.} along the island perimeters), one can still write: $TOF \approx k_{r} \theta_A^{x'} p_B$, which leads to the same general dependence for $E_{app}^{TOF}$ (Eq. \ref{E_app_temkin}).

In comparison, our formulation leads to: $TOF = M_{r} k_{r}$, where $M_{r}$ is the multiplicity of the local configuration where the recombination reaction $A + B_{(g)} \rightarrow C$ can be performed. Thus, disregarding the small contribution $E_{r}^{k^0}$, the apparent activation energy is given by: $E_{app}^{TOF} \approx E_{r}^{k} + E_{r}^{M}$. This way, $E_{app}^{TOF}$ differs from $E_{r}^{k}$ due to the configurational entropy contribution, $E_{r}^{M}$, which contains the actual change with temperature in the multiplicity of the local configuration where the recombination reaction can be performed. More generally, even if the RDS cannot be clearly assigned to any particular elementary reaction, the proposed multiplicity approach allows describing any regime of Eq. \ref{ER_eq}, especially for the study of configurational correlations appearing beyond the mean field approximation in systems with a spatial representation.

\section{Conclusions}
\label{Conclusions}

Focusing on the description of heterogeneous catalysis beyond the mean field approximation, the traditional formulation of the turnover frequency ($TOF$) in terms of the coverage by certain reaction intermediates is generalized by considering the multiplicity of each elementary reaction. Directly characterizing the number of precisely those surface sites involved in each elementary reaction, the multiplicities enable determining the changes experienced in configurational entropy with temperature. This allows formulating the probability of observing any particular elementary reaction, thus providing a complete understanding of the relative importance of every reaction in the overall network. In addition, it allows identifying the Rate Determining Step (RDS), if it exists, as well as the Rate Controlling Steps (RCSs). In this manner, monitoring the multiplicities provides a straightforward alternative to computationally-expensive approaches based on the Degree of Rate Control ($\chi_{\alpha}$) and/or the Degree of Rate Sensitivity  ($\xi_{\alpha}$).

The use of the multiplicities also allows formulating a simple expression to describe the temperature dependence of the apparent activation energy of the $TOF$ ($E_{app}^{TOF}$). Even in the simplest case, when $E_{app}^{TOF}$ remains constant within some temperature range, we show that $E_{app}^{TOF}$ does not correspond to the largest elementary activation energy available in the system, as still believed by some researchers. In fact, $E_{app}^{TOF}$ does not even correspond to the elementary activation energy of the RDS, when it exists, as also amply believed. In addition to the elementary activation energy of the RDS, $E_{app}^{TOF}$ contains an important, unbound configurational entropy contribution from the temperature dependence of the multiplicity of the dominating reaction ({\it i.e.} the coverage for those surface sites participating in the RDS). Due to this contribution, $E_{app}^{TOF}$ may depart from a constant value even when a single RDS is controlling the overall reaction.

In comparison, the traditional Temkin formulation of $E_{app}^{TOF}$ in terms of the formation enthalpies (or adsorption heats) of one or several intermediates in typical Langmuir-Hinshelwood and/or Eley-Rideal mechanisms is limited in practice by difficulties in determining the required reaction orders.
Similarly, alternative formulations of $E_{app}^{TOF}$ in terms of sensitivities (Eqs. \ref{E_app_using_sensitivities} and \ref{E_app_jorgensen_3}) also suffer in practice from difficulties in determining the actual sensitivities as well as from underlying assumptions about the existence and mathematical form of an overall prefactor. Altogether, our results strongly indicate that monitoring the surface morphology should allow a deeper understanding of heterogeneous catalysis as an alternative to focusing on the determination of reaction orders and/or sensitivities.

\section*{Supporting Information Avail-able}
 A PDF file is provided with the following content: (\ref{Apparent_activation_energy_in_the_Langmuir_Hinshelwood_model}) Apparent activation energy in the Langmuir-Hinshelwood model, (\ref{Description_of_the_elementary_reactions}) Description of the elementary reactions: \ref{Oxidation_of_CO} Oxidation of CO [with Tables \ref{Table_Processes} and \ref{tb_rates}], and \ref{Selective_oxidation_of_NH_3} Selective oxidation of NH$_3$ [with Table \ref{tb_rates_NH3}], (\ref{computational_method}) Computational method, (\ref{Comparison_to_previous_results_for_additional_models}) Comparison to previous $TOF$ results for  additional models: \ref{Oxidation_of_CO_using_models_II_III_and_IV} Oxidation of CO using models II, III and IV [with Figure \ref{Figure1S}], and \ref{Comp_Selective_oxidation_of_NH_3} Selective oxidation of NH$_3$ [with Figure \ref{Figure2S}], (\ref{Wrong_apparent_activation_energies_based_on_the_Temkin_formulation}) Wrong apparent activation energies based on the Tempkin formulation [with Examples S1 and S2], (\ref{Cut_offs_in_the_proximity}) Cut-offs in the proximity $\sigma_{\alpha}^{TOF}$, (\ref{Multiplicity_analysis_for_additional_models}) Multiplicity analysis for additional models: \ref{Multiplicity_Oxidation_of_CO_using_models_II_III_and_IV} Oxidation of CO using models II, III and IV [with Figures \ref{Figure3S} and \ref{Figure4S}], and \ref{Multiplicity_Selective_oxidation_of_NH3} Selective oxidation of NH$_3$ [with Figure \ref{Figure5S}], (\ref{Rate_Determining_Step_for_model_I}) Rate Determining Step for model I.

\section*{Acknowledgments}
We are thankful to technical contributions by K. Valencia-Guinot in the computational implementation during the initial stage of the study as part of her Final Degree Assignment (TFG, UPV/EHU). We acknowledge support by the 2015/01 postdoctoral contract by the DIPC. The KMC calculations were performed on the ATLAS supercomputer in the DIPC.

\bibliography{gosalvez_bib}

\end{document}


\title[A microscopic perspective on heterogeneous catalysis - SupInfo]
  {Supporting information: \\
  A microscopic perspective on heterogeneous catalysis}

\author{Miguel~A.~Gosalvez$^{1,2,3}$}
\email[]{miguelangel.gosalvez@ehu.es \\ \\ {\bf References:} }

\author{Joseba~Alberdi-Rodriguez$^{2}$}

\affiliation{$^1$
Dept. of Materials Physics, University of the Basque Country UPV/EHU, 20018 Donostia-San Sebasti\'an, Spain
}
\affiliation{$^2$
Donostia International Physics Center (DIPC), 20018 Donostia-San Sebasti\'an, Spain
}
\affiliation{$^3$
Centro de F\'isica de Materiales CFM-Materials Physics Center MPC, centro mixto CSIC -- UPV/EHU, 20018 Donostia-San Sebasti\'an, Spain
}

\date{\today}

\maketitle

\section{Apparent activation energy in the Langmuir-Hinshelwood model}
\label{Apparent_activation_energy_in_the_Langmuir_Hinshelwood_model}

In a typical Langmuir-Hinshelwood mechanism,
\begin{equation}
\begin{tabular}{ccccc}
$A_{(g)}$ & & $\frac{1}{2} {B_2}_{(g)}$ & & $AB_{(g)}$ \\
${\scriptstyle k_{d}^{A}} \updownarrows {\scriptstyle k_{a}^{A}p_{A}}$ & & ${\scriptstyle k_{d}^{B}} \updownarrows {\scriptstyle k_{a}^{B}p_{B}}$ & & ${\scriptstyle k_{d}^{AB}} \updownarrows {\scriptstyle k_{a}^{AB}p_{AB}}$ \\
$A$ & $+$ & $B$ & $\xrightarrow[RDS]{k_3}$ & $AB$ 
\end{tabular},
\end{equation}
molecules $A$, $B_2$ and $AB$ with partial pressures $p_{A}$, $p_{B}$ and $p_{AB}$ compete for adsorption on the same surface sites and the reaction between $A$ and $B$ adsorbates generates the adspecies $AB$ at a rate $k_3 \propto e^{-E_{3}^{k} \beta}$, which is the Rate Determining Step (RDS). [Here, $\beta=1/k_BT$.] This means that $k_3$ is much smaller than the adsorption and desorption rate constants of the reactants and products ($k_{a}^{X}p_{X}$ and $k_{d}^{X}$, with $X=A, B, AB$). Assuming the adsorbates $A$, $B$ and $AB$ are highly mobile and freely intermix (random homogeneous mixing or mean field approximation), the rate of production of $AB$ per unit area is traditionally described as: $TOF = k_3 \theta_{A} \theta_{B}$, where the coverages are written out assuming Langmuir-like adsorption-desorption equilibrium for $A$, $B$ and $AB$: $\theta_{A} = K_{A}p_{A} / ( 1 + K_{A}p_{A} + \sqrt{K_{B}p_{B}} + K_{AB}p_{AB})$, $\theta_{B} = \sqrt{K_{B}p_{B}} / ( 1 + K_{A}p_{A} + \sqrt{K_{B}p_{B}} + K_{AB}p_{AB})$ and $\theta_{AB} = \sqrt{K_{AB}p_{AB}} / ( 1 + K_{A}p_{A} + \sqrt{K_{B}p_{B}} + K_{AB}p_{AB})$, where $K_{X} = k_{a}^{X} / k_{d}^{X} \propto e^{\Delta H_{X} \beta}$ is the equilibrium constant for adsorption-desorption of molecule $X$, with $\Delta H_{X} = E_{d}^{X} - E_{a}^{X}$ the formation enthalpy (or heat of adsorption) of $X$ \cite{Chorkendorff2003, Lynggaard2004, Bond2006}. Here, $E_{a}^{X}$ ($E_{d}^{X}$) is the atomistic activation energy for adsorption (desorption) of $X$ and the temperature dependence of $K_{X}$ can be easily obtained by considering that $k_{a}^{X} \propto e^{-E_a^{X} \beta}$ and $k_{d}^{X} \propto e^{-E_d^{X} \beta}$.

If $A$ is strongly adsorbed and both $B$ and $AB$ are weakly adsorbed, traditionally one obtains: $TOF = k_3 (K_{A}p_{A})^{-1}(K_{B}p_{B})^{1/2} \propto e^{-E_{3}^{k} \beta} e^{-\Delta H_{A} \beta} e^{\frac{1}{2} \Delta H_{B} \beta} $. Since by definition we also have that $TOF \propto e^{-E_{app}^{TOF} \beta}$, the apparent activation energy is identified as: $E_{app}^{TOF} = E_{3}^{k} + \Delta H_{A} - \frac{1}{2} \Delta H_{B}$. In turn, if $B_2$ is strongly adsorbed and both $A$ and $AB$ are weakly adsorbed, one obtains: $TOF = k_3 (K_{A}p_{A})(K_{B}p_{B})^{-1/2} \propto e^{-E_{3}^{k} \beta} e^{\Delta H_{A} \beta} e^{-\frac{1}{2} \Delta H_{B} \beta} $ and $E_{app}^{TOF} = E_{3}^{k} - \Delta H_{A} + \frac{1}{2} \Delta H_{B}$. Similarly, if both $A$ and $B$ are weakly adsorbed and $AB$ is strongly adsorbed, one obtains: $TOF = k_3 (K_{A}p_{A})(K_{B}p_{B})^{1/2}(K_{AB}p_{AB})^{-2}$ and $E_{app}^{TOF} = E_{3}^{k} - \Delta H_{A} - \frac{1}{2} \Delta H_{B} + 2 \Delta H_{AB}$. Thus, in general, for some suitable range of pressure and temperature, one may use the phenomenological Power Rate Law, $TOF = k_3 (K_{A}p_{A})^{x}(K_{B}p_{B})^{y}(K_{AB}p_{AB})^{z}$, where $x$, $y$ and $z$ are the reaction orders for $A$, $B$ and $AB$, respectively, which leads to the Temkin formula \cite{Chorkendorff2003, Lynggaard2004, Bond2006}: $E_{app}^{TOF} = E_{3}^{k} -x \Delta H_{A} -y \Delta H_{B} -z \Delta H_{AB}$.

Our formalism (as proposed in Section \ref{Theory} of the main report) agrees completely with these descriptions, although we substitute $\theta_{A} \theta_{B}$ by $M_3$ in the expression for the $TOF$ ({\it i.e.} $TOF=k_3 M_3$, where $M_3  \propto e^{-E_{3}^{M} \beta}$ is the multiplicity for the recombination process) and focus on determining $M_3$ instead of making assumptions on its dependence on pressure and temperature. This is useful when the homogeneous mixing approximation fails and/or the adsorption-desorption equilibria for $A$ and/or $B$ and/or $AB$ do not hold. We obtain $E_{app}^{TOF} = E_{3}^{k} + E_{3}^{M} + E_{3}^{k^0} + E_{3}^{M^0}$ (see Eq. \ref{EappTOF_6} of the main report), where $E_{3}^{k^0}$ and $E_{3}^{M^0}$ are usually small while $E_{3}^{M}$ contains the temperature dependence of $M_3$ in the same way as $-x \Delta H_{A} -y \Delta H_{B}$ carries that dependence for $\theta_{A} \theta_{B} \sim (K_{A}p_{A})^{x}(K_{B}p_{B})^{y}$ in the Temkin formulation.

If instead the adsorption of $B_2$ is the RDS,
\begin{equation}
\begin{tabular}{ccccc}
$A_{(g)}$ & & $\frac{1}{2} {B_2}_{(g)}$ & & $AB_{(g)}$ \\
${\scriptstyle k_{d}^{A}} \updownarrows {\scriptstyle k_{a}^{A}p_{A}}$ & & ${\scriptstyle RDS} \downarrow {\scriptstyle k_{a}^{B}p_{B}}$ & & ${\scriptstyle k_{d}^{AB}} \updownarrows {\scriptstyle k_{a}^{AB}p_{AB}}$ \\
$A$ & $+$ & $B$ & $\underset{k_{-3}}{\overset{k_3}{\rightleftarrows}}$ & $AB$ 
\end{tabular},
\end{equation}
traditionally one will write: $TOF = k_{a}^{B}p_{B}\theta_{*}^2$, where $\theta_{*}$ is the coverage by all empty sites and $\theta_{*}^2$ describes the coverage by all empty pairs of sites in the homogeneous mixing approximation. Since the adsorption of $B_2$ is the RDS, traditionally one assumes adsorption-desorption equilibrium for $A$ and $AB$, thus leading to the Langmuir isotherm: $\theta_{A} = K_{A}p_{A}/(1+K_{A}p_{A}+K_{AB}p_{AB})$, $\theta_{AB} = K_{AB}p_{AB}/(1+K_{A}p_{A}+K_{AB}p_{AB})$ and $\theta_{*} = 1/(1+K_{A}p_{A}+K_{AB}p_{AB})$. If $A$ ($AB$) is strongly (weakly) adsorbed, then $\theta_{*} \approx (K_{A}p_{A})^{-1}$ and one obtains: $TOF = k_{a}^{B}p_{B}(K_{A}p_{A})^{-2}$ with $E_{app}^{TOF} = E_{a}^{B} + 2 \Delta H_{A}$. If $AB$ ($A$) is strongly (weakly) adsorbed, then $\theta_{*} \approx (K_{AB}p_{AB})^{-1}$ and one gets: $TOF = k_{a}^{B}p_{B}(K_{AB}p_{AB})^{-2}$ with $E_{app}^{TOF} = E_{a}^{B} + 2 \Delta H_{AB}$. As before, in general we can write the phenomenological expression $TOF = k_{a}^{B}p_{B} (K_{A}p_{A})^{x}(K_{AB}p_{AB})^{z}$ and, thus, $E_{app}^{TOF} = E_{a}^{B} -x \Delta H_{A} -z \Delta H_{AB}$.

Similarly, if the adsorption of $A$ is the RDS,
\begin{equation}
\begin{tabular}{ccccc}
$A_{(g)}$ & & $\frac{1}{2} {B_2}_{(g)}$ & & $AB_{(g)}$ \\
${\scriptstyle RDS} \downarrow {\scriptstyle k_{a}^{A}p_{A}}$ & & ${\scriptstyle k_{d}^{B}} \updownarrows {\scriptstyle k_{a}^{B}p_{B}}$ & & ${\scriptstyle k_{d}^{AB}} \updownarrows {\scriptstyle k_{a}^{AB}p_{AB}}$ \\
$A$ & $+$ & $B$ & $\underset{k_{-3}}{\overset{k_3}{\rightleftarrows}}$ & $AB$ 
\end{tabular},
\end{equation}
traditionally one will write: $TOF = k_{a}^{A}p_{A}\theta_{*}$. Considering the adsorption-desorption equilibrium for $B$ and $AB$ leads to the Langmuir isotherm: $\theta_{B} = \sqrt{K_{B}p_{B}}/(1+\sqrt{K_{B}p_{B}}+K_{AB}p_{AB})$, $\theta_{AB} = K_{AB}p_{AB}/(1+\sqrt{K_{B}p_{B}}+K_{AB}p_{AB})$ and $\theta_{*} = 1/(1+\sqrt{K_{B}p_{B}}+K_{AB}p_{AB})$. If $B$ ($AB$) is strongly (weakly) adsorbed, then $\theta_{*} \approx (K_{B}p_{B})^{-1/2}$ and one obtains: $TOF = k_{a}^{A}p_{A}(K_{A}p_{A})^{-1/2}$ with $E_{app}^{TOF} = E_{a}^{B} + \frac{1}{2} \Delta H_{A}$. If $AB$ ($B$) is strongly (weakly) adsorbed, then $\theta_{*} \approx (K_{AB}p_{AB})^{-1}$ and one gets: $TOF = k_{a}^{A}p_{A}(K_{AB}p_{AB})^{-1}$ with $E_{app}^{TOF} = E_{a}^{B} + \Delta H_{AB}$. In general, as previously, we can write the phenomenological expression $TOF = k_{a}^{A}p_{A} (K_{B}p_{B})^{y}(K_{AB}p_{AB})^{z}$ and, thus, $E_{app}^{TOF} = E_{a}^{B} -y \Delta H_{B} -z \Delta H_{AB}$.

Finally, if the desorption of $AB$ is the RDS,
\begin{equation}
\begin{tabular}{ccccc}
$A_{(g)}$ & & $\frac{1}{2} {B_2}_{(g)}$ & & $AB_{(g)}$ \\
${\scriptstyle k_{d}^{A}} \updownarrows {\scriptstyle k_{a}^{A}p_{A}}$ & & ${\scriptstyle k_{d}^{B}} \updownarrows {\scriptstyle k_{a}^{B}p_{B}}$ & & ${\scriptstyle k_{d}^{AB}} \uparrow {\scriptstyle RDS}$ \\
$A$ & $+$ & $B$ & $\underset{k_{-3}}{\overset{k_3}{\rightleftarrows}}$ & $AB$ 
\end{tabular},
\end{equation}
traditionally one will write: $TOF = k_{d}^{AB}\theta_{AB} = k_{d}^{AB} K_{3}  \theta_{A} \theta_{B}$, where we have considered the equilibrium in the recombination reaction ($r_3 = k_{3} \theta_{A} \theta_{B} - k_{-3} \theta_{AB} = 0$), which gives: $\theta_{AB} = K_{3}  \theta_{A} \theta_{B}$, with $K_{3} = \frac{k_{3}}{k_{-3}}$. As previously, considering the adsorption-desorption equilibrium for $A$ and $B$ leads to the Langmuir isotherm: $\theta_{A} = K_{A}p_{A}/(1+K_{A}p_{A}+\sqrt{K_{B}p_{B}})$, $\theta_{B} = \sqrt{K_{B}p_{B}}/(1+K_{A}p_{A}+\sqrt{K_{B}p_{B}})$ and $\theta_{*} = 1/(1+K_{A}p_{A}+\sqrt{K_{B}p_{B}})$. If $A$ ($B$) is strongly (weakly) adsorbed, then one gets: $TOF = k_{d}^{AB}K_{3}(K_{A}p_{A})^{-1}(K_{B}p_{B})^{1/2}$ with $E_{app}^{TOF} = E_{d}^{AB} - \Delta H_{3} + \Delta H_{A} - \frac{1}{2}\Delta H_{B}$, where $\Delta H_3 = E_{-3}^{k} - E_{3}^{k}$, with $E_{3}^{k}$ ($E_{-3}^{k}$) the activation energy for the forward (backward) recombination reaction. If $B$ ($A$) is strongly (weakly) adsorbed, then one gets: $TOF = k_{d}^{AB}K_{3}(K_{A}p_{A})(K_{B}p_{B})^{-1/2}$ with $E_{app}^{TOF} = E_{d}^{AB} - \Delta H_{3} - \Delta H_{A} + \frac{1}{2}\Delta H_{B}$. As before, we can write the general expression $TOF = k_{d}^{AB} K_{3}(K_{A}p_{A})^{x}(K_{B}p_{B})^{y}$ and, thus, $E_{app}^{TOF} = E_{d}^{AB} - \Delta H_{3} -x \Delta H_{A} -y \Delta H_{B}$.

\section{Description of the elementary reactions}
\label{Description_of_the_elementary_reactions}

\subsection{Oxidation of CO}
\label{Oxidation_of_CO}

Table \ref{Table_Processes} provides the 21 elementary reactions considered in models I, I-bis, II, III and IV for the oxidation of CO on RuO$_2$(110). The data for models I-bis, II and III were collected in one publication by Hess et al. \cite{Hess2012}, based on the work by Reuter and Scheffler \cite{Reuter2006}, Seitsonen and Over \cite{Seitsonen2009}, and Kiejna et al. \cite{Kiejna2006}, respectively. Model I corresponds to the original report by Reuter and Scheffler \cite{Reuter2006}, where (i) the final values (used in their KMC simulations) regarding the activation energies for the four recombination processes differ from those collected by Hess et al. in model I-bis (which is thus discarded in this study), and (ii) some attempt frequencies $k_{\alpha}^{0}$ were determined differently from models II, III and IV (and the discarded I-bis), as described in those reports and summarized below. In turn, model IV corresponds to our implementation of the parameter set reported by Farkas et al. \cite{Farkas2012}. Based on experiment, this model additionally contains repulsion between nearest neighbor (NN) COs located at C sites, which leads to several differentiated processes (rows 22 through 29). Depending on temperature and pressure, some of these models are dominated by adsorption-desorption processes while others are dominated by diffusion events. 

In all four models the adsorption barrier is zero ($E_{\alpha_{\downarrow}}^{k} = 0$ eV, $\alpha_{\downarrow} = 1, 2, 3, 4$). Considering $k_{\alpha_{\downarrow}}^0$ is the attempt frequency from kinetic gas theory (the number of collisions per site per unit time), the adsorption rate constant is:
 \begin{equation}
   k_{\alpha_{\downarrow}} = k_{\alpha_{\downarrow}}^0  e^{-E_{\alpha_{\downarrow}}^{k} / k_BT}, \; \; \alpha_{\downarrow} = 1, 2, 3, 4,
   \label{eq_adsorption}
 \end{equation}
where $k_{\alpha_{\downarrow}}^0 = s \cdot \frac{P_{\rm X} A_s} {\sqrt{2 \pi m_{\rm X} k_B T}}$, X = CO or O$_2$ indicates the gas species, $s$ is the sticking coefficient (1/2 for model I and 1 for models II - IV), $P_{\rm X}$ is the partial pressure for species X, $A_s$ is the area assigned to the adsorption site (10.03 {\AA}$^2$ for both B and C sites), and $m_X$ is the atomic weight for species X ($m_{\rm CO}=28$ g/mol and $m_{\rm O_2}=32$ g/mol).

Since the adsorption of O$_2$ requires two nearest neighbor empty sites, every empty site having at least one empty neighbor is assigned an adsorption rate for atomic O ($k_{V \rightarrow O}$) that is half the adsorption rate for molecular O$_2$ ($k_{V_{2} \rightarrow O_{2}}$): $k_{V \rightarrow O} = \frac{1}{2} k_{V_{2} \rightarrow O_{2}}$. Here, $k_{V_{2} \rightarrow O_{2}} = k_{O_{2}\downarrow}$, as given in Eq. \ref{eq_adsorption}. Accordingly, when a process with rate $k_{V \rightarrow O}$ is selected during a simulation, the adsorption of one molecule (two atoms) is performed.

In this context, $M_{V \rightarrow O} k_{V \rightarrow O}$ ($= 2 M_{V_2 \rightarrow O_2} \frac{1}{2} k_{V_2 \rightarrow O_2} = M_{V_2 \rightarrow O_2} k_{V_2 \rightarrow O_2}$) is the total adsorption rate of O$_2$ molecules per active site, where we have used the fact that the multiplicity of empty site pairs ($M_{V_2 \rightarrow O_2}$) is half the multiplicity of empty sites having at least one empty neighbor ($M_{V \rightarrow O}$): $M_{V_2 \rightarrow O_2} = \frac{1}{2} M_{V \rightarrow O}$. Since the total adsorption rate of O atoms per active site is twice the total adsorption rate of O$_2$ molecules per active site, $2M_{V \rightarrow O} k_{V \rightarrow O}$ is assigned to the total adsorption rate of O atoms per active site. Similarly, $M_{V \rightarrow O}k_{V \rightarrow O} / R$ ($= M_{V_2 \rightarrow O_2} k_{V_2 \rightarrow O_2} / R$) is the probability to observe the adsorption of a molecule and $2M_{V \rightarrow O}k_{V \rightarrow O} / R$ is the probability to observe the adsorption of an atom. Thus, the probability of adsorbing an O atom is twice that of adsorbing an O$_2$ molecule.

In all plots of the report, the label $V \rightarrow O$ refers to the adsorption of atomic O. Thus, for the plots showing the temperature dependence of the total rate per active site for each elementary reaction ($M_{\alpha}k_{\alpha}$ {\it vs} $\beta$) [{\it i.e.} Fig. \ref{Figure04}(e) of the main text and Figs. \ref{Figure3S}(e)-\ref{Figure4S}(e) of this Supporting Information] we display $2M_{V \rightarrow O}k_{V \rightarrow O}$ ({\it i.e.} the total rate of adsorption of O atoms per active site). Similarly, for the plots showing the reaction probabilities ($\omega_{\alpha}^{R}$ {\it vs} $\beta$) [{\it i.e.} Fig. \ref{Figure05}(a) of the main text and Figs. \ref{Figure3S}(g)-\ref{Figure4S}(g) of this Supporting Information], we display $2M_{V \rightarrow O}k_{V \rightarrow O} / R$ ({\it i.e.} the probability of adsorption of O atoms).

\begin{table*}[htb]
\centering
\caption{Elementary reactions, indicating the attempt frequency ($k_{\alpha}^{0}$, 1/s) and activation energy ($E_{\alpha}^{k}$, eV or KJ/mol) used in four different models for the same reaction mechanism (oxidation of CO on RuO$_2$(110)): I. Reuter \cite{Reuter2006} / I-bis. (discarded) \cite{Hess2012, Reuter2006}, II. Seitsonen \cite{Hess2012, Seitsonen2009}, III. Kiejna \cite{Hess2012, Kiejna2006}, and IV. Farkas \cite{Farkas2012}. Model IV contains repulsion between nearest neighbor (NN) COs located at C sites, which leads to several differentiated reactions (rows 22 through 29).}
\label{Table_Processes}
{\tiny
\begin{tabular}{llllllllll}
  $\alpha$ & \begin{tabular}[c]{@{}l@{}}Type \end{tabular} & Process                     & \begin{tabular}[c]{@{}l@{}}Attempt freq. \\ (1/s)\end{tabular} & \begin{tabular}[c]{@{}l@{}}I. Reuter\\ (eV)\end{tabular} & \begin{tabular}[c]{@{}l@{}}I-bis. (discarded) \\ (eV)\end{tabular} & \begin{tabular}[c]{@{}l@{}}II. Seitsonen\\ (eV)\end{tabular} & \begin{tabular}[c]{@{}l@{}}III. Kiejna\\ (eV)\end{tabular} & \begin{tabular}[c]{@{}l@{}}IV. Farkas\\ ($\frac{kJ}{mol}$ [eV])\end{tabular} \\ \hline
1   & Adsorption                                              & $V_B \rightarrow CO_B$           & Eq. \ref{eq_adsorption}                                             & 0                                                              & 0                                                     & 0                                                        & 0                                                                    & 0   \\
2   & Adsorption                                              & $V_C \rightarrow CO_C$           & Eq. \ref{eq_adsorption}                                             & 0                                                              & 0                                                     & 0                                                        & 0                                                                    & 0   \\
3   & Adsorption                                              & $V_B \rightarrow O_B$ (at least one vacant NN)            & Eq. \ref{eq_adsorption}                                             & 0                                                              & 0                                                     & 0                                                        & 0                                                                    & 0   \\
4   & Adsorption                                              & $V_C \rightarrow O_C$ (at least one vacant NN)           & Eq. \ref{eq_adsorption}                                             & 0                                                              & 0                                                     & 0                                                        & 0                                                                   & 0   \\
5   & Desorption                                              & $CO_B\rightarrow V_B$        & Eq. \ref{eq_desorption}                                                                  & 1.6                                                      & 1.6                                                           & 1.85                                                  & 1.69                                                                                     & 193 [2.00] \\
6   & Desorption                                              & $CO_C\rightarrow V_C$        & Eq. \ref{eq_desorption}                                                                 & 1.3                                                    & 1.3                                                              & 1.32                                                  &  1.31                                                                                  & 129 [1.34]$^b$ \\
7   & Desorption                                              & $O_B+O_B\rightarrow V_B+V_B$ & Eq. \ref{eq_desorption}                                                                 & 4.6                                                     & 4.6                                                             & 4.82                                                  & 4.66                                                                                     & 414 [4.29] \\
8   & Desorption                                              & $O_B+O_C\rightarrow V_B+V_C$ & Eq. \ref{eq_desorption}                                                                 & 3.3                                                     & 3.3                                                              & 3.3                                                   & 3.19                                                                                    & 291 [3.02]\\
9   & Desorption                                              & $O_C+O_C\rightarrow V_C+V_C$ & Eq. \ref{eq_desorption}                                                                 & 2.0                                                     & 2.0                                                             & 1.78                                                  & 1.72                                                                                    & 168 [1.74]\\
10   & Recombination                                                & $CO_B+O_B\rightarrow CO_2$   & $k_BT/h$ $^a$                                                           & 1.5                                                     & 1.54                                                            & 1.4                                                   & 1.48                                                                      & 133 [1.38]\\
11  & Recombination                                                & $CO_B+O_C\rightarrow CO_2$   & $k_BT/h$ $^a$                                                           & 0.8                                                     & 0.76                                                            & 0.6                                                   & 0.61                                                                       & 91 [0.94]\\
12  & Recombination                                                & $CO_C+O_B\rightarrow CO_2$   & $k_BT/h$ $^a$                                                           & 1.2                                                     & 1.25                                                            & 0.74                                                  & 0.99                                                                      & 89 [0.92]$^b$ \\
13  & Recombination                                                & $CO_C+O_C\rightarrow CO_2$   & $k_BT/h$ $^a$                                                           & 0.9                                                     & 0.89                                                            & 0.71                                                  & 0.78                                                                      & 89 [0.92]$^b$ \\
14  & Diffusion                                               & $CO_B\rightarrow CO_B$       & $k_BT/h$                                                            & 0.6                                                      & 0.6                                                            & 0.7                                                   & 0.6                                                                                  & 87 [0.90]\\
15  & Diffusion                                               & $CO_B\rightarrow CO_C$       & $k_BT/h$                                                            & 1.6                                                      & 1.6                                                            & 2.06                                                  & 1.6                                                                                 & 122 [1.26]\\
16  & Diffusion                                               & $CO_C\rightarrow CO_B$       & $k_BT/h$                                                            & 1.3                                                      & 1.3                                                            & 1.4                                                   & 1.3                                                                                 & 58 [0.60]$^b$ \\
17  & Diffusion                                               & $CO_C\rightarrow CO_C$       & $k_BT/h$                                                            & 1.7                                                      & 1.7                                                            & 1.57                                                  & 1.7                                                                               & 106 [1.10]$^b$ \\
18  & Diffusion                                               & $O_B \rightarrow O_B$        & $k_BT/h$                                                            & 0.7                                                      & 0.7                                                            & 0.9                                                   & 0.7                                                                                 & 87 [0.90]\\
19  & Diffusion                                               & $O_B \rightarrow O_C$        & $k_BT/h$                                                            & 2.3                                                      & 2.3                                                            & 1.97                                                  & 2.3                                                                               & 191 [1.98]\\
20  & Diffusion                                               & $O_C \rightarrow O_B$        & $k_BT/h$                                                            & 1.0                                                      & 1.0                                                             & 0.7                                                   & 1.0                                                                               & 68 [0.70]\\
21  & Diffusion                                               & $O_C \rightarrow O_C$        & $k_BT/h$                                                            & 1.6                                                      & 1.6                                                            & 1.53                                                  & 1.6                                                                                 & 106 [1.10]\\ \hline
22  & Desorption                                              & $CO_C\rightarrow V_C$        (1 NN $CO_C$) &                                                      &                                                                &                                                       &                                                          &                                                                                    &  129-10.6/2 = 123.7 [1.28]  \\
23  & Desorption                                              & $CO_C\rightarrow V_C$        (2 NN $CO_C$) &                                                      &                                                                &                                                       &                                                          &                                                                                  &  129-10.6 = 118.4 [1.23]  \\
24  & Recombination                                                & $CO_C+O_B\rightarrow CO_2$   (1 NN $CO_C$) &                                                       &                                                                &                                                       &                                                          &                                                                                     & 89--10.6/2 = 83.7 [0.87]   \\
25  & Recombination                                                & $CO_C+O_B\rightarrow CO_2$   (2 NN $CO_C$) &                                                      &                                                                &                                                       &                                                          &                                                                                   &  89--10.6 = 78.4 [0.81]  \\
26  & Recombination                                                & $CO_C+O_C\rightarrow CO_2$   (1 NN $CO_C$) &                                                       &                                                                &                                                       &                                                          &                                                                                    & 89--10.6/2 = 83.7 [0.87]    \\
27  & Diffusion                                               & $CO_C\rightarrow CO_B$       (1 NN $CO_C$) &                                                        &                                                                &                                                       &                                                          &                                                                                   &  58--10.6/2 = 52.7 [0.55]  \\
28  & Diffusion                                               & $CO_C\rightarrow CO_B$       (2 NN $CO_C$) &                                                      &                                                                &                                                       &                                                          &                                                                                       &  58--10.6 = 47.4 [0.49]  \\
29  & Diffusion                                               & $CO_C\rightarrow CO_C$       (1 NN $CO_C$) &                                                       &                                                                &                                                       &                                                          &                                                                                 & 106-10.6/2 = 100.7 [1.04]
\end{tabular}
\begin{flushleft}
$^a$ In model I, the attempt frequency for recombination is $\frac{1}{2} k_BT/h$ (instead of $k_BT/h$, as used in the other models). See Ref. \cite{Reuter2006} for details.

$^b$ In model IV, repulsion of 10.6 kJ/mol per $CO_C$ nearest neighbor (NN) is included, as described in rows 22 through 29.

\end{flushleft}
}
\end{table*}

The desorption rate constant is computed to satisfy detailed balance (or microreversibility) with respect to the reverse reaction (adsorption). The used expression is (see Eqs. (9) and (13) in \cite{Reuter2006}, with $\Delta E_{st,i}^{\rm ad} = 0$ and $q_{st,i}^{\rm vib} \approx 1$ or, equivalently, see Eq. A2 in \cite{Temel2007}, where we believe that the argument in the $\exp()$ function should be preceded by a negative sign):
\begin{equation}
   k_{\alpha_{\uparrow}} = k_{\alpha_{\downarrow}}^0 e^{- ( E_{\alpha_{\uparrow}}^{k}+\mu_{\rm X} ) / k_BT}, \; \; \alpha_{\uparrow} = 5,...,9,22,23, 
   \label{eq_desorption}
 \end{equation}
where $k_{\alpha_{\downarrow}}^0 = s \cdot \frac{P_{\rm X} A_s} {\sqrt{2 \pi m_{\rm X} k_B T}}$ is the attempt frequency for the reverse adsorption reaction (Eq. \ref{eq_adsorption}), $E_{\alpha_{\uparrow}}^{k}$ is the activation barrier for desorption and $\mu_{\rm X}$ is the chemical potential for species X  (= CO or O$_2$):
 \begin{equation}
   \mu_X = - k_B T \log\left(\frac{k_BT}{P_X} q_t^X q_r^X q_v^X\right). 
   \label{eq_mu}
 \end{equation}
Here, $q_t^X$, $q_r^X$ and $q_v^X$ are the translational, rotational and vibrational partition functions, assuming an ideal mixture of diatomic molecules (see Eq. (8) in \cite{Reuter2006} and the text after Eq. A2 in \cite{Temel2007}):
 \begin{equation}
   q_t^X = \left(\frac{2\pi m_X k_BT}{h^2}\right)^{3/2},
   \label{eq_q_t}
 \end{equation}
 \begin{equation}
   q_r^X =\frac{8\pi^2I_{\rm X} k_BT}{\sigma_{\rm X} h^2},
   \label{eq_q_r}
 \end{equation}
 \begin{equation}
   q_v^X = \frac{1}{1-e^{-h \nu_{\rm X} / k_BT}},
   \label{eq_q_v}
 \end{equation}
where $I_{\rm X} = \frac{m_1^{\rm X} m_2^{\rm X}}{m_1^{\rm X} + m_2^{\rm X}} R_{\rm X}^2$, with $m_1^{\rm X}$ and $m_2^{\rm X}$ the masses of the two atoms in the molecule, and $R_{\rm X}$ the distance between them (1.13 {\AA} for CO and 1.21 {\AA} for O$_2$), $\sigma_{\rm X}$ is the symmetry number (we use 0.98 for CO and 1.32 for O$_2$), and $\nu_{\rm X}$ is the vibrational frequency (we use $6.5 \times 10^{13}$ Hz for CO and $4.7 \times 10^{13}$ Hz for O$_2$).

The recombination rate constants are computed according to:
 \begin{equation}
   k_{\alpha} = k_{\alpha}^0 e^{ -E_{\alpha}^{k} / k_BT}, \alpha = 10,...,13,24,...,26
   \label{eq_recombination}
 \end{equation}
where $k_{\alpha}^0 = g \cdot \frac{k_BT}{h}$, with $g=1$ for models II - IV and $g=\frac{1}{2}$ for model I (see Ref. \cite{Reuter2006}). Similarly, using $k_{\alpha}^0 = \frac{k_BT}{h}$ the diffusion rate constants are computed according to:
 \begin{equation}
   k_{\alpha} = k_{\alpha}^0 e^{ -E_{\alpha}^{k} / k_BT},  \alpha = 14,...,21,27,...,29.
   \label{eq_diffusion}
 \end{equation}
 
Because the reaction mechanism assumes that CO$_2$ is immediately desorbed after recombination, the decomposition of CO$_2$ admolecules on the surface is disregarded in all four models and, thus, there is no need to consider microreversibility for recombination. On the other hand, the collection of activation energies used for diffusion are such that the diffusion rates comply with detailed balance.

In summary, while desorption and diffusion are formulated identically in all four models, adsorption and recombination differ in model I, due to using a different sticking coefficient (1/2 instead of 1) for adsorption and a different prefactor ($\frac{1}{2} \frac{k_BT}{h}$ instead of $\frac{k_BT}{h}$) for recombination. For completeness, particular values of the rate constants are shown in Table \ref{tb_rates} for models I and II at representative temperatures and pressures.
\begin{table*}[ht!]
\centering
\caption{Rate constants at three representative temperatures (in K) for model I ($p_{\rm CO} = 1$ atm, $p_{\rm O_{2}} = 2$ atm) and model II ($p_{\rm CO} = 1 \times 10^{-10}$ bar, $p_{\rm O_{2}} = 2 \times 10^{-10}$ bar).}
\label{tb_rates}
{\tiny
\begin{tabular}{lll|ccc|ccc}
  $\alpha$ & \begin{tabular}[c]{@{}l@{}}Type \end{tabular} & Reaction                     & \begin{tabular}[c]{@{}l@{}} \end{tabular} & \begin{tabular}[c]{@{}l@{}}I. Reuter\end{tabular} & \begin{tabular}[c]{@{}l@{}}  \end{tabular} & \begin{tabular}[c]{@{}l@{}} \end{tabular} & \begin{tabular}[c]{@{}l@{}}II. Seitsonen \end{tabular} &  \\ 
   &     &                              & 450 K                   & 550 K                 & 650 K                & 300 K                & 340 K                & 375 K                 \\ \hline
1  & Adsorption         & $V_B \rightarrow CO_B$           & $2.4\cdot 10^{+08}$    & $2.2\cdot 10^{+08}$ & $2.0\cdot 10^{+08}$ & $5.8\cdot 10^{-02}$ & $5.4\cdot 10^{-02}$ & $5.2\cdot 10^{-02}$ \\
2  & Adsorption         & $V_C \rightarrow CO_C$           & $2.4\cdot 10^{+08}$    & $2.2\cdot 10^{+08}$ & $2.0\cdot 10^{+08}$ & $5.8\cdot 10^{-02}$ & $5.4\cdot 10^{-02}$ & $5.2\cdot 10^{-02}$ \\
3  & Adsorption         & $V_B \rightarrow O_B$ (at least one vacant NN)           & $1.1\cdot 10^{+08}$    & $1.0\cdot 10^{+08}$ & $9.3\cdot 10^{+07}$ & $2.7\cdot 10^{-02}$ & $2.5\cdot 10^{-02}$ & $2.4\cdot 10^{-02}$ \\
4  & Adsorption         & $V_C \rightarrow O_C$ (at least one vacant NN)           & $1.1\cdot 10^{+08}$    & $1.0\cdot 10^{+08}$ & $9.3\cdot 10^{+07}$ & $2.7\cdot 10^{-02}$ & $2.5\cdot 10^{-02}$ & $2.4\cdot 10^{-02}$ \\
5  & Desorption         & $CO_B\rightarrow V_B$        & $3.9\cdot 10^{-01}$    & $1.3\cdot 10^{+03}$ & $3.8\cdot 10^{+05}$ & $1.6\cdot 10^{-14}$ & $1.0\cdot 10^{-10}$ & $5.1\cdot 10^{-08}$ \\
6  & Desorption         & $CO_C\rightarrow V_C$        & $8.9\cdot 10^{+02}$    & $7.2\cdot 10^{+05}$ & $8.1\cdot 10^{+07}$ & $1.3\cdot 10^{-05}$ & $7.5\cdot 10^{-03}$ & $6.8\cdot 10^{-01}$ \\
7  & Desorption         & $O_B+O_B\rightarrow V_B+V_B$ & $2.2\cdot 10^{-34}$    & $9.6\cdot 10^{-25}$ & $4.9\cdot 10^{-18}$ & $4.6\cdot 10^{-64}$ & $2.3\cdot 10^{-54}$ & $1.4\cdot 10^{-47}$ \\
8  & Desorption         & $O_B+O_C\rightarrow V_B+V_C$ & $8.1\cdot 10^{-20}$    & $7.8\cdot 10^{-13}$ & $5.9\cdot 10^{-08}$ & $1.6\cdot 10^{-38}$ & $7.7\cdot 10^{-32}$ & $3.8\cdot 10^{-27}$ \\
9  & Desorption         & $O_C+O_C\rightarrow V_C+V_C$ & $2.9\cdot 10^{-05}$    & $6.4\cdot 10^{-01}$ & $7.1\cdot 10^{+02}$ & $5.4\cdot 10^{-13}$ & $2.6\cdot 10^{-09}$ & $1.0\cdot 10^{-06}$ \\
10  & Reaction           & $CO_B+O_B\rightarrow CO_2$   & $7.4\cdot 10^{-05}$    & $1.0\cdot 10^{-01}$ & $1.6\cdot 10^{+01}$ & $1.9\cdot 10^{-11}$ & $1.3\cdot 10^{-08}$ & $1.2\cdot 10^{-06}$ \\
11 & Reaction           & $CO_B+O_C\rightarrow CO_2$   & $5.1\cdot 10^{+03}$    & $2.7\cdot 10^{+05}$ & $4.2\cdot 10^{+06}$ & $5.2\cdot 10^{+02}$ & $9.0\cdot 10^{+03}$ & $6.7\cdot 10^{+04}$ \\
12 & Reaction           & $CO_C+O_B\rightarrow CO_2$   & $1.7\cdot 10^{-01}$    & $5.8\cdot 10^{+01}$ & $3.4\cdot 10^{+03}$ & $2.3\cdot 10^{+00}$ & $7.6\cdot 10^{+01}$ & $8.9\cdot 10^{+02}$ \\
13 & Reaction           & $CO_C+O_C\rightarrow CO_2$   & $3.9\cdot 10^{+02}$    & $3.2\cdot 10^{+04}$ & $7.1\cdot 10^{+05}$ & $7.4\cdot 10^{+00}$ & $2.1\cdot 10^{+02}$ & $2.2\cdot 10^{+03}$ \\
14 & Diffusion          & $CO_B\rightarrow CO_B$       & $1.8\cdot 10^{+06}$    & $3.6\cdot 10^{+07}$ & $3.0\cdot 10^{+08}$ & $1.1\cdot 10^{+01}$ & $3.0\cdot 10^{+02}$ & $3.1\cdot 10^{+03}$ \\
15 & Diffusion          & $CO_B\rightarrow CO_C$       & $1.1\cdot 10^{-05}$    & $2.5\cdot 10^{-02}$ & $5.3\cdot 10^{+00}$ & $1.5\cdot 10^{-22}$ & $2.1\cdot 10^{-18}$ & $1.6\cdot 10^{-15}$ \\
16 & Diffusion          & $CO_C\rightarrow CO_B$       & $2.6\cdot 10^{-02}$    & $1.4\cdot 10^{+01}$ & $1.1\cdot 10^{+03}$ & $1.9\cdot 10^{-11}$ & $1.3\cdot 10^{-08}$ & $1.2\cdot 10^{-06}$ \\
17 & Diffusion          & $CO_C\rightarrow CO_C$       & $8.6\cdot 10^{-07}$    & $3.0\cdot 10^{-03}$ & $8.9\cdot 10^{-01}$ & $2.6\cdot 10^{-14}$ & $3.8\cdot 10^{-11}$ & $6.2\cdot 10^{-09}$ \\
18 & Diffusion          & $O_B \rightarrow O_B$        & $1.4\cdot 10^{+05}$    & $4.4\cdot 10^{+06}$ & $5.1\cdot 10^{+07}$ & $4.7\cdot 10^{-03}$ & $3.2\cdot 10^{-01}$ & $6.3\cdot 10^{+00}$ \\
19 & Diffusion          & $O_B \rightarrow O_C$        & $1.6\cdot 10^{-13}$    & $9.6\cdot 10^{-09}$ & $2.0\cdot 10^{-05}$ & $5.0\cdot 10^{-21}$ & $4.5\cdot 10^{-17}$ & $2.6\cdot 10^{-14}$ \\
20 & Diffusion          & $O_C \rightarrow O_B$        & $5.9\cdot 10^{+01}$    & $7.9\cdot 10^{+03}$ & $2.4\cdot 10^{+05}$ & $1.1\cdot 10^{+01}$ & $3.0\cdot 10^{+02}$ & $3.1\cdot 10^{+03}$ \\
21 & Diffusion          & $O_C \rightarrow O_C$        & $1.1\cdot 10^{-05}$    & $2.5\cdot 10^{-02}$ & $5.3\cdot 10^{+00}$ & $1.2\cdot 10^{-13}$ & $1.5\cdot 10^{-10}$ & $2.1\cdot 10^{-08}$ \\ 

\end{tabular}}
\end{table*}

\begin{table*}[ht!]
\centering
\caption{Elementary reactions, indicating the attempt frequency ($k_{\alpha}^{0}$, 1/s) and activation energy ($E_{\alpha}^{k}$, eV) used in the reaction mechanism for the selective oxidation of NH$_3$ on RuO$_2$(110) \cite{Hong2010}. All reactions occur only at/between C sites.}
\label{tb_rates_NH3}
{\tiny
\begin{tabular}{llllll}
  $\alpha$ & \begin{tabular}[c]{@{}l@{}}Type \end{tabular} & Process                     & \begin{tabular}[c]{@{}l@{}}Attempt freq. \\ (1/s)\end{tabular} & \begin{tabular}[c]{@{}l@{}}Act. energy\\ (eV)\end{tabular}  \\ \hline
1   & Adsorption                                              & $V_C \rightarrow NH_3$      		  				& Eq. \ref{eq_adsorption}  	& 0.0         \\
2   & Desorption					& $NH_3 \rightarrow V_C$						& $10^{13}$ 				& 1.46   \\
3   &  Adsorption                                             & $V_C + V_C \rightarrow O + O$      				& Eq. \ref{eq_adsorption}           & 0.0         \\
4   & Desorption					& $O + O \rightarrow V_C + V_C$					& $10^{13}$				& 1.26   \\
5   & Recombination (abstraction)		& $NH_3 + O \rightarrow NH_2 + OH$  				& $10^{13}$				& 0.55   \\
6   & Recombination (abstraction)		& $NH_2 + OH \rightarrow NH + V_C + H_{2}O(g)$  	& $10^{13}$				& 0.27   \\
7   & Recombination (abstraction)		& $NH + OH \rightarrow N + V_C + H_{2}O(g)$ 	 	& $10^{13}$				& 0.0   \\
8   & Recombination (abstraction)		& $NH + O \rightarrow N + OH$ 		 			& $10^{13}$				& 0.0   \\
9   & Recombination 				& $N + O \rightarrow NO + V_C$ 	 				& $10^{13}$				& 0.14   \\
10   & Desorption					& $N + N \rightarrow V_C + V_C + N_2(g)$		 	& $10^{13}$				& 0.27   \\
11   & Desorption					& $NO  \rightarrow V_C + NO(g)$ 				 	& $10^{13}$				& 1.49   \\
12   & Diffusion						& $N  \rightarrow N$ 						 	& $10^{13}$				& 0.96   \\
13   & Diffusion						& $O  \rightarrow O$ 						 	& $10^{13}$				& 0.93   \\
14   & Diffusion						& $OH  \rightarrow OH$ 						 	& $10^{13}$				& 1.12   \\
15   & Recombination (abstraction)		& $NH_2 + O \rightarrow NH + OH$			 	 	& $10^{13}$				& 1.0   \\
16   & Recombination				& $NH + OH \rightarrow NH_2 + O$ 				 	& $10^{13}$				& 0.0   \\
17   & Recombination				& $NH_2 + OH \rightarrow NH_3 + O$ 			 	& $10^{13}$				& 0.26   \\
18   & Recombination				& $N + OH \rightarrow NH + O$ 			 		& $10^{13}$				& 0.9   \\

\end{tabular}}
\end{table*}

Since the attempt frequencies (or prefactors) for adsorption, desorption, diffusion and recombination depend on temperature, we can directly determine their effective energies, $E_{\alpha}^{k^{0}} =  - \frac{ d ln(k_{\alpha}^{0}) }{ d \beta }$, required in Eq. \ref{EappTOF_5} of the report. Here, $\beta = 1 / k_B T$. For adsorption (Eq. \ref{eq_adsorption}) we have: $k_{\alpha}^{0} = s \cdot \frac{P_{\rm X} A_s} {\sqrt{2 \pi m_{\rm X} k_B T}} \propto \beta^{1/2}$. Thus:
\begin{align}
E_{\alpha}^{k^{0}} & =  - \frac{1}{k_{\alpha}^{0}} \frac{ d k_{\alpha}^{0} }{ d \beta } =  - \frac{k_B T}{2}, \; \; \alpha = 1, 2, 3, 4 . \label{E_alpha_ads_k0}
\end{align}
Similarly, for desorption (Eq. \ref{eq_desorption}) the overall prefactor is $k_{\alpha}^{0} = s \cdot \frac{P_{\rm X} A_s} {\sqrt{2 \pi m_{\rm X} k_B T}} e^{- \mu_{\rm X} / k_BT}$, where $\mu_{\rm X}$ depends on $\beta=1/k_B T$ according to Eqs. \ref{eq_mu} - \ref{eq_q_v}. Thus:
\begin{align}
\hspace{-3mm}
E_{\alpha}^{k^{0}} & = 3 k_B T + \frac{h \nu_{\rm X} e^{-h \nu_{\rm X} / k_BT}}{1-e^{-h \nu_{\rm X} / k_BT}}, \; \; \alpha = 5,...,9,22,23  . \label{E_alpha_des_k0}
\end{align}
Similarly, inspection of Eqs. \ref{eq_recombination} - \ref{eq_diffusion} for recombination and diffusion gives the prefactor as $k_{\alpha}^{0} \propto \beta^{-1}$. Thus, the effective energies are:
\begin{align}
E_{\alpha}^{k^{0}} & =  k_B T, \; \; \alpha = 10,...,21, 24,...,29. \label{E_alpha_recNdif_k0}
\end{align}

This study considers the coverages of certain collections of sites as the multiplicities for the various processes. For the reaction mechanism introduced above, we have the following. For diffusion ($A_X \rightarrow A_Y$, where $A = CO, O$ and $X,Y = B, C$), the multiplicity is equal to the coverage of all empty sites of type $Y$ surrounding all the $X$ sites populated by $A$. Similarly, for recombination ($CO_{\rm X} + O_{\rm Y} \rightarrow CO_{2}$, where $X,Y = B, C$), $M_{\alpha}$ is equal to the coverage by all NN pairs of $CO_{\rm X}$ and $O_{\rm Y}$ adparticles. In turn, for the five desorption types, $M_{\alpha}$ equals, respectively, the coverage by CO$_B$, $CO_C$ and three NN pairs of adsorbed O (O$_B$-O$_B$, O$_B$-O$_C$ and O$_C$-O$_C$). Finally, for the adsorption of CO (O$_2$) the multiplicity is equal to the coverage by all empty sites (all NN pairs of empty sites).

\subsection{Selective oxidation of NH$_3$}
\label{Selective_oxidation_of_NH_3}

Table \ref{tb_rates_NH3} shows the reaction mechanism consisting of 18 elementary reactions proposed by Hong {\it et al.} in order to describe the selective oxidation of NH$_3$ on RuO$_2$(110) \cite{Hong2010}. All elementary reactions occur only at C sites and the attempt frequencies are taken to be $10^{13}$ Hz, except for the adsorption reactions ($\alpha = 1,3$), where Eq. \ref{eq_adsorption} is used (with the sticking coefficient equal to 1) \cite{Hong2010}. In the implementation by Hong {\it et al.} the desorption of NH$_3$ and $NO$ ($\alpha = 2$ and 11, respectively) considers lateral interactions (repulsion) in such a manner that the desorption rate is given by $k_{\alpha} = k_{\alpha}^0 e^{-E_{\alpha}^k / k_B T} e^{-\epsilon_{\alpha} \theta_{\alpha} / k_B T}$, where $\theta_{\alpha}$ represents the coverage by NH$_3$ and NO, respectively, $\epsilon_{\alpha} = 0.34$ and 0.16 eV for NH$_3$ and NO, respectively, and $k_{\alpha}^0 = 10^{13}$ Hz for both. After implementing this feature, we observed that: (i) The lateral interactions effectively introduce a large number of elementary activation energies as a function of the local coverage around the desorbing NH$_3$/NO molecules. Correspondingly, the multiplicities for the desorption reactions of NH$_3$ and NO should be split into additional sub-multiplicities (one for each identified elementary activation energy). However, this requires a rather sophisticated programming effort while it is believed to add little value from a physical/chemical perspective, simply splitting the number of contributions that explain the actual value of the apparent activation energy. This is specially notable considering that (ii) the lateral interactions modify the behavior of the system only marginally, as shown in section S4.2 (see Fig. \ref{Figure2S}(a)), while repulsive lateral interactions are already explicitly taken into account in model IV for the oxidation of CO, where the number of additional elementary activation energies (and multiplicities) is small enough so that the splitting of the various contributions can still be visualized reasonably well (see section S7.1). Since the effect of lateral interactions is already considered in one model we do not feel the need to include it in the case of the oxidation of NH$_3$.

\vspace{5 mm}
\section{Computational method}
\label{computational_method}

{\em KMC}. The KMC simulations are performed using a typical lattice-gas model with the rejection-free, time-dependent implementation \cite{Reuter2006, Voter2007, Chatterjee2007, Reuter2011, Jansen2012, Temel2007, Gosalvez2017}. Every time step ($k$) starts by updating time as $t^{k+1} = t^{k} + \Delta t$, where $\Delta t = -\log(u)/\hat{r}_{}$ is the inverse of the instantaneous total rate, $\hat{r}_{}=\hat{r}_{a}+\hat{r}_{d}+\hat{r}_{h}+\hat{r}_{r}$, with $\hat{r}_{a}$, $\hat{r}_{d}$, $\hat{r}_{h}$ and $\hat{r}_{r}$ the (instantaneous) total adsorption rate, total desorption rate, total hop rate and total recombination rate, respectively. The factor $-\log(u)$, where $u$ is a uniform random number in $(0,1]$, enforces the correct Poisson distribution for the time steps, with a mean value of $1$. 
All instantaneous total rates ($\hat{r}$ and $\hat{r}_g$, with $g=a,d,h,r$) are simply related to the instantaneous total rates per active site ($\hat{R}$ and $\hat{R}_g$): $\hat{R}=\hat{r} / \hat{s}$ and $\hat{R}_g = \hat{r}_g / \hat{s}$, where $\hat{s}$ the number of active sites.
After updating $t$, the next reaction type (adsorption, desorption, diffusion or recombination) is selected by performing a linear search (LS) amongst $\hat{r}_{a}$, $\hat{r}_{d}$, $\hat{r}_{h}$ and $\hat{r}_{r}$ \cite{Chatterjee2007, Gosalvez_AtomisticMethods}. Once one of the four main reaction types has been chosen, say $\hat{r}_{x}$, one particular elementary reaction is selected by performing either a LS or a binary search (BS) amongst the rate constants contained in $\hat{r}_{x}$ \cite{Chatterjee2007, Gosalvez_AtomisticMethods}. Note that $\hat{r}_{x}$ typically contains the rate constants of many elementary reactions for various reaction types. The use of LS or BS is automatically selected by the program, depending on the number of rate constants $n$ contained in $\hat{r}_{x}$. In particular, LS is performed if $n \le 100$ and BS is used otherwise. 
Once an elementary reaction has been selected, it is executed, thus modifying the neighborhoods of the origin and/or end sites. As a result, the corresponding rate constants and total rates ($\hat{r}_{a}$, $\hat{r}_{d}$, $\hat{r}_{h}$, $\hat{r}_{r}$ and $\hat{r}_{}$) are updated. 
In this manner, the simulation is continued by incrementing time, selecting a new elementary reaction, executing it, and updating the neighborhoods until the simulation is finished (see {\em Termination} below). 

{\em Steady state}. The steady state is reached after a transient from the chosen initial state (see {\em Intial State} below). 
The steady state is characterized by the fact that the instantaneous coverage of any adspecies fluctuates with time about a constant value.
This includes four adspecies ($\hat{\theta}_{CO}^{B}$, $\hat{\theta}_{CO}^{C}$, $\hat{\theta}_{O}^{B}$ and $\hat{\theta}_{O}^{C}$) for the case of the oxidation of CO, and seven adspecies ($\hat{\theta}_{NH_3}$, $\hat{\theta}_{NH_2}$, $\hat{\theta}_{NH}$, $\hat{\theta}_{NO}$, $\hat{\theta}_{N}$, $\hat{\theta}_{O}$ and $\hat{\theta}_{OH}$, all at C sites only) for the oxidation of NH$_3$. Thus, in the steady state the tendency for any of these coverages is to become independent of time and the correlation coefficient $R^2$ of any computed linear regression between coverage and time should become 0. On the other hand, before reaching the steady state, even if the dependence between coverage and time is not linear, the correlation coefficient $R^2$ will necessarily deviate from 0. Based on this, we sample the various coverages every $E = 10^5$ executed elementary reactions and mark the onset of the steady state as follows: (i) For the case of the oxidation of CO, the steady state starts when the four $R^2$ coefficients of the linear regressions become less than 0.1 simultaneously for the last $P$ sampled coverages, where $P=20\sqrt{L_x L_y}$, with $L_x L_y$ the total number of catalyst sites. For the typical size of the simulations ($30 \times 30$, see {\em Size} below) this gives $P=600$. (ii) For the oxidation of NH$_3$, the system is considered to enter the steady state when the seven coverages satisfy simultaneously the condition $| \hat{\theta}_{\rm MAX}^{X} - \hat{\theta}_{\rm MIN}^{X} | \le 0.05$, where $\hat{\theta}_{\rm MAX}^{X}$ and $\hat{\theta}_{\rm MIN}^{X}$ are the maximum and minimum values of the coverage for adspecies $X$ for the last $P$ sampled coverages, where $P=10\sqrt{L_x L_y}$. For the typical size of the simulations ($30 \times 40$, see {\em Size} below) this gives $P=346$. These criteria are rather useful, since the total number of events (including adsorption, desorption, diffusion and recombination) required to reach the steady state varies by orders of magnitude, depending on the physical model, the temperature and the partial pressures (of CO and O$_2$, for the oxidation of CO, and of NH$_3$ and O$_2$, for the oxidation of NH$_3$). Some models are overwhelmingly dominated by adsorption and desorption reactions while others are dominated by diffusion reactions. And this depends on temperature and pressure. Thus, it is difficult to estimate beforehand the total number of executed elementary reactions required to enter the steady state. The use of the previous criterion provides a robust procedure to simplify data collection, especially regarding the need of performing thousands of simulations for different models at different temperatures and pressures. 

{\em Termination}. After the onset of the steady state the simulated time is set to 0 and the simulation is continued until Z molecules of the target product/s are generated, at which point the simulation is terminated.  We use $Z=1000$ molecules of CO$_2$ in the case of the oxidation of CO and $Z=250$ molecules of $NO$ and $N_2$ (distributed in any manner amongst the two species) in the case of the oxidation of NH$_3$. At this moment, the value of the simulated time $t$ is stored and averaged over $K$ simulations (see {\em Size} below). The $TOF$ is determined using the expression: $TOF = \frac{Z}{L_x L_y} \langle t \rangle$, where $L_x L_y$ is the total number of catalyst sites and $\langle t \rangle$ is the average time.

{\em Initial state}. Simulations were performed with different initial states ({\it e.g.} O-terminated, CO-terminated, random with 50\% O-terminated + 50\% CO-terminated, all empty, etc...) and the obtained steady states were confirmed to be essentially identical. 

{\em Acceleration}. Although we are aware of various acceleration algorithms to increase the computational efficiency of the KMC simulations \cite{Chatterjee2007, DeVita2005}, we have avoided them on purpose to eliminate any chance of affecting the analysis of the apparent activation energy.

{\em Size}. {\em Oxidation of CO:} The simulations were performed on systems with $L_x L_y = $ 30$\times$30, 60$\times$60, and 100$\times$100 active sites and repeated $K$ times to obtain ensemble averages of all quantities, with $K=10$. Any error bars indicated in the main text correspond to the standard deviations of the corresponding variable amongst the $K$ runs. As expected, on going from $L_x = 30$ to $100$ we observe the same overall behavior with a reduction in the fluctuations in all variables and a huge increase in computational time. In other words, $L_x = 30$ provides similar results to $60$ and $100$, for a fraction of the computational effort. The reported results correspond to $L_x L_y = 30 \times 30$ (900 active sites). This is larger than in previous studies ($20 \times 20$) \cite{Reuter2006}. {\em Oxidation of NH$_3$}: We use $K=10$ and $L_x L_y = 30 \times 40$. Since in this system the elementary reactions take place only on C sites, this makes a total of $15 \times 40 = 600$ active sites. This is the same size used by Hong. {\it et al.} \cite{Hong2010} (confirmed by private communication).

{\em Temperature and pressure}. We use a wide range of temperatures and pressures. {\em Oxidation of CO:} $T = 250-750$ K and $p=1 \times 10^{-10} - 2 \times 10^{0}$ bar. {\em Oxidation of NH$_3$:} $T = 455-590$ K and $p=(0.5 - 20) \times 10^{-7}$ mbar.

\begin{figure*}[htb!]
  \begin{center}
    \includegraphics[width=1.8\columnwidth]{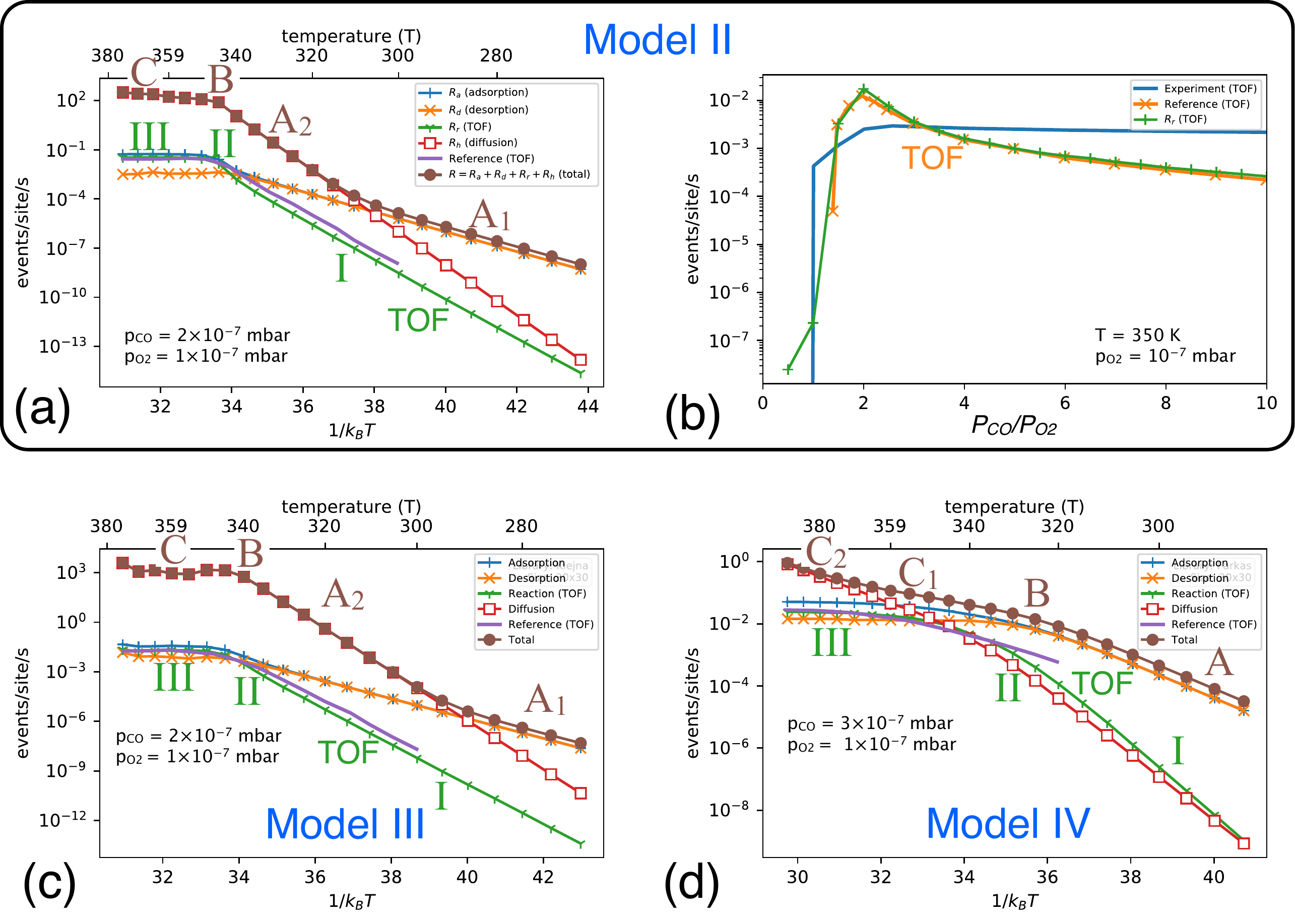}
  \end{center}
  \caption{Typical results for models II, III and IV for the oxidation of CO on RuO$_2$\{110\}: 
    (a) Arrhenius plot for the total rates per active site $R_{a}$, $R_{d}$, $R_{h}$, $R_{r}$ ($= TOF$), and $R_{}=R_{a}+R_{d}+R_{h}+R_{r}$ {\it vs} inverse temperature $\beta=1/k_B T$ for model II. (b) CO pressure dependence of $R_{r}$ ($= TOF$) for model II. (c)-(d) Same as frame (a), now for models III and IV, respectively. Reference $TOF$ data: (a)-(c) Ref. \cite{Hess2012}, (d) Ref. \cite{Farkas2012}. Experimental data in frame (b): Ref. \cite{Wang2002}.
}
  \label{Figure1S}
\end{figure*}

\section{Comparison to previous results for additional models}
\label{Comparison_to_previous_results_for_additional_models}

\subsection{Oxidation of CO using models II, III and IV}
\label{Oxidation_of_CO_using_models_II_III_and_IV}

Fig. \ref{Figure1S}(a) shows the temperature dependence of $R_{}$, $R_{a}$, $R_{d}$, $R_{h}$, and $R_{r}$ ($= TOF$) as well as the corresponding $TOF$ data obtained by Hess {\it et al.} \cite{Hess2012} for model II at $p_{\rm CO} = 2 \times 10^{-7}$ mbar and $p_{\rm O_{2}} = 10^{-7}$ mbar. The corresponding pressure dependence of the $TOF$ for $T=350$ K and $p_{\rm O_{2}} = 10^{-7}$ mbar is shown in Fig. \ref{Figure1S}(b). In turn, Figs. \ref{Figure1S}(c)-(d) show the temperature dependence of $R_{}$, $R_{a}$, $R_{d}$, $R_{h}$, and $R_{r}$ ($= TOF$) as well as the corresponding $TOF$ data obtained by Hess {\it et al.} \cite{Hess2012} and Farkas {\it et al.} \cite{Farkas2012} for models III and IV, respectively. The pressure dependence of the $TOF$ for model II in Fig. \ref{Figure1S}(b) is practically identical, while a small, horizontal shift is observed in the temperature dependence for both models II and III in Figs. \ref{Figure1S}(a) and \ref{Figure1S}(c), presumably due to our improved steady state detection. Regarding model IV in fig. \ref{Figure1S}(d), our $TOF$ departs from the reference results at low temperature ($\beta > 35$). This is probably due to differences in the details of the implementation of repulsion, which we may have carried out differently from Ref. \cite{Farkas2012}. Overall, comparison of six $TOF$ curves (considering Fig. \ref{Figure02}of the main text and Fig. \ref{Figure1S} of this Supporting Information) strongly indicates that our implementation of the KMC method and the reaction mechanism is correct. 

\subsection{Selective oxidation of NH$_3$}
\label{Comp_Selective_oxidation_of_NH_3}

\begin{figure*}[htb!]
  \begin{center}
    \includegraphics[width=2.0\columnwidth]{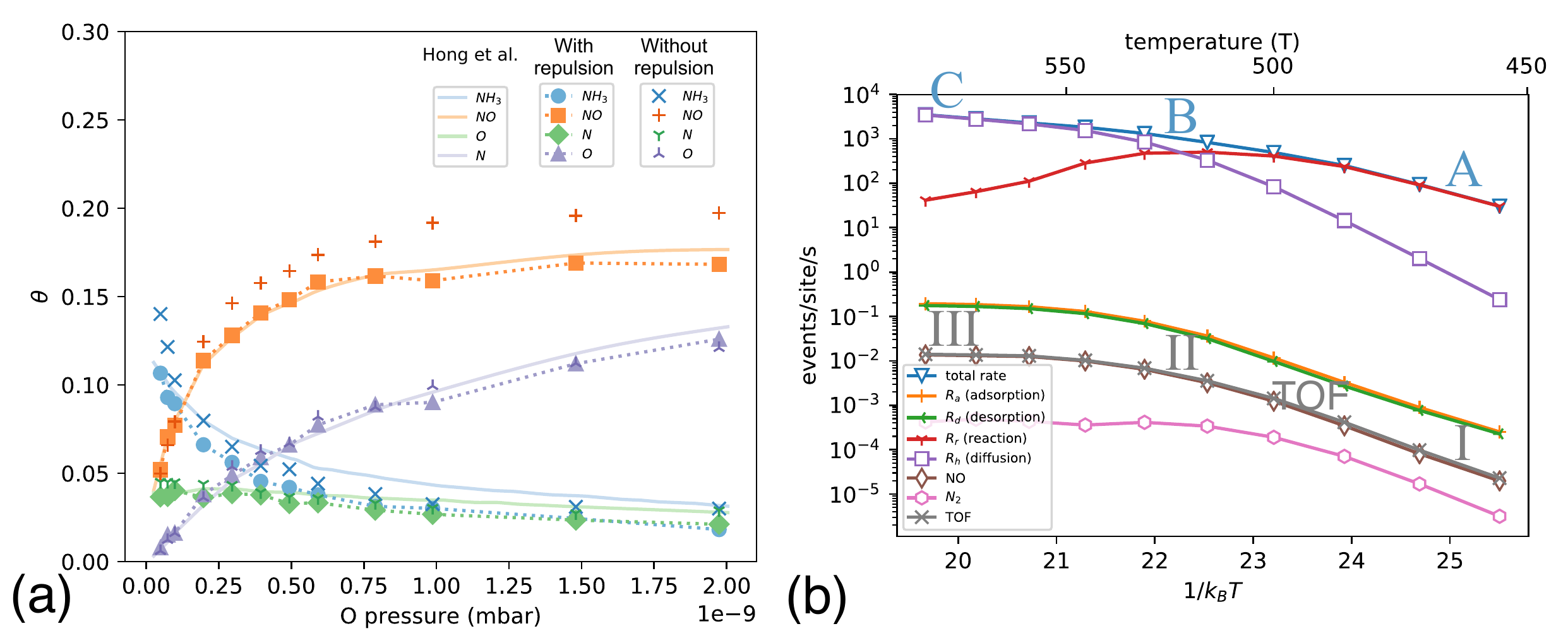}
  \end{center}
  \caption{Typical results for selective oxidation of NH$_3$ on RuO$_2$\{110\}: (a) Coverage {\it vs} pressure for various surface intermediates at 530 K, as implemented in this study, with and without lateral interactions (repulsion) in the desorption reactions of NH$_3$ and NO. Reference data obtained with lateral interactions are shown from Hong {\it et al.} \cite{Hong2010}. (b) Arrhenius plot for the total rates per active site $R_{a}$, $R_{d}$, $R_{r}$, $R_{h}$, and $R_{}=R_{a}+R_{d}+R_{h}+R_{r}$ {\it vs} inverse temperature $\beta=1/k_B T$ at $p_{NH_3}=0.1 \times 10^{-7}$ mbar and $p_{O_2} = 1.5 \times 10^{-7}$ mbar, as obtained in this study (no lateral interactions). The total desorption rates per active site ($M_{\alpha} k_{\alpha}$) for NO and N$_2$, as well as their sum ($TOF$), are also shown.
}
  \label{Figure2S}
\end{figure*}

Fig. \ref{Figure2S}(a) compares our results for the coverage of various adspecies as a function of the O pressure, as obtained with and without lateral interactions at 530 K. The figure also includes the corresponding results from Hong. {\it et al.} (with lateral interactions). The results strongly indicate that our implementation of the KMC method and the reaction mechanism is correct.
Fig. \ref{Figure2S}(b) shows the temperature dependence of $R_{}$, $R_{a}$, $R_{d}$, $R_{h}$, $R_{r}$ and the total desorption rates per active site ($M_{\alpha} k_{\alpha}$) for NO and N$_2$, as well as the sum of the last two ($TOF$) at $p_{NH_3}=0.1 \times 10^{-7}$ mbar and $p_{O_2} = 1.5 \times 10^{-7}$ mbar, as obtained without lateral interactions for the desorption of NH$_3$ and NO. No reference data is available for the temperature dependence.

\section{Wrong apparent activation energies based on the Temkin formulation}
\label{Wrong_apparent_activation_energies_based_on_the_Temkin_formulation}

Although the Temkin formulas derived in Section \ref{Apparent_activation_energy_in_the_Langmuir_Hinshelwood_model} ($E_{app}^{TOF} = E_{3}^{k} -x \Delta H_{A} -y \Delta H_{B}$ when the recombination of $A$ and $B$ is the RDS, $E_{app}^{TOF} = E_{a}^{A} -y \Delta H_{B}$ when the adsorption of $A$ is the RDS, etc...) have a strong mathematical basis, the following examples show that, in practice, these expressions may result in wrong apparent activation energies due to difficulties in determining the reaction orders ($x$, $y$, etc...) that multiply the adsorption heats / formation enthalpies. The present study stresses the perspective that the apparent activation energy includes configurational entropy contributions (see Eq. \ref{EappTOF_6} of the main report), rather than the traditional adsorption heats / formation enthalpies.

{\em Example S1.} Let us focus on model I at $T_4$, where recombination type $CO_C + O_C \rightarrow CO_2$ is the RDS (see Fig. \ref{Figure04}(f) of the main report) and $A$=$CO$ ($B_2$=$O_2$) is weakly (strongly) adsorbed (see Fig. \ref{Figure05}(b) of the main report). Traditionally, in the mean-field approximation one will write (see Section \ref{Apparent_activation_energy_in_the_Langmuir_Hinshelwood_model}): $TOF = k_{CO_C + O_C \rightarrow CO_2} \theta_{CO} \theta_{O} = k_{CO_C + O_C \rightarrow CO_2} (K_{CO}p_{CO})(K_{O}p_{O})^{-1/2}$ and $E_{app}^{TOF} = E_{CO_C + O_C \rightarrow CO_2}^{k} - \Delta H_{CO} + \frac{1}{2} \Delta H_{O}$. Thus, using the values in Table \ref{Table_Processes}, we obtain: $E_{app}^{TOF} = 0.9 - 1.3 + \frac{1}{2} 2.0 = 0.6$ eV. Here, we have considered that desorption of CO occurs dominantly from C sites ($\omega_{\alpha}^{R} \sim 0.5$ in Fig. \ref{Figure05}(a) of the main report, with $E_{CO_C \rightarrow V_C}^{k} = 1.3$ eV in Table \ref{Table_Processes}), which gives $\Delta H_{CO} = E_{CO_C \rightarrow V_C}^{k}  - E_{V \rightarrow CO}^{k} = 1.3 - 0 = 1.3$ eV. In comparison, CO desorption from B sites is negligible ($\omega_{\alpha}^{R} \sim 10^{-5}$ in Fig. \ref{Figure05}(a) of the main report, with $E_{CO_B \rightarrow V_B}^{k} = 1.6$ eV in Table \ref{Table_Processes}). Similarly, desorption of $O$ occurs dominantly from C sites ($\omega_{\alpha}^{R} \sim 0.1$ in Fig. \ref{Figure05}(a) of the main report, with $E_{O_C + O_C \rightarrow V_C + V_C}^{k} = 2.0$ eV in Table \ref{Table_Processes}), which gives $\Delta H_{O} = E_{O_C + O_C \rightarrow V_C + V_C}^{k}  - E_{V \rightarrow O}^{k} = 2.0 - 0 = 2.0$ eV. Since $E_{app}^{TOF} \approx 1.3$ eV at $T_4$ according to Fig. \ref{Figure05}(a) of the main report, the Temkin value of 0.6 eV fails by about 0.7 eV. This cannot be explained only by the fact that the Temkin expression for $E_{app}^{TOF}$ neglects the term $E_{CO_C + O_C \rightarrow CO_2}^{k^0} = k_BT$ (Eq. \ref{E_alpha_recNdif_k0}), since this term is small ($\approx 0.06$ eV at $T_4$). The error can be assigned to the failure of random mixing, due to the presence of strong adsorbate correlations at high coverage (of $O$). In this case, the adsorption of $CO$ is not random, occurring preferentially at $C$ sites, as a result of preferential desorption of $CO$ and $O_2$ from $C$ sites. Thus, substituting $M_{CO_C + O_C \rightarrow CO_2}$ by $\theta_{CO} \theta_{O} \approx (K_{CO}p_{CO})(K_{O}p_{O})^{-1/2}$ is a poor approximation in this system, due to the failure of random mixing. The present study shows that $E_{app}^{TOF}$ is described accurately when $M_{CO_C + O_C \rightarrow CO_2}$ is determined correctly.

{\em Example S2.}  When the RDS in the Langmuir-Hinshelwood model is the adsorption of $B_2$, traditionally one will write (see Section S1): $TOF = k_{a}^{B}p_{B}\theta_{*}^2$, where $\theta_{*}$ is the coverage by all empty sites and $\theta_{*}^2$ describes the coverage by all empty pairs of sites in the homogeneous mixing approximation. Since the adsorption of $B_2$ is the RDS, traditionally one assumes adsorption-desorption equilibrium for $A$, thus leading to the Langmuir isotherm: $\theta_{A} = K_{A}p_{A}/(1+K_{A}p_{A})$ and $\theta_{*} = 1/(1+K_{A}p_{A})$. If $A$ is strongly adsorbed, then $\theta_{A} \approx 1$ (large) and $\theta_{*} \approx (K_{A}p_{A})^{-1}$ (small), which is the situation at low temperatures for models I-IV, with $A=CO$ and $B_2=O_2$. Then, traditionally one obtains: $TOF = k_{a}^{B}p_{B}(K_{A}p_{A})^{-2}$ and, thus, $E_{app}^{TOF} = E_{a}^{B} + 2 \Delta H_{A}$. Focusing on the temperature range between $T_1$ and $T_2$ for model I  (see Figs. 4(d)-(f) and 5(a-b) of the main report), we obtain: $E_{app}^{TOF} = E_{V \rightarrow O}^{k} + 2 \Delta H_{CO} = 0 + 2 \times 1.3 = 2.6$ eV. Here, we have used the fact that the activation energy for adsorption of $O_2$ is zero ($E_{V \rightarrow O}^{k} = 0$ eV in Table \ref{Table_Processes}, as for any other adsorption process in models I-IV) and desorption of CO occurs dominantly from C sites ($\omega_{\alpha}^{R} \sim 0.5$ in \ref{Figure05}(a) of the main report, with $E_{CO_C \rightarrow V_C}^{k} = 1.3$ eV in Table \ref{Table_Processes}), which gives $\Delta H_{CO} = E_{CO_C \rightarrow V_C}^{k}  - E_{V_C \rightarrow CO_C}^{k} = 1.3 - 0 = 1.3$ eV. In comparison, CO desorption from B sites is negligible ($\omega_{\alpha}^{R} \sim 10^{-4}$ in Fig. \ref{Figure05}(a) of the main report, with $E_{CO_B \rightarrow V_B}^{k} = 1.6$ eV in Table \ref{Table_Processes}). Since $E_{app}^{TOF} \approx 2.87$ eV between $T_1$ and $T_2$ according to Fig. \ref{Figure04}(d) of the main report, the Temkin value of 2.6 eV fails by about 0.27 eV. This is due to the inadequacy of the approximation to describe the coverage by all empty pairs of sites using $\theta_{*}^2$, which ultimately is due to the failure of the random mixing approximation. The present study shows that $E_{app}^{TOF}$ is described accurately when the coverage for this collection of sites is determined correctly as the multiplicity for the adsorption of $O_2$.

\section{Cut-offs in the proximity $\sigma_{\alpha}^{TOF}$}
\label{Cut_offs_in_the_proximity}

\begin{figure*}[htb!]
  \begin{center}
    \includegraphics[width=1.45\columnwidth]{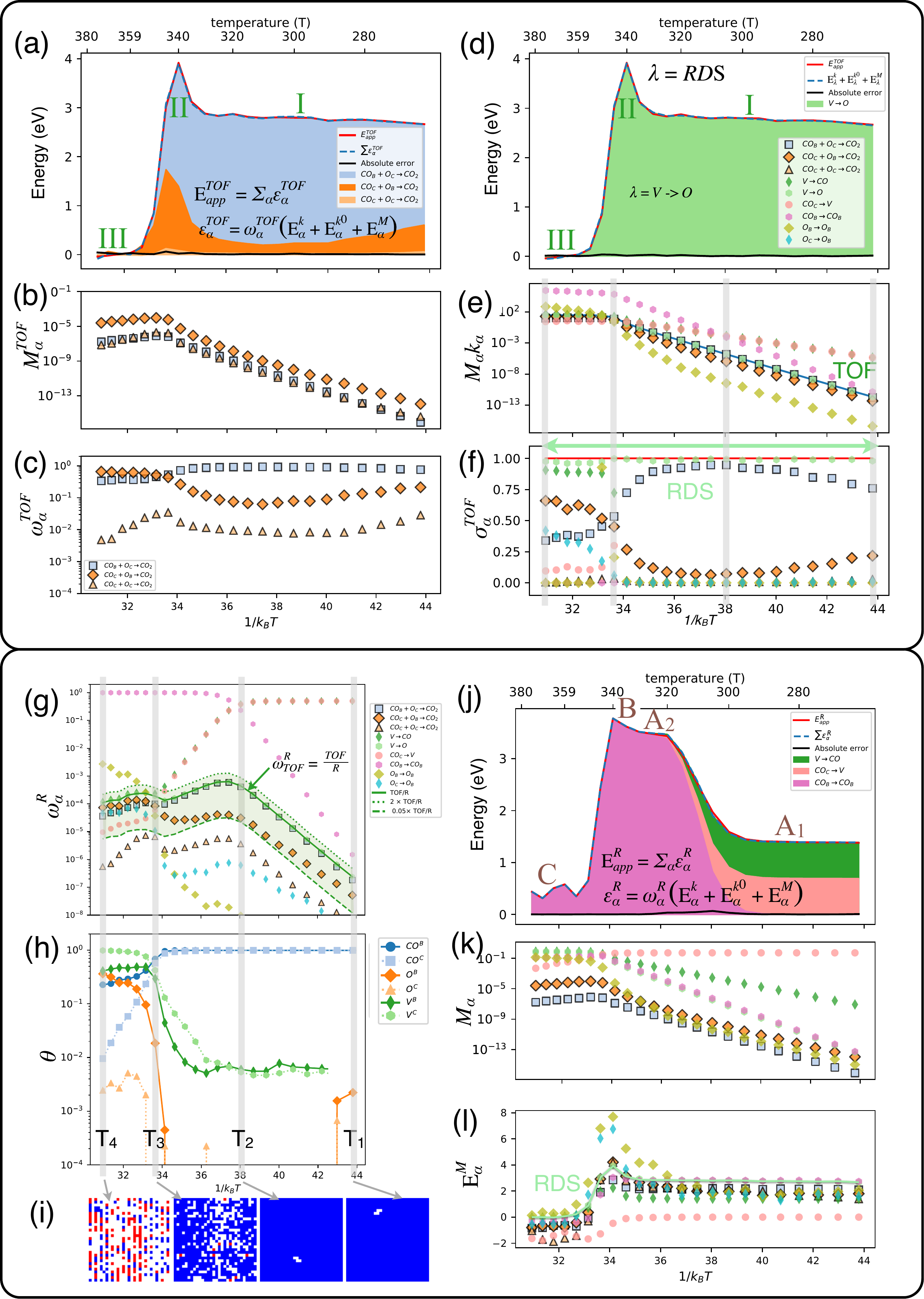}
  \end{center}
  \caption{
  Temperature dependence for model II (oxidation of CO on RuO$_2$\{110\}): 
  (a) Apparent activation energy ($E_{app}^{TOF}$) for the $TOF$ in Fig. \ref{Figure1S}(a). $E_{app}^{TOF}$ is described well by $\sum_{\alpha \in \{ x \}} \epsilon_{\alpha}^{TOF}$, where $\epsilon_{\alpha}^{TOF} = \omega_{\alpha}^{TOF} (E_{\alpha}^{k} + E_{\alpha}^{k^{0}} + E_{\alpha}^{M})$. The absolute error $|E_{app}^{TOF} -  \sum_{\alpha \in \{ x \}} \epsilon_{\alpha}^{TOF} |$ is also plotted. 
  (b), (c) Multiplicities ($M_{\alpha}^{TOF}$) and probabilities ($\omega_{\alpha}^{TOF}$) for those elementary reactions explicitly contributing to the $TOF$, respectively. 
  (d) Same as (a), now describing $E_{app}^{TOF}$ as $E_{\lambda}^{k} + E_{\lambda}^{k^{0}} + E_{\lambda}^{M}$ for the RDS. The absolute error $|E_{app}^{TOF} -  (E_{\lambda}^{k} + E_{\lambda}^{k^{0}} + E_{\lambda}^{M}) |$ is also plotted.  
  (e) $M_{\alpha} k_{\alpha}$ for any elementary reaction with probability $\omega_{\alpha}^{R} \ge 10^{-8}$ at any temperature. The $TOF$ is matched by $M_{\lambda}k_{\lambda}$ for some $\lambda$ within some range of temperature .
  (f) Proximity to the $TOF$ ($\sigma_{\alpha}^{TOF}$), enabling the assignment of the RDS at every temperature.
     (g) reaction probabilities $\omega_{\alpha}^{R}$ ($\ge 10^{-8}$). 
     (h) Coverage for all adspecies ($\theta_{\rm CO_{B}}$, $\theta_{\rm CO_{C}}$, $\theta_{\rm O_{B}}$, $\theta_{\rm O_{C}}$, $\theta_{\rm V_{B}}$ and $\theta_{\rm V_{C}}$ $\ge 10^{-4}$). 
     (i) Morphology snapshots at various temperatures.
  (j) Apparent activation energy ($E_{app}^{R}$) for the total rate per active site $R$ in Fig. \ref{Figure1S}(a). $E_{app}^{R}$ is described well by $\sum_{\alpha \in \{ e \}} \epsilon_{\alpha}^{R}$, where $\epsilon_{\alpha}^{R} = \omega_{\alpha}^{R} (E_{\alpha}^{k} + E_{\alpha}^{k^{0}} + E_{\alpha}^{M})$. The absolute error $|E_{app}^{R} -  \sum_{\alpha \in \{ e \}} \epsilon_{\alpha}^{R} |$ is also plotted. 
  (k), (l) Multiplicities ($M_{\alpha}$) and effective configurational energies ($E_{\alpha}^{M}$) for any elementary reaction with probability $\omega_{\alpha}^{R} \ge 10^{-8}$ at any temperature, respectively. $E_{\alpha}^{M}$ valid for both frames (b) and (k).
}
  \label{Figure3S}
\end{figure*}

Fig.~\ref{Figure05}(a) of the main report and Figs. \ref{Figure3S}(g), \ref{Figure4S}(g) and \ref{Figure5S}(g) of this Supporting Information display the probability to observe any elementary reaction ($\omega_{\alpha}^{R} = M_{\alpha} k_{\alpha} / R$) together with the probability to observe any reaction explicitly contributing to the $TOF$, $\omega_{TOF}^{R} = TOF / R$. In addition, these figures display two more curves, namely, $2\omega_{TOF}^{R}$ and $0.05\omega_{TOF}^{R}$. Considering the definition of the proximity $\sigma_{\alpha}^{TOF}$ in Eq. \ref{sigma_definition} of the main text, any elementary reaction with probability $\omega_{\alpha}^{R}$ between $\omega_{TOF}^{R}$ and $2\omega_{TOF}^{R}$ will lead to proximity values between 1 and 0. Likewise, if $\omega_{\alpha}^{R}$ falls between $\omega_{TOF}^{R}$ and $0.05\omega_{TOF}^{R}$ the proximity will lie between 1 and 0.05.
 
Although the definition of the proximity implies the use of an upper cutoff ($2\omega_{TOF}^{R}$, beyond which the sensitivity is 0), no actual lower cutoff is used. In practice, any reaction with probability $< 0.05\omega_{TOF}^{R}$ will occur so rarely (with respect to the TOF) that the reaction itself becomes irrelevant, thus justifying the use of $0.05\omega_{TOF}^{R}$ as a visual lower cut-off in Fig. \ref{Figure05}(a) of the main report and Figs. \ref{Figure3S}(g), \ref{Figure4S}(g) and \ref{Figure5S}(g) of this Supporting Information. Regarding the upper cut-off, a general value $c \cdot \omega_{TOF}^{R}$ with $c > 1$ can be used by modifying the definition of the proximity to: $\sigma_{\alpha}^{TOF} = 1 - min(c-1,\delta_{\alpha}^{TOF})/(c-1)$. While we use $c=2$, $c=4-10$ will leave Fig. \ref{Figure05}(a) of the main report and Figs. \ref{Figure3S}(f), \ref{Figure4S}(f) and \ref{Figure5S}(f) of this Supporting Information essentially unchanged, only modifying the left or right half of the sensitivity spikes, {\it i.e.} that half corresponding to the reaction probabilities falling between $\omega_{TOF}^{R}$ and $c\cdot \omega_{TOF}^{R}$.

\section{Multiplicity analysis for additional models}
\label{Multiplicity_analysis_for_additional_models}

\subsection{Oxidation of CO using models II, III and IV}
\label{Multiplicity_Oxidation_of_CO_using_models_II_III_and_IV}

Fig. \ref{Figure1S}(a) shows that model II is dominated by adsorption and desorption at low temperatures (region $A_{1}$ for $R$), while diffusion becomes the leading reaction above $\sim$ 305 K (regions $A_{2}$, $B$ and $C$ for $R$). Although $R$ displays four regions, the curves for $R_{a}$, $R_{d}$, $R_{h}$ and $R_{r}$ exhibit three regions, labelled as I, II and III for the $TOF$ ($=R_{r}$). Here, region II displays a larger slope than region I, as evidenced in the corresponding derivative, shown in Fig. \ref{Figure3S}(a). As in the main report, the analysis of the apparent activation energy of the $TOF$ performed here for model II, as displayed in Figs. \ref{Figure3S}(a)-(c), concludes that $E_{app}^{TOF}$ is accurately explained by Eq. \ref{EappTOF_5} of the report, with the absolute error remaining $ \lesssim 0.062$ eV across all regions. Similarly, the analysis of  $E_{app}^{TOF}$ based on determining the RDS, as shown in Figs. \ref{Figure3S}(d)-(f), concludes that also Eq. \ref{EappTOF_6} of the report explains accurately the atomistic origin of $E_{app}^{TOF}$, with the absolute error remaining $ \lesssim 0.036$ eV across all regions. In this model, the RDS is assigned to the adsorption of O atoms ($V \rightarrow O$) in all three regions.

As already found in the main report, the mere observation of a linear Arrhenius behavior (in region I, see Fig. \ref{Figure1S}(a)) does not imply that $E_{app}^{TOF}$ ($\approx 2.82$ eV) can be assigned to the elementary reaction with largest activation energy (4.82 eV, for $O_{\rm B} + O_{\rm B} \rightarrow V_{\rm B} + V_{\rm B}$) nor to only the elementary activation energy of the RDS, $E_{\lambda}^{k}$ (= 0 eV, for $\lambda = V \rightarrow O$), since this will neglect the configurational contribution, $E_{\lambda}^{M} \approx 2.82$ eV, which in this case fully explains the value of $E_{app}^{TOF}$.

Figs. \ref{Figure3S}(g)-(i) provide detailed information about the relative competition between the different elementary reactions. The situation at $T_{1}$ is similar to that for model I, involving equilibrated adsorption and desorption of CO ($V \rightarrow CO$ and $CO_{\rm C} \rightarrow V$, with probability $\omega_{\alpha}^{R} \sim 0.5$), diffusion ($CO_{\rm B} \rightarrow CO_{\rm B}$, with $\omega_{\alpha}^{R} \sim 10^{-6}$), adsorption of O ($V \rightarrow O$, with $\omega_{\alpha}^{R} \sim 2.5 \times 10^{-7}$), and recombination ($CO_{\rm B} + O_{\rm C} \rightarrow CO_{2}$, with $\omega_{\alpha}^{R} \sim 2 \times 10^{-7}$). The next probable reaction, $CO_{\rm C} + O_{\rm B} \rightarrow CO_{2}$, has roughly four times lower probability ($\omega_{\alpha}^{R} \sim 5 \times 10^{-8}$) than $CO_{\rm B} + O_{\rm C} \rightarrow CO_{2}$ and, thus, can be neglected. As in the similar context for model I at $T_1$, the adsorption of O ($V \rightarrow O$) is the RDS. This agrees with Fig. \ref{Figure3S}(f).

At $T_2$, the description from $T_1$ remains essentially valid, although now diffusion ($CO_{\rm B} \rightarrow CO_{\rm B}$, with $\omega_{\alpha}^{R} \sim 0.2$) is almost as probable as the adsorption and desorption of CO ($V \rightarrow CO$ and $CO_{\rm C} \rightarrow V$, with $\omega_{\alpha}^{R} \sim 0.4$). As for $T_1$, recombination occurs essentially through the $CO_{\rm B} + O_{\rm C} \rightarrow CO_{2}$ route (now with $\omega_{\alpha}^{R} \sim 4 \times 10^{-7}$) and the adsorption of O ($V \rightarrow O$, with $\omega_{\alpha}^{R} \sim 4 \times 10^{-4}$) remains the RDS, in agreement with Fig. \ref{Figure3S}(f).

At $T_3$, the picture has changed significantly. Now, the diffusion of CO along the B rows dominates the activity of the system ($CO_{\rm B} \rightarrow CO_{\rm B}$, with probability $\sim 1.0$). The next most probable reaction is the adsorption of CO ($V \rightarrow CO$, with $\omega_{\alpha}^{R} \sim 2.5 \times 10^{-4}$), followed by the adsorption of O ($V \rightarrow O$, with $\omega_{\alpha}^{R} \sim 2 \times 10^{-4}$), two recombinations ($CO_{\rm B} + O_{\rm C} \rightarrow CO_{2}$ and $CO_{\rm C} + O_{\rm B} \rightarrow CO_{2}$, with $\omega_{\alpha}^{R} \sim 10^{-4}$ and $\sim 8 \times 10^{-5}$, respectively), the desorption of CO ($CO_{\rm C} \rightarrow V$, with $\omega_{\alpha}^{R} \sim 5 \times 10^{-5}$), and diffusion of O ($O_{\rm B} \rightarrow O_{\rm B}$, with $\omega_{\alpha}^{R} \sim 3 \times 10^{-5}$). Any other elementary reaction is significantly less probable. Thus, the situation is as follows. If CO is adsorbed on a B site, recombination has to wait until an O is adsorbed on a C site. On the other hand, if CO is adsorbed on a C site, there is a small chance that adsorption occurs next to an existing O$_B$, thus leading to recombination, but in most cases recombination has to wait until an O is adsorbed on a B site. In other words, the system is ready for recombination as soon as O is adsorbed on either B or C sites. Thus, the adsorption of O ($V \rightarrow O$) is the RDS, as shown in Fig. \ref{Figure3S}(f).

Finally, at $T_4$, we have a rather different situation. Compared to the super-frequent random diffusion of CO along the B rows ($CO_{\rm B} \rightarrow CO_{\rm B}$, with $\omega_{\alpha}^{R} \sim 1$ ), the next most-probable elementary reaction is the diffusion of O, also along the B rows ($O_{\rm B} \rightarrow O_{\rm B}$, with $\omega_{\alpha}^{R} \sim 3 \times 10^{-3}$), while the rest of the reactions are executed with much lower probabilities, in the range $10^{-4}$ to $10^{-5}$. Since the C rows are essentially empty (see Fig. \ref{Figure3S}(h)), in relative terms, the adsorptions of CO and O (both predominantly at C sites) occur rather frequently ($V \rightarrow CO$ and $V \rightarrow O$, respectively, with $\omega_{\alpha}^{R} \sim 10^{-4}$ for both). In turn, inter-row diffusion of O ($O_{\rm C} \rightarrow O_{\rm B}$, with $\omega_{\alpha}^{R} \sim 5 \times 10^{-5}$) has become comparable to the two recombination reactions ($CO_{\rm C} + O_{\rm B} \rightarrow CO_{2}$ and $CO_{\rm B} + O_{\rm C} \rightarrow CO_{2}$, with probabilities $\sim 8 \times 10^{-5}$ and $\sim 4 \times 10^{-5}$, respectively) while the desorption of CO has become relatively infrequent ($CO_{\rm C} \rightarrow V$, with $\omega_{\alpha}^{R} \sim 10^{-5}$). Thus, the situation is as follows. Minor adsorption of both CO and O at B sites essentially restores their overall coverage, compensating their desorption as CO$_2$. On the other hand, after the adsorption of CO at a C site, recombination is attempted many times (and eventually occurs) as many O atoms pass by, diffusing along the left and right neighbor B rows. Similarly, after the adsorption of O at a C site, recombination is also attempted many times, eventually occurring with one of the many CO molecules passing by as they diffuse along either neighboring B row. In this manner, the system is rather sensitive to the actual values of the recombination rates for $CO_{\rm C} + O_{\rm B} \rightarrow CO_{2}$ and $CO_{\rm B} + O_{\rm C} \rightarrow CO_{2}$, as shown in Fig. \ref{Figure3S}(f). However, quantitatively, Fig. \ref{Figure3S}(f) shows that it is the adsorption of CO ($V \rightarrow CO$) and, especially, the adsorption of O ($V \rightarrow O$) that must be considered as the RDS. The special role of $V \rightarrow O$ (as the true RDS) can be understood from the fact that, once adsorbed, an O$_C$ has a sizable chance to change row and become $O_B$ due to diffusion ($O_{\rm C} \rightarrow O_{\rm B}$). Thus, the adsorption of O on C sites contributes indirectly to the recombination of type $CO_{\rm C} + O_{\rm B} \rightarrow CO_{2}$, in addition to contributing directly to $CO_{\rm B} + O_{\rm C} \rightarrow CO_{2}$. This shows how the proposed formalism allows understanding the assignment of the RDS based on $\sigma_{\alpha}^{TOF} \approx 1$.

\begin{figure*}[htb!]
  \begin{center}
    \includegraphics[width=1.45\columnwidth]{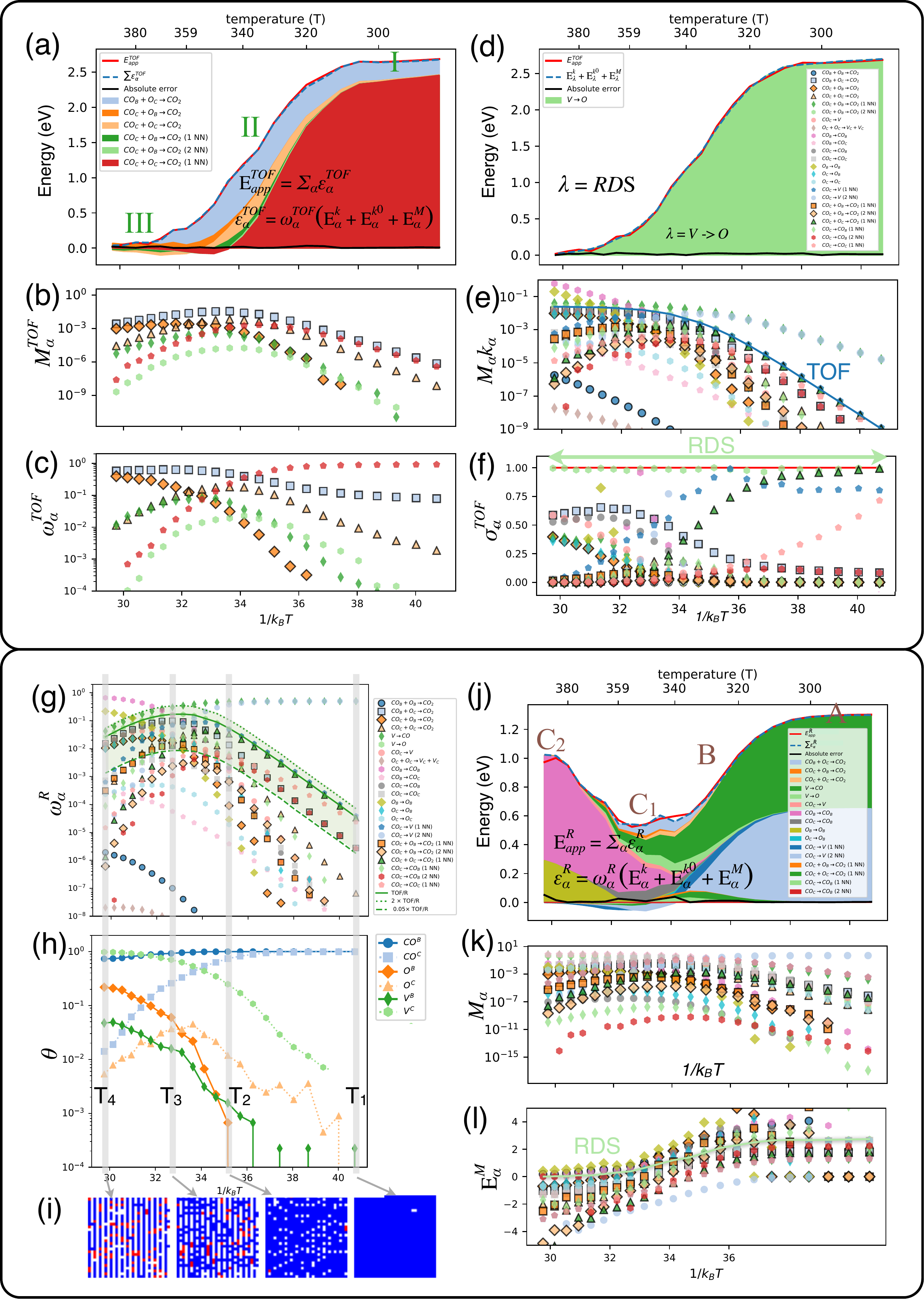}
  \end{center}
  \caption{Same as Fig. \ref{Figure3S}, now for model IV in relation to the $TOF$ and total rate $R$ shown in Fig. \ref{Figure1S}(d).}
  \label{Figure4S}
\end{figure*}

Lastly, Figs. \ref{Figure3S}(j)-(l) confirm that the overall reaction for model II is dominated by adsorption and desorption at the lower temperatures of region $A_{1}$ while diffusion already dominates at the higher temperatures of region $A_{2}$, fully prevailing in both regions $B$ and $C$. As indicated in the last paragraph of Section "Theory" of the main text, the $TOF$ will be sensitive to variations in the rates of these processes due to their ability to scale the time increment $\Delta t \propto 1/\hat{r}_{}$. The analysis of the apparent activation energy of the total rate per active site $R$, displayed in Fig. \ref{Figure3S}(j), concludes that $E_{app}^{R}$ is accurately explained by Eq. \ref{Eapp_e_1} of the report, with the absolute error remaining $ \lesssim 0.068$ eV across all regions. 

Regarding model III, comparison of Figs. \ref{Figure1S}(a) and \ref{Figure1S}(c) shows that models II and III behave essentially the same, except for the fact that: (i) the cross-over between regions $A_1$ and $A_2$ is located at lower temperature in model III, and (ii) region C displays small fluctuations around a constant value in model III, instead of the small--but steady--increase observed in model II. Due to these similarities, model III does not provide any novelty with respect to model II and its detailed analysis is skipped.

Model IV considers the presence of repulsion between nearest neighbor COs located at C sites, which leads to several differentiated elementary reactions (rows 22 through 29 in Table \ref{Table_Processes}). Thus, compared to models I through III, model IV involves a larger number of atomistic activation energies $E_{\alpha}^{k}$. In spite of this, Fig. \ref{Figure1S}(d) shows that, qualitatively, model IV shares some similarities with models II and III, with adsorption and desorption dominating at low temperature and diffusion leading the activity at high temperature. In fact, the total rates per active site for adsorption ($R_{a}$), desorption ($R_{d}$), diffusion ($R_{h}$) and recombination ($R_{r} = TOF$) can also be broken into three regions (labelled I, II and III for the $TOF$) while the total rate per active site $R_{}$ displays four regions ($A$, $B$, $C_1$ and $C_2$), in this case due to diffusion overtaking adsorption and desorption at high temperature, while the corresponding cross-over takes place at low temperature for models II and III.

Despite the larger number of elementary reactions, the analysis of the apparent activation energy of the TOF, as shown in Figs. \ref{Figure4S}(a)-(c), concludes that $E_{app}^{TOF}$ is accurately explained by Eq. \ref{EappTOF_5} of the report, with a small absolute error $ \lesssim 0.033$ eV across all regions. Similarly, the analysis of $E_{app}^{TOF}$ based on finding the RDS, as shown in Figs. \ref{Figure4S}(d)-(f), concludes that also Eq. \ref{EappTOF_6} of the report explains accurately the temperature dependence of $E_{app}^{TOF}$, with the absolute error also remaining $ \lesssim 0.032$ eV across all regions. In spite of the complexity of the model, the RDS is clearly assigned to the adsorption of O atoms ($V \rightarrow O$) in the complete range of explored temperatures.

As before, the observation of a linear Arrhenius behavior (in region I) does not imply that $E_{app}^{TOF}$ ($\approx 2.65$ eV) can be assigned to the elementary reaction with largest activation energy (4.29 eV, for $O_{\rm B} + O_{\rm B} \rightarrow V_{\rm B} + V_{\rm B}$) nor to only the elementary activation energy of the RDS, $E_{\lambda}^{k}$ (= 0 eV, for $\lambda = V \rightarrow O$), since this will neglect the important configurational contribution, $E_{\lambda}^{M} \approx 2.65$ eV, which fully explains the value of $E_{app}^{TOF}$ also in model IV.

In turn, Figs. \ref{Figure4S}(g)-(i) provide detailed information about the relative competition between the different elementary reactions. The situation at $T_{1}$ through $T_{4}$ is very similar to that for models II and III, and thus we refrain from giving all the details. Overall, the reaction at $T_{1}$ and $T_{2}$ occurs with equilibrated adsorption and desorption of CO, and a dominating recombination ($CO_{\rm C} + O_{\rm C} \rightarrow CO_2 (1 NN)$), which is triggered as soon as the adsorption of O takes place. Thus, $V \rightarrow O$ is the RDS. At $T_{3}$ the same picture is valid, with also $V \rightarrow O$ as the RDS, but now the dominating recombination is $CO_{\rm B} + O_{\rm C} \rightarrow CO_2$ and, the adsorption and desorption of CO are not (one-to-one) equilibrated anymore (equilibration is through the overall network of reactions). Finally, at $T_{4}$, the predominant recombination ($CO_{\rm B} + O_{\rm C} \rightarrow CO_2$) occurs soon after the adsorption of O at a C site ($V \rightarrow O$), which reacts with one of the many CO molecules passing by as they diffuse along either neighboring B row ($CO_{\rm B} \rightarrow CO_{\rm B}$). In probability space, the other possible recombination ($CO_{\rm C} + O_{\rm B} \rightarrow CO_2$) lies about 4 times below $V \rightarrow CO$, thus making the adsorption of CO less critical than the adsorption of O. In this manner, $V \rightarrow O$ is the RDS, in agreement with Fig. \ref{Figure4S}(f). Thus, at $T_{4}$ the picture is very similar to that for models II and III, with $V \rightarrow CO$ having a less significant role.

Lastly, Figs. \ref{Figure4S}(j)-(l) confirm that the overall reaction for model IV is dominated by adsorption and desorption events at low temperature (regions $A$ and $B$, dominated by the adsorption and desorption of CO) while the diffusion reactions dominate at high temperature (region $C_{2}$, with diffusion of both CO and O along the B rows). In region $C_{1}$, there is a complex mixture of elementary reactions with relative relevance for the overall catalytic reaction. Nevertheless, in terms of the $TOF$, the important reactions in that region are displayed in Fig. \ref{Figure4S}(f). Fig. \ref{Figure4S}(j) shows that the apparent activation energy of the total rate per active site $R$, $E_{app}^{R}$, is accurately explained by Eq. \ref{Eapp_e_1} of the report, with the absolute error remaining $ \lesssim 0.051$ eV across all regions.

\vspace{5 mm}
\subsection{Selective oxidation of NH$_3$}
\label{Multiplicity_Selective_oxidation_of_NH3}

\begin{figure*}[htb!]
  \begin{center}
    \includegraphics[width=1.45\columnwidth]{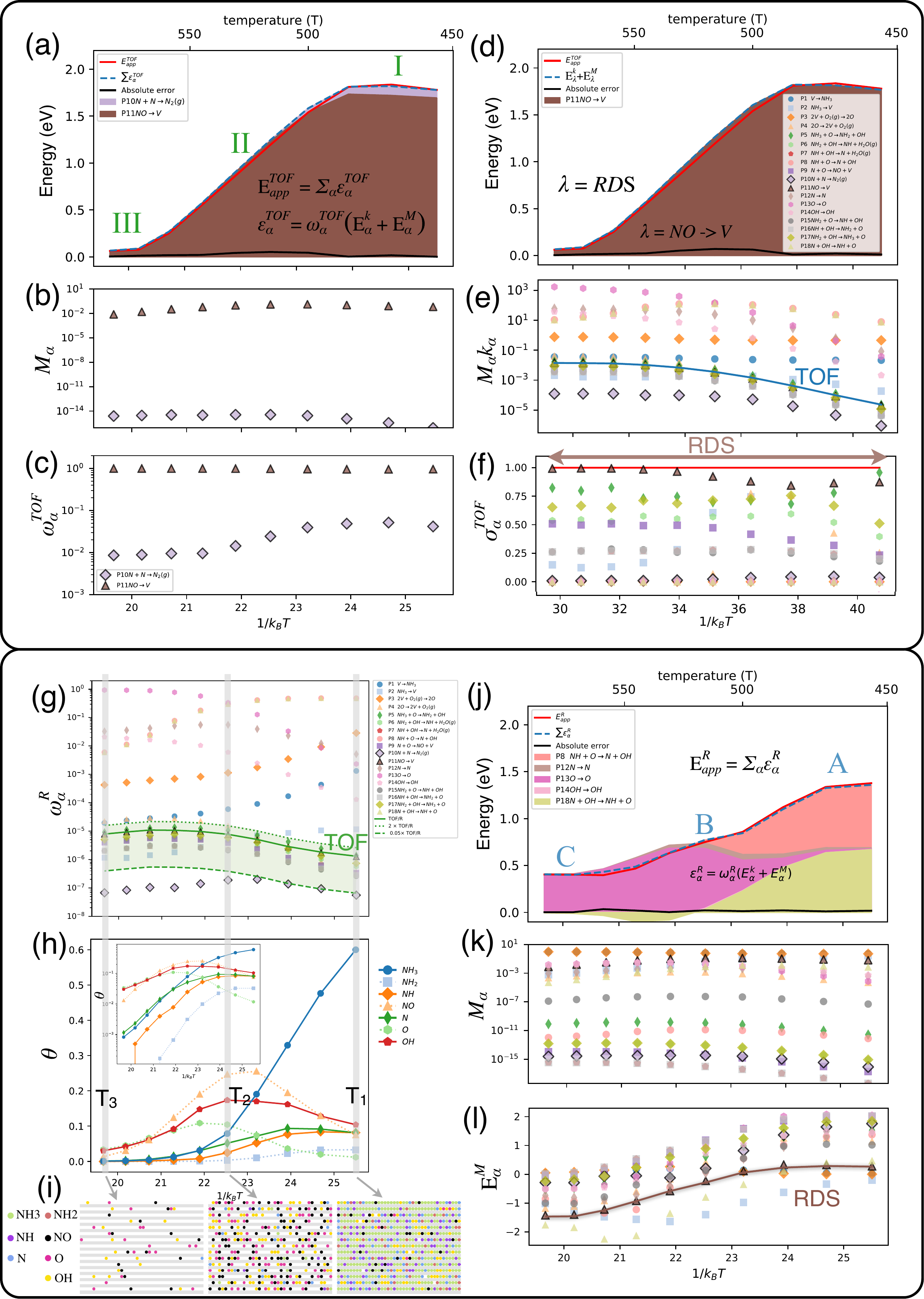}
  \end{center}
  \caption{Same as Figs. \ref{Figure3S} and \ref{Figure4S}, now for the selective oxidation of NH$_3$ on RuO$_2$\{110\} in relation to the $TOF$ and total rate $R$ shown in Fig. \ref{Figure2S}(b)}
  \label{Figure5S}
\end{figure*}

Regarding our results for Hong {\it et al.}'s reaction mechanism for the oxidation of NH$_3$ on RuO$_2$\{110\} \cite{Hong2010}, Fig. \ref{Figure2S}(b) shows how the overall reaction is dominated by recombinations at low temperature (region A) and diffusion hops at high temperature (region C), with a clear crossover at around 530 K (region B). According to this figure, the adsorption and desorption reactions occur much less frequently (roughly about one adsorption/desorption every 10$^4$ recombinations/hops). In particular, the desorption of the target products (NO and N$_2$) occurs even less frequently (approximately about one desorption of NO every 10$^5$ recombinations/hops and roughly one desorption of N$_2$ every 10$^6$ recombinations/hops). Thus, the two reactions explicitly contributing to the $TOF$ in this system (the desorption of NO and the formation-and-direct-desorption of $N_2$) occur rather infrequently. This is clearly reflected by the fact that, in probability space, the $TOF$ appears roughly in the range between $10^{-5}$ and $10^{-6}$ for the considered temperature span (see Fig. \ref{Figure5S}(g)). 

As with previous models, assuming negligible temperature dependence of the multiplicity prefactors ($E_{\alpha}^{M^{0}}=0$), the analysis of the apparent activation energy of the $TOF$ performed in Figs. \ref{Figure5S}(a)-(c) concludes that $E_{app}^{TOF}$ is accurately explained by Eq. \ref{EappTOF_5} of the main report, the absolute error remaining $ \lesssim 0.052$ eV across all regions. Note that, in Hong {\it et al.}'s model the rate prefactors are constant and, thus, $E_{\alpha}^{k^0} = 0$ for most processes (except for the adsorption of NH$_3$ and $O_2$, which have no contribution to Eq. \ref{EappTOF_5}). Similarly, the analysis of  $E_{app}^{TOF}$ based on determining the RDS, as shown in Figs. \ref{Figure5S}(d)-(f), concludes that also Eq. \ref{EappTOF_6} of the main report explains accurately the atomistic origin of $E_{app}^{TOF}$, the absolute error remaining $ \lesssim 0.071$ eV across all regions in this case. Again, in Hong {\it et al.}'s model $E_{\lambda}^{k^0} = 0$ for the RDS.

According to Fig. \ref{Figure5S}(f), we conclude that (i) the desorption of NO (reaction P11) is the Rate Determining Step (RDS) in all three regions, and (ii) there is a wide variety of Rate Controlling Steps (RCSs), {\it i.e.} the $TOF$ is strongly sensitive to the rate constants of many elementary reactions. These include H abstraction reactions and their reverse processes (P5 and P17; P15 and P16; and P6, which has no reverse reaction in Hong {\it et al.}'s mechanism) and the formation of NO (P9). Note that the corresponding formation (and direct desorption) of N$_2$ is not a RCS, since this reaction represents a tiny contribution to the TOF. Finally, we note that also the reactions for which $w_{\alpha}^R \approx 1$ need to be considered as RCSs, since they affect the $TOF$ by scaling the total rate and, thus, time (see the last paragraph of Section \ref{Theory} of the main report). Accordingly, considering Fig. \ref{Figure5S}(g) we conclude that also the abstraction reaction P8 and its reverse P18 are RCSs, especially below 530 K, as well as the diffusion reaction P13, especially above 530 K. In fact, the strong dependence of the total rate per active site on the elementary reactions P8, P18 and P13 is clearly reflected in Fig. \ref{Figure5S}(j).

Considering Fig. \ref{Figure5S}(g), the previous information about the RDS and RCSs can be used to draw a simple picture about the overall catalytic reaction at any particular temperature (such as $T_{1}$, $T_{2}$ and $T_{3}$ in Figs. \ref{Figure5S}(g)-(i)). Namely, the reaction takes place as a cascade of abstraction reactions (between adsorbed NH$_3$/NH$_2$/NH and adsorbed O/OH), sequentially stripping the H atoms until bare N is present at the surface, where it recombines with either adsorbed O (to form NO, which is desorbed later) or with itself (to form N$_2$, which is desorbed immediately). Although having a relatively low energy barrier (0.27 eV), the formation of N$_2$ occurs rarely (see Fig. \ref{Figure5S}(g)) due to the low chance for two N atoms to meet each other as nearest neighbors (very low $M_{\alpha}$ for this recombination reaction, as shown in Fig. \ref{Figure5S}(b),(k)). On the contrary, having the largest energy barrier (1.49 eV), the desorption of NO occurs relatively frequently (see Fig. \ref{Figure5S}(g)) due to the large chance for the N atoms to meet O atoms as nearest neighbors (very high $M_{\alpha}$ for this reaction, as shown in Fig. \ref{Figure5S}(b),(k)). Nevertheless, the desorption of NO is relatively infrequent with respect to the other rate controlling reactions (H abstractions, formation of NO and the diffusion of O), thus justifying its role as RDS. All of the aspects described here are in excellent agreement with the analysis of the overall reaction presented by Hong {\it et al.}, as summarized in Fig. 2(b) of Ref. \cite{Hong2010}.

\vspace{5 mm}
\section{Rate Determining Step for model I}
\label{Rate_Determining_Step_for_model_I}

Following the main report, let us use $\xi_{\alpha}^{}$ to refer to the {\em degree of rate sensitivity}, as defined in Refs. \cite{Dumesic2008, Meskine2009}. Similarly, as previously considered in Refs. \cite{Lynggaard2004, Meskine2009, Hess2017, Campbell1994}, let us refer to the {\em degree of rate control} as $\chi_{\alpha_*}^{}=\xi_{\alpha_+}^{} + \xi_{\alpha_-}^{}$, where $\alpha_*$ designates the combined forward-and-backward reaction. Regarding the Rate Determining Step (RDS) for model I (see Fig. \ref{Figure04}(f) of the main report), our results agree with the data presented in Fig. 5 of Ref. \cite{Meskine2009}. Furthermore, considering the $\xi_{\alpha}^{}$ data displayed in Fig. 5 of Ref. \cite{Meskine2009} for the adsorption of CO at C sites and the desorption of CO from C sites, we conclude that the two curves are essentially the same, but have opposite signs. Thus, by summing them to obtain $\chi_{\alpha_*}^{}$, one gets essentially $\chi_{\alpha_*}^{} \approx 0$ for the adsorption-and-desorption of CO in the whole range of temperature. This means that the RDS cannot be assigned to the adsorption/desorption of CO in this system.

On the other hand, by summing the $\xi_{\alpha}^{}$ curves for the adsorption and desorption of O$_2$ shown in Fig. 5 of Ref. \cite{Meskine2009} the resulting $\chi_{\alpha*}^{}$ becomes 0 below $\sim 1.7 \times 10^{-3}$ K$^{-1}$ and stands as the only non-zero curve above $1.8 \times 10^{-3}$ K$^{-1}$, with value $\sim 0.5$ at low temperatures. Similarly, the recombination $O_{\rm C} + CO_{\rm C} \rightarrow CO_2$ remains as the only process with non-zero $\chi_{\alpha*}^{}$ value below $\sim 1.7 \times 10^{-3}$ K$^{-1}$, also with value 0.5. Assuming the value $\chi_{\alpha*}^{} \approx 0.5$ may be treated as $\chi_{\alpha*}^{} \approx 1$ (perhaps due to a factor of 2 somewhere in the equations/analysis of Ref. \cite{Meskine2009}), the RDS will correspond to (i) the adsorption/desorption of O$_2$ above $1.8 \times 10^{-3}$ K$^{-1}$ and (ii) the recombination of $CO_{\rm C}$ and $O_{\rm C}$ below $\sim 1.7 \times 10^{-3}$ K$^{-1}$, which would be in excellent agreement with our result, as shown in Fig. \ref{Figure04}(d) of the main report. Note that our data are clearer, presumably due to the lack of any additional processing in our case. 

This designation of the RDS in region I to the adsorption of O in both studies is in conflict with the assignment of the apparent activation energy (2.85 eV) to the desorption of CO from C sites in Ref. \cite{Meskine2009} ($ E_{app}^{TOF} \approx \xi_{\lambda}^{} \Delta E_{\lambda} \approx 2 \times 1.3$ = 2.6 eV, resulting in an error of $0.25$ eV). In fact, such assignment contradicts the first paragraph of this section, which concludes that the RDS ($\lambda$) cannot be assigned to the desorption of CO (nor to its adsorption). According to Example S2 in Section \ref{Wrong_apparent_activation_energies_based_on_the_Temkin_formulation}, the value
$2 \times 1.3$ eV corresponds to the contribution $x \Delta H_{CO}$ with $x \approx 2$, due to the approximate dependence $TOF \propto \theta_{*}^2$, where $\theta_{*}^2$ describes the coverage by all empty pairs of sites in the homogeneous mixing approximation. Thus, the sensitivity value $ \xi_{\lambda}^{} \approx 2 $ for the desorption of CO in Fig. 5 of Ref. \cite{Meskine2009} might be related to the reaction order $ x \approx 2 $ for CO. Although the adsorption-desorption equilibrium for CO is a good approximation in this system, the accurate determination of the reaction order and/or the sensitivity seems a difficult task.

From our perspective, the desorption of CO from C sites ($CO_{\rm C} \rightarrow V_{\rm C}$) plays an important role in this system, essentially controlling the total rate per active site $R_{}=r/s$ in combination with the adsorption of CO at C sites ($V \rightarrow CO$) ($\omega_{\alpha}^{R} \sim 0.5$ for both processes in Fig. \ref{Figure05}(a) of the main report; see also Fig. \ref{Figure05}(d)). Thus, the two processes affect the $TOF$ by scaling the time increment $\Delta t \propto 1/\hat{r}_{}$ (see last paragraph of Section \ref{Theory} of the main report). However, neither the adsorption of CO nor its desorption are the RDS.

When the actual multiplicity for the adsorption of $O$ is carefully monitored, Fig. \ref{Figure04}(d) of the main report shows that $E_{app}^{TOF}$ is explained with great accuracy in all regions. This strongly indicates that monitoring the surface morphology should allow a deeper understanding of heterogeneous catalysis as an alternative to focusing on the determination of reaction orders and/or sensitivities.

\bibliography{gosalvez_bib}